\documentclass{article}
\usepackage{amssymb}

\usepackage{bookstyle,graphicx,epsf,bm,amsmath,amssymb}

\newcommand{\beq}{\begin{equation}}
\newcommand{\eeq}{\end{equation}}
\newcommand{\beqa}{\begin{eqnarray}}
\newcommand{\eeqa}{\end{eqnarray}}

\newcommand{\PRB}{Phys. Rev. B~}
\newcommand{\PRL}{Phys. Rev. Lett.~}

\begin{document}

\author{Dmitrii L. Maslov}
\runauthor{Dmitrii L. Maslov}
\address{Department of Physics, University of Florida,
P. O. Box 118441, Gainesville, FL 32611-8440, USA}
\title{Fundamental aspects of electron correlations and quantum transport in
one-dimensional systems}
\runtitle{Fundamental aspects of electron correlations and quantum transport}
\maketitle

\begin{abstract}
Some aspects of physics on interacting fermions in 1D are
discussed in a tutorial-like manner. We begin by  showing that the
non-analytic corrections to the Fermi-liquid forms of
thermodynamic quantities result from essentially 1D collisions
embedded into a higher-dimensional phase space. The role of these
collisions increases progressively as dimensionality is reduced
until, finally, they lead to a breakdown of the Fermi liquid in
1D. An exact solution of the Tomonaga-Luttinger model, based on
the Ward identities, is reviewed in the fermionic language.
Tunneling in a 1D interacting systems is discussed first in terms
of the scattering theory for interacting fermions and then via
bosonization. Universality of conductance quantization in
disorder-free quantum wires is discussed along with the breakdown
of this universality in the spin-incoherent case. A difference
between charge (universal) and thermal (non-universal)
conductances is explained in terms of Fabry-Perrot resonances of
charge plasmons.

\end{abstract}

%\frontmatter
%\mainmatter
\clearpage
%%%%%%%%%%%%%%%%%%%%%%%%%%%%
%%%%%%%%%%INTRO
%%%%%%%%%%%%%%%%%%%%%%%%%%%%%
\section{Introduction}
The theory of interacting fermions in one dimension (1D) has
survived several metamorphoses. From what seemed to be a purely
mathematical exercise up until the 60s, it had evolved into a
practical tool for predicting and describing phenomena in
conducting polymers and organic compounds--which were \emph{the
}1D systems of the 70s. Beginning from the early 90s, when the
progress in nanofabrication led to creation of artificial 1D
structures--quantum wires and carbon nanotubes, the theory of 1D
systems started its expansion into the domain of mesoscopics; this
trend promises to continue in the future. Given that there is
already quite a few excellent reviews and books on the subject
\cite{solyom}-\cite{giamarchi_book} , I should probably begin with
an explanation as to what makes this review different from the
others. First of all, it is not a review but--being almost a
verbatim transcript of the lectures given at the 2004  Summer
School in Les Houches--rather a tutorial on some (and  definitely
not all) aspects of 1D physics. A typical review on the subject
starts with describing the Fermi Liquid (FL) in higher dimensions
with an aim of emphasizing the differences between the FL and its
1D counter-part --Luttinger Liquid (LL). My goal--if defined only
after the manuscript was written--was rather to highlight the
\emph{similarities }between higher-D and 1D systems. The progress
in understanding of 1D systems has been facilitated tremendously
and advanced to a greater detail, as compared to higher
dimensions, by the availability of exact or asymptotically exact
methods (Bethe Ansatz, bosonization, conformal field theory),
which typically do not work too well above 1D.  The downside part
of this progress is that 1D effects, being studied by specifically
1D methods, look somewhat special and not obviously related to
higher dimensions. Actually, this is not true. Many effects that
are viewed as hallmarks of 1D physics, \emph{e.g., }the
suppression of the tunneling conductance by the electron-electron
interaction and the infrared catastrophe, do have higher-D
counter-parts and stem from essentially the same physics. For
example, scattering at Friedel oscillations caused by tunneling
barriers and impurities is responsible for zero-bias tunneling
anomalies in all dimensions \cite{YGM,altshuler}. The difference
is in the magnitude of the effect but not in its qualitative
nature. Following the tradition, I also start with the FL in Sec.
\ref{sec:fl}, but  the main message of this Section is that the
difference between $D=1$ and $D>1$ is not all that dramatic. In
particular, it is shown that the well-known non-analytic
corrections to the FL forms of thermodynamic quantities (such as a
venerable $T^{3}\ln T$-correction to the linear-in-$T$ specific
heat in 3D) stem from rare events of essentially 1D collisions
embedded into a higher-dimensional phase space. In this approach,
the difference between $D=1$ and $D>1$ is quantitative rather than
qualitative: as the dimensionality goes down, the
phase space has difficulties  suppressing the small-angle and $%
2k_{F}-$scattering events, which are responsible for
non-analyticities. The special point when these events go out of
control and start to dominate the physics happens to be in 1D.
This theme is continued in Sec.\ref{sec:ygm}, where scattering
from a single impurity embedded into a 1D system is analyzed in
the fermionic language, following the work by Yue, Matveev,
Glazman \cite{YGM}. The drawback of this approach--the
perturbative treatment of the interaction--is  compensated by the
clarity of underlying physics. Another feature which makes these
notes different from the rest of the literature in the field  is
that  the description goes in terms of the
original fermions for quite a while (Secs.\ref{sec:fl} through \ref{sec:ygm}%
) , whereas the weapon of choice of all 1D studies--bosonization--
is invoked only at a later stage (Sec. \ref{sec:bosonization} and
beyond). The rationale--again, a post-factum one--is two-fold.
First, 1D systems in a mesoscopic environment--which are the main
real-life application discussed here-- are invariably coupled to
the outside world via leads, gates, etc. As the outside world is
inhabited by real fermions, it is  sometimes easier to think of,
\emph{e.g., } both the interior and exterior a\emph{\ }quantum
wire coupled to reservoirs in terms of the same elementary
quasi-particles. Second, after 40 years or so of bosonization,
what could have been studied within a model of fermions with
\emph{linearized} dispersion and not too strong interaction--and
this is when bosonization works--was probably studied. (As all
statements of this kind, this is one is also an exaggeration.) The
last couple of years are characterized by a growing interest in
either the effects that do not occur in a model with linearized
dispersion, \emph{e.g., }Coulomb drag due to small-momentum
transfers \cite{drag} and energy relaxation, or situations when
strong Coulomb repulsion does not permit linearization of the
spectrum at any energies \cite {matveev,zvonarev,balents}.
Experiment seems to indicate that the Coulomb repulsion is strong
in most systems of interest, thus studies of a strongly-coupled
regime are quite timely. Once the assumption of the linear
spectrum is abandoned, the beauty of a bosonized description is by
and large lost,  and one might as well come back to original
fermions. Sec.\ref{sec:bosonization} is devoted to transport in
quantum wires, mostly in the absence of impurities. The
universality of conductance quantization is explained in some
detail, and is followed by a brief discussion of the recent result
by Matveev \cite{matveev}, who showed that incoherence in the spin
sector leads to a breakdown of the universality at higher
temperatures (Sec. \ref{sec:spin_incoh}). Also, a difference in
charge (universal) and thermal (non-universal)
transport--emphasized by Fazio, Hekking, and Khmelnitskii
\cite{fazio}-- in addressed in Sec.\ref {sec:thermal}. What is
missing is a discussion of transport in a disordered (as opposed
to a single-impurity) case. However, this canonically difficult
subject, which involves an interplay between localization and
interaction, is perhaps not ready for a tutorial-like discussion
at the moment. (For a recent development on this subject, see
Ref.\cite{gornyi}.)

Even a brief inspection of these notes shows that the choice
between making them comprehensive or self-contained was made for
the latter with a focus on a relatively small number of topics. It
is quite easy to see what is missing: there is no discussion of
lattice effects, bosonization is introduced without the Klein
factors, the sine-Gordon model is not treated in depth, chiral
Luttinger liquids are not discussed at all, and the list goes on.
The discussion of the experiment is scarce and perfunctory.
However, the few subjects that are discussed are provided with
quite a detailed--perhaps somewhat excessively detailed--
treatment, so that a reader may not feel a need to consult the
reference list too often. For the same reason, the notes also
cover such canonical procedures as the perturbative
renormalization group in the fermionic language (Sec.
\ref{sec:RG}) and elementary bosonization (Sec. \ref
{sec:bosonization}), which are discussed in many other sources and
a reader already familiar with the subject is encouraged to skip
them.

Also, a relatively small number of references (about one per page
on average) indicates once again that this is \emph{not }a review.
The choice of cited papers is subjective and the reference list in
no way pretends to represent a comprehensive bibliography to the
field. My apologies in advance to those whose contributions to the
field I have failed to acknowledge here.

$\hbar=k_B=1$ through out the notes, unless specified otherwise.
%%%%%%%%%%%%%%%%%%%%%%%%%%%%%%%%%%%%%%
%%%%%%%%%%LECTURE 1
%%%%%%%%%%%%%%%%%%%%%%%%%%%%%%%%%%%%%%%
\section{Non-Fermi liquid features of Fermi liquids: 1D physics in higher
dimensions}
\label{sec:fl}

One often hears the statement that, by and large, a Fermi liquid
(FL) is just a Fermi gas of weakly interacting quasi-particles;
the only difference being the renormalization of the essential
parameters (effective mass, $g-$ factor) by the interactions. What
is missing here is that the similarity between the FL and Fermi
gas holds only for leading terms in the expansion of the
thermodynamic quantities (specific heat $C(T)$, spin
susceptibility $\chi _{s}$, etc.) in the energy (temperature) or
spatial (momentum) scales. Next-to-leading terms, although
subleading, are singular (non-analytic) and, upon a closer
inspection, reveal a rich physics of essentially 1D scattering
processes, embedded into a high-dimensional phase space.

In this chapter, I will discuss the difference between ``normal''
processes which lead to the leading, FL forms of thermodynamic
quantities and ``rare'' 1D processes which are responsible for the
non-analytic behavior. We will see that the role of these rare
processes increases as the dimensionality is reduced and,
eventually, the rare processes become the norm in 1D, where the FL
breaks down.

In a Fermi gas, thermodynamic quantities form regular, analytic
series as function of either temperature, $T,$ or the inverse
spatial scale (bosonic
momentum $q)$ of an inhomogeneous magnetic field. For $T\ll E_{F},$ where $%
E_{F}$ is the Fermi energy, and $q\ll k_{F},$ where $k_{F}$ is the
Fermi momentum, we have
\begin{subequations}
\begin{eqnarray}
C(T)/T &=&\gamma +aT^{2}+bT^{4}+\dots ;  \label{fg1} \\
\chi _{s}\left( T,q=0\right) &=&\chi _{s}^{0}(0)+cT^{2}+dT^{4}+\dots ;
\label{fg2} \\
\chi _{s}\left( T=0,q\right) &=&\chi _{s}^{0}(0)+eq^{2}+fq^{4}+\dots ,
\label{fg3}
\end{eqnarray}
\end{subequations}
where $\gamma $ is the Sommerfeld constant, $\chi _{s}^{0}$ is the static,
zero-temperature spin susceptibility (which is finite in the Fermi gas), and
$a\dots f$ are some constants. Even powers of $T$ occur because of the
approximate particle-hole symmetry of the Fermi function around the Fermi
energy and even powers of $q$ arise because of the analyticity requirement \footnote{%
The expressions presented above are valid in all dimensions, except for $%
D=2$ with quadratic dispersion. There, because the density of
states (DoS) does not depend on energy, the
leading correction to the $\gamma T-$ term in $C(T)$ is exponential in $%
E_{F}/T$ and $\chi _{s}$ does not depend on $q$ for $q\leq 2k_{F}.$ However,
this anomaly is removed as soon as we take into account a finite bandwidth of
the electron spectrum,
upon which the universal ($T^{2n}$ and $q^{2n})$ behavior of the series is
restored.}

Our knowledge of the interacting systems comes from two sources.
For a system with repulsive interactions, one can assume that as
long as the strength of the interaction does not exceed some
critical value, none of the symmetries (translational invariance,
time-reversal, spin-rotation, etc.), inherent to the original
Fermi gas, are broken. In this range, the FL theory is supposed to
work. However, the FL theory is an asymptotically low-energy
theory by construction, and it is really suitable only for
extracting the leading terms, corresponding to the first terms in
the Fermi-gas expressions (\ref{fg1}-\ref{fg3}). Indeed, the free
energy of a FL as an ensemble of quasi-particles interacting in a
pair-wise manner can be written as \cite{pines}

\begin{equation}
F-F_{0}=\sum_{k}\left( \epsilon _{k}-\mu \right) \delta n_{k}+\frac{1}{2}%
\sum_{k,k^{\prime }}f_{k,k^{\prime }}\delta n_{k}\delta n_{k^{\prime
}}+O\left( \delta n_{k}^{3}\right) ,  \label{fe_fl}
\end{equation}
where $F_{0}$ is the ground state energy, $\delta n_{k}$ is the deviation of
the fermion occupation number from its ground-state value, and $%
f_{k,k^{\prime }}$ is the Landau interaction function. As $\delta
n_{k}$ is of the order of $T/E_{F},$ the free energy is at most
quadratic in $T,$ and therefore the corresponding $C(T)$ is at
most linear in $T.$ Consequently, the FL theory--at least, in the
conventional formulation--claims only that
\begin{eqnarray*}
C^{\ast }(T)/T &=&\gamma ^{\ast }; \\
\chi _{s}^{\ast }\left( T,q\right)  &=&\chi _{s}^{\ast }\left( 0\right) ,
\end{eqnarray*}
where $\gamma ^{\ast }$ and $\chi _{s}^{\ast }\left( 0\right) $ differ from
the corresponding Fermi-gas values, and does not say anything about
higher-order terms \footnote{%
Strictly speaking, non-analytic terms in $C(T)$ can be obtained
from the free energy (\ref{fe_fl}) by taking into account the
non-analytic terms in the quasi-particle spectrum, see Ref.
\cite{amit}b.}.

\begin{figure}[tbp]
\begin{center}
\epsfxsize=0.7 \columnwidth
%\epsffile{fig1lanie.eps}
\epsffile{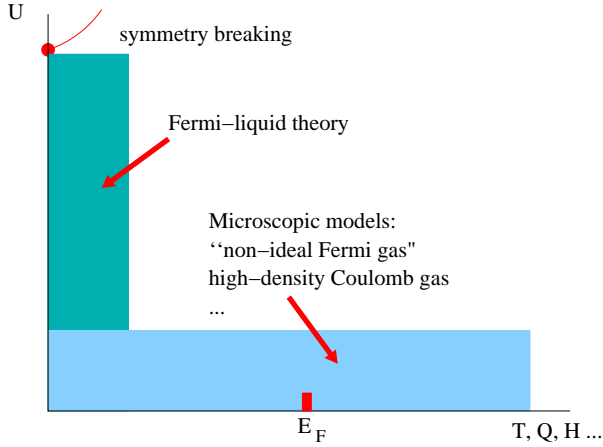}
\end{center}
\caption{Combined ``diagram of knowledge". x-axis: energy scale
(given by temperature $T$, bosonic momentum $Q$, magnetic field
$H$) in appropriate units. y-axis: interaction strength. Fermi
liquid works for not necessarily weak interactions (but smaller
than the critical value for an instability of the ground state
denoted by the red dot) but at the lowest energy scales.
Microscopic models work for weak interactions but for arbitrary
energies.} \label{fig:diagram}
\end{figure}

Higher-order terms in $T$ or $q$ can be obtained within microscopic models
which specify particular interaction and, if an exact solution is
impossible--which is always the case in higher dimensions-- employ some kind
of a perturbation theory. Such an approach is complementary to the FL: the
former nominally works for weak interactions \footnote{%
Some results of the perturbation theory can be rigorously extended
to an infinite order in the interaction, and most of them can be
guaranteed to hold even if the interactions are not weak.} but at
arbitrary temperatures, whereas FL works both for weak and strong
interactions, up to some critical value corresponding to an
instability of some kind, \emph{e.g.}, a ferromagnetic transition,
but only in the low-temperature limit. In the $\{$ temperature,
interaction$\}$ plane, the validity regions of these two
approaches are two strips running along the two axes (cf.
Fig.~\ref{fig:diagram}). For weak interactions and at low
temperatures, the regions should overlap.

Microscopic models (Fermi gas with weak repulsion, Coulomb gas in
the high-density limit, electron-phonon interaction, paramagnon
model, etc.) show that the higher-order terms in the specific heat
and spin susceptibility are non-analytic functions of $T$ and $q$
\cite
{eliashberg,doniach,brinkman,amit,bedell,belitz,marenko,millis,chm_long,pepin,dassarma,zerosound,gal_chub}%
. For example,
\begin{subequations}
\begin{eqnarray}
C(T)/T &=&\gamma _{3}-\alpha _{3}T^{2}\ln T\text{\textrm{\ (3D);}}
\label{2d3d_1} \\
C(T)/T &=&\gamma _{2} -\alpha _{2}T\text{\textrm{\  (2D) ;}}  \label{2d3d_2} \\
\chi _{s}\left( q\right) &=&\chi _{s}(0)+\beta _{3}q^{2}\ln q^{-1}\text{%
\textrm{\  (3D);}}  \label{2d3d_3} \\
\chi _{s}\left( q\right) &=&\chi _{s}(0)+\beta _{2}\left| q\right| \,\text{%
\textrm{\  (2D),}}  \label{2d3d}
\end{eqnarray}
\end{subequations}
where all coefficients are positive for the case of repulsive electron-electron
interaction \footnote{%
Notice that not only the functional forms but also the \textbf{sign }of the $%
q-$ dependent term in the spin susceptibility is different for
free and interacting systems. ``Wrong'' sign of the $q-$ dependent
corrections has far-reaching consequences for quantum critical
phenomena. For example, it precludes a possibility of a
second-order, homogeneous quantum ferromagnetic phase transition
in an itinerant system \cite{belitzRMP}. What is possible is
either a first-order transition or ordering at finite $q$, {\emph
e.g.} helical structure. In 1D, a homogeneous ferromagnetic state
is forbidden by the Lieb-Mattis theorem \cite{lieb}, which states
that the ground state of 1D fermions, interacting via
spin-independent but otherwise arbitrary forces, is non-magnetic.
One could speculate that the non-analyticities in higher
dimensions indicate the existence of a higher-$D$ version of the
Lieb-Mattis theorem. Certainly, this does not mean that
ferromagnetism does not exist in higher dimensions (it is hard to
deny the existence of, \emph{e.g.}, iron). However, ferromagnetism
may not exist in {\emph models} dealing \emph{only} with itinerant
electrons in continuum.}

As seen from Eqs.~(\ref{2d3d_1}-\ref{2d3d}), the non-analyticities
become stronger as the dimensionality is reduced. The strongest
non-analyticity occurs in 1D, where-- as far as single-particle
properties are concerned--the FL breaks down:
\begin{eqnarray*}
C(T)/T &=&\gamma _{1}+\alpha _{1}\ln T\text{\textrm{\ \ (1D);}} \\
\chi \left( q\right) &=&\chi _{0}+\beta _{1}\ln \left| q\right| \mathrm{\ \
(1D).}
\end{eqnarray*}

It turns out that the evolution of the non-analytic behavior with the
dimensionality reflects an increasing role of special, almost 1D scattering
processes in higher dimensions. Thus non-analyticities in higher dimensions
can be viewed as precursors of 1D physics for $D>1.$

It is easier to start with the non-analytic behavior of a
single-particle property, the self-energy, which can be related to
the thermodynamic quantities via standard means \cite{agd} (see
also appendix A). Within the Fermi liquid,
\begin{subequations}
\begin{eqnarray}
\mathrm{Re}\Sigma ^{R}\left( \varepsilon ,k\right) &=&-A\varepsilon +B\xi
_{k}+\dots  \label{resigma} \\
-\mathrm{Im}\Sigma ^{R}\left( \varepsilon ,k\right) &=&C(\varepsilon
^{2}+\pi ^{2}T^{2})+\dots  \label{imsigma}
\end{eqnarray}
\end{subequations}
Expressions (\ref{resigma}) and (\ref{imsigma}) are equivalent to two
statements: i) quasi-particles have a finite effective mass near the Fermi
level

\[
m^{\ast }=m_{0}\frac{A+1}{B+1},
\]
and ii) damping of quasiparticles is weak: the level width is much smaller
than the typical quasi-particle energy

\[
\Gamma =-2\mathrm{Im}\Sigma ^{R}\left( \varepsilon ,k\right) \propto \max
\left\{ \left| \varepsilon \right| ^{2},T^{2}\right\} \ll \left| \varepsilon
\right| ,T.
\]

\begin{figure}[tbp]
\begin{center}
\epsfxsize=0.6 \columnwidth
\epsffile{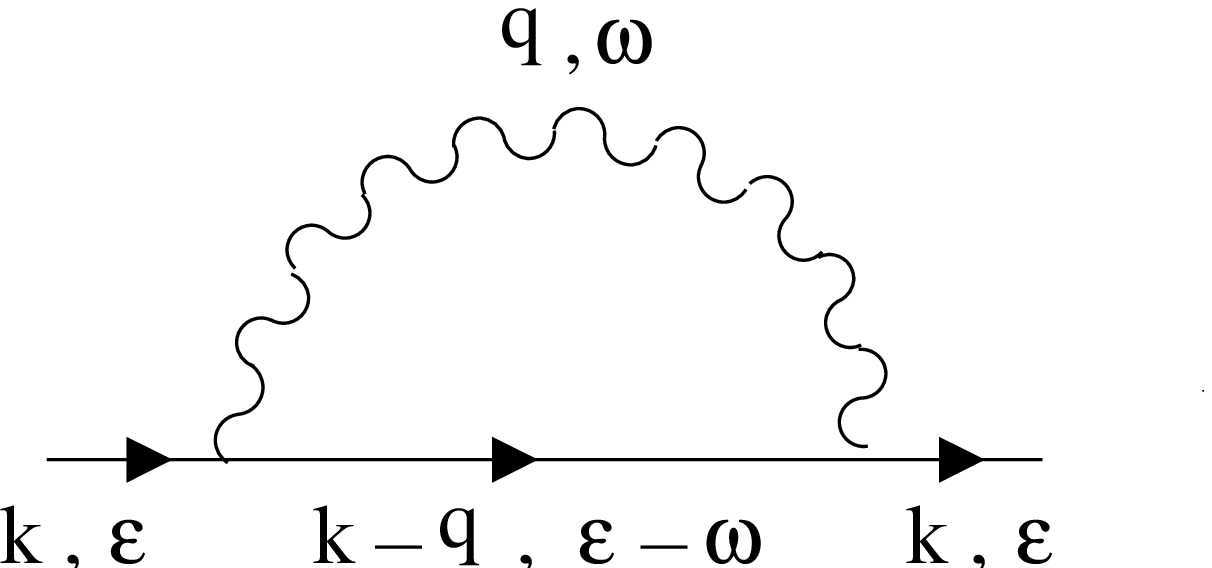}
\end{center}
\caption{ Self-energy to first order in the interaction with a
dynamic bosonic field.} \label{fig:sigma1order}
\end{figure}

    Landau's argument for the $\varepsilon ^{2}$ (or $T^{2})$ behavior of $%
\mathrm{Im}\Sigma ^{R}$ relies on the Fermi statistics of quasiparticles and
on the assumption that the effective interaction is screened at large
distances \cite{agd}. It requires two conditions. One condition is obvious:
the temperature has to be much smaller than the degeneracy temperature $%
T_{F}=k_{F}v_{F}^{\ast }$, where $v_{F}^{\ast }$ is the renormalized Fermi
velocity. The other condition is less obvious: \emph{\ }it requires
inter-particle scattering to be dominated by processes with large
(generically, of order $k_{F})$ momentum transfers. Once these two
conditions are satisfied, the self-energy assumes a universal form, Eqs.~(%
\ref{resigma}) and ( \ref{imsigma}), \emph{regardless of a specific type of
the interaction (e-e, e-ph) and dimensionality}. To see this, let's have a
look at $\mathrm{Im}\Sigma ^{R}\left( \varepsilon ,k\right) $ due to the
interaction with some ``boson'' (Fig.~\ref{fig:sigma1order}).

The wavy line in Fig.\ref{fig:sigma1order} can be, \emph{e.g.,} a
dynamic Coulomb interaction, phonon propagator, etc. On the mass
shell ($\varepsilon =\xi _{k}$) at $T=0$ and for $\varepsilon >0,$
we have \footnote{%
To get Eq.~(\ref{imsigma_1}), one can start with the Matsubara
form of diagram Fig.~\ref{fig:sigma1order}, convert the Matsubara
sums into the contour integrals, use the dispersion relation
\[
D^{R}(\varepsilon )=\frac{1}{\pi }\int_{-\infty }^{\infty
}d\varepsilon
^{\prime }\frac{\text{Im}D^{R}\left( \varepsilon ^{\prime }\right) }{%
\varepsilon ^{\prime }-\varepsilon -i0^{+}},
\]
which is valid for any retarded function, and take the limit
$T\rightarrow 0. $}
\begin{equation}
\mathrm{Im}\Sigma ^{R}\left( \varepsilon \right) =-\frac{2}{\left( 2\pi
\right) ^{D+1}}\int_{0}^{\varepsilon }d\omega \int d^{d}q\text{\textrm{Im}}%
G^{R}\left( \varepsilon -\omega ,\mathbf{k}-\mathbf{q}\right) \text{\textrm{%
Im}}V^{R}\left( \omega ,\mathbf{q}\right).  \label{imsigma_1}
\end{equation}
The constraint on energy transfers ($0<\omega <\varepsilon $) is a
direct manifestation of the Pauli principle which limits the
number of accessible energy levels. In real space and time,
$V\left( r,t\right) $ is a propagator of some field which has a
classical limit (when the occupation numbers of all modes are
large). Therefore, $V\left( r,t\right) $ is a real function, hence
Im$V$ is an odd function of $\omega .$ I will make this fact
explicit writing Im$V$ as
\[
\mathrm{Im}V^{R}\left( \omega ,q\right) =\omega W\left( \left| \omega
\right| ,q\right) .
\]
Now, suppose that we integrate over $q$ and the result does not depend on $%
\omega $. Then we immediately get
\[
-\mathrm{Im}\Sigma ^{R}\left( \varepsilon \right) \sim
C\int_{0}^{\varepsilon }d\omega \omega \sim C\varepsilon ^{2},
\]
where $C$ is the result of the $q-$ integration which contains all
the
information about the interaction. Once we got the $\varepsilon ^{2}$-form for $%
\mathrm{Im}\Sigma ^{R}\left( \varepsilon \right) ,$ the $\varepsilon $- term
in $\mathrm{Re}\Sigma ^{R}\left( \varepsilon \right) $ follows immediately
from the Kramers-Kronig transformation, and we have a Fermi-liquid form of
the self-energy regardless of a particular interaction and dimensionality.
Thus a sufficient condition for the Fermi liquid is the \emph{separability }%
of the frequency and momentum integrations, which can only happen if the
energy and momentum transfers are decoupled.

Now, what is the condition for separability? As a function of $q,$ $W$ has
at least two characteristic scales. One is provided by the internal
structure of the interaction (screening wavevector for the Coulomb
potential, Debye wavevector for electron-phonon interaction, etc.) or by $%
k_{F},$ whichever is smaller. This scale (let's call it $Q)$ does
not depend on $\omega .$ Moreover, as $\left| \omega \right| $ is
bounded from above by $\varepsilon ,$ and we are interested in the
limit $\varepsilon \rightarrow 0,$ one can safely assume that
$Q\gg \left| \omega \right| /v_{F}.$ The role of $Q$ is just to
guarantee the convergence of the momentum integral in the
ultraviolet, that is, to ensure that for $q\gg Q$ the integrand
falls off rapidly enough. Any physical interaction will have this
property as larger momentum transfer will have smaller weight. The
other scale is $\left| \omega \right| /v_{F}.$ Now, let's
summarize this by re-writing Im$V$ in the following scaling form
\[
\mathrm{Im}V^{R}\left( \omega ,q\right) =\omega \frac{1}{Q^{D}}U\left( \frac{%
\left| \omega \right| }{v_{F}Q},\frac{q}{Q}\right) ,
\]
where $U$ is a dimensionless function and the factor $Q^{-D}$ was
singled out to keep the units right.

\begin{figure}[tbp]
\begin{center}
\epsfxsize=1.0 \columnwidth
\epsffile{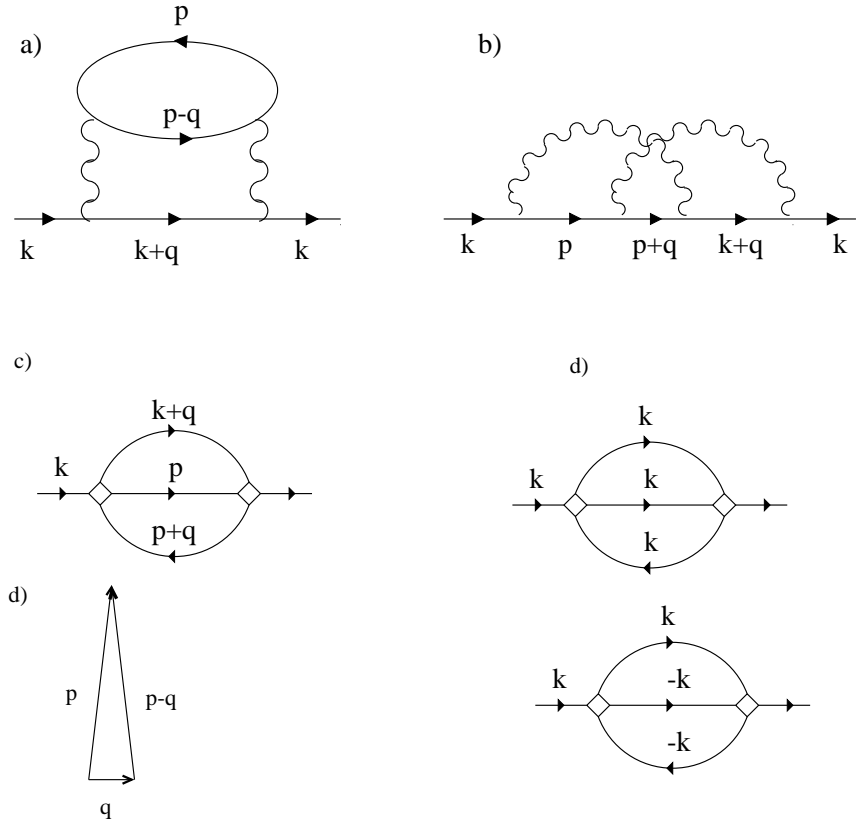}
\end{center}
\caption{ a) and b) Non-trivial second order diagrams for the
self-energy. c) Same diagrams as in a) and b) re-drawn as a single
``sunrise" diagram. d) Diagrams relevant for non-analytic terms in
the self-energy. e) Kinematics of
scattering in a polarization bubble: the dynamic part $\Pi\propto \protect%
\omega/v_Fq$ comes from the processes in which the internal fermionic
momentum (${\vec p}$) is almost perpendicular to the external bosonic one (${%
\vec q}$).}
\label{fig:selfenergy_2nd}
\end{figure}
In the perturbation theory, the Green's function in
(\ref{imsigma_1}) is a free one. Assuming the free-electron
spectrum $\xi _{k}=(k^{2}-k_{F}^{2})/2m$,
\[
\mathrm{Im}G^{R}\left( \varepsilon -\omega ,\vec{k}-\vec{q}\right) =-\pi
\delta \left( \varepsilon -\omega -\xi _{k}+\vec{v}_{k}\cdot \vec{q}%
-q^{2}/2m\right).
\]
On the mass shell,
\[
\mathrm{Im}G^{R}\left( \varepsilon -\omega ,\vec{k}-\vec{q}\right)
|_{\varepsilon =\xi _{k}}=-\pi \delta \left( \omega -\vec{v}_{k}\cdot \vec{q}%
+q^{2}/2m\right).
\]
The argument of the delta-function simply expresses the energy and momentum
conservation for a process $\varepsilon \rightarrow \varepsilon -\omega ,%
\vec{k}\rightarrow \vec{k}-\vec{q}.$ The angular integral involves only the
delta-function. For any $D,$ this integral gives
\[
\langle \delta \left( \dots \right) \rangle _{\Omega }=\frac{1}{v_{F}q}%
A_{D}\left( \frac{\omega +q^{2}/2m}{v_{F}q}\right) ,
\]
where $v_{k}$ was replaced by $v_{F}$ because all the action takes
place near the Fermi surface. For $D=3$ and $D=2$,
\begin{eqnarray*}
A_{3}\left( x\right) &=&2\theta (1-|x|); \\
A_{2}\left( x\right) &=&\frac{2\theta \left( 1-\left| x\right| \right) }{%
\sqrt{1-x^{2}}}.
\end{eqnarray*}
The constraint on the argument of $A_{D}$ is purely geometric: the magnitude
of the cosine of the angle between $\vec{k}$ and $\vec{q}$ has to be less
then one. For power-counting purposes, function $A_{D}$ has a dimensionality
of 1. Therefore, its only role is to provide a lower cut-off for the
momentum integral. Then, by power counting
\begin{equation}
\mathrm{Im}\Sigma ^{R}\left( \varepsilon \right) \sim \frac{1}{v_{F}Q^{D}}%
\int_{0}^{\varepsilon }d\omega \omega \int_{q\geq \left| \omega \right|
/v_{F}}dqq^{D-2}U\left( \frac{\left| \omega \right| }{v_{F}q},\frac{q}{Q}%
\right) .  \label{imsigma_2}
\end{equation}
Now, if the integral over $q$ is dominated by $q\sim Q$ and is
convergent in the \emph{infrared}, one can put $\omega =0$ in this
integral. After this step, the integrals over $\omega $ and $q$
decouple. The $\omega -$ integral gives $\varepsilon^{2}$
regardless of the nature of the interaction and dimensionality
whereas the $q-$ integral supplies a prefactor which entails all
the details of the interaction
\[
\mathrm{Im}\Sigma ^{R}\left( \varepsilon \right) =C_{D}\frac{\varepsilon ^{2}%
}{v_{F}Q}.
\]
For example, for a screened Coulomb interaction in the weak-coupling
(high-density) limit $Q=\kappa $, where $\kappa $ is the screening
wavevector, we have in 3D
\[
-\mathrm{Im}\Sigma ^{R}\left( \varepsilon \right) =\frac{\pi ^{2}}{64}\frac{%
\kappa }{k_{F}}\frac{\varepsilon ^{2}}{E_{F}}.
\]
Now we can formulate a sufficient (but not necessary) condition
for the Fermi-liquid behavior. It will occur whenever if
kinematics of scattering is such that the typical momentum
transfers are determined by some internal and, what is crucial,
$\omega -$ independent scale, whereas the energy transfers are of
order of the quasi-particle energy (or temperature). Excluding
special situations, such as the high-density limit of the Coulomb interaction, $%
Q $ is generically of order of the ultraviolet range of the problem $\sim
k_{F}.$ In other words, isotropic scattering guarantees a $\varepsilon ^{2}$%
- behavior. Small-angle scattering with typical angles of order $\varepsilon
/v_{F}\ll Q\ll k_{F}$ gives this behavior as well.

The $\varepsilon ^{2}-$ result seems to be quite general under the
assumptions made. When and why these assumptions  are violated?

A long-range interaction, associated with small-angle scattering,
is known to destroy the FL. For example, transverse long-range
(current-current \cite {current} or gauge \cite{gauge})
interactions, which--unlike the Coulomb one--are not screened by
electrons, lead to the breakdown of the Fermi liquid. However, the
current-current interaction is of the relativistic origin and
hence does the trick only at relativistically small energy scales,
whereas the gauge interaction occurs only under special
circumstances, such as near half-filling or for composite
fermions. What about a generic case when nothing of this kind
happens? It turns out that even if the bare interaction is of the
most benign form, \emph{e.g.}, a delta-function in real space,
there are deviations from a (perceived) FL behavior. These
deviations get amplified as the system dimensionality is lowered,
and, eventually, lead to a complete breakdown of the FL in 1D.

A formal reason for the deviation from the FL-behavior is that the
argument which led us to the $\varepsilon ^{2}$-term is good only
in the leading order in $\omega /qv_{F}.$ Recall that the angular
integration gives us $q^{-1}$ factors in all dimensions, and, to
arrive at the $\varepsilon ^{2}$ result we put $\omega =0$ in functions $%
A_{D}$ and $U.$ If we want to get a next term in $\varepsilon ,$
then we need to expand $U$ and $A$ in $\omega .$ Had such
expansions generated regular series, Im$\Sigma ^{R}$ would have
also formed regular series in $\varepsilon ^{2}$: Im$\Sigma
^{R}=a\varepsilon ^{2}+b\varepsilon ^{4}+c\varepsilon ^{6}+\dots
.$ However, each factor of $\omega $ comes with $q^{-1}$, so that
no matter how high the dimensionality is, at some order of $\omega
/v_{F}q$ we are bound to have an infrared divergence.

\subsection{Long-range effective interaction}

\begin{figure}[tbp]
\begin{center}
\epsfxsize=1.0 \columnwidth
\epsffile{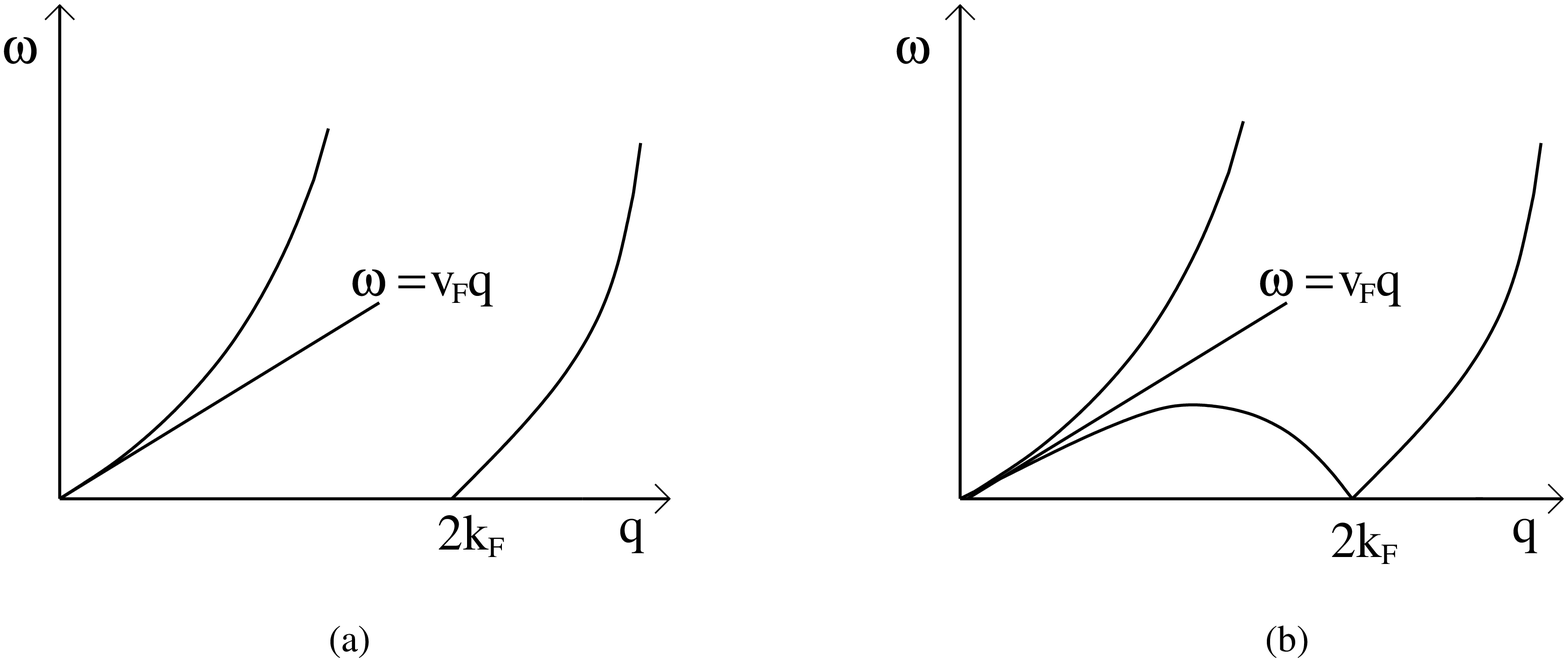}
\end{center}
\caption{Particle-hole continua for $D>1$ (left) and $D=1$
(right). For the 1D case, only half of the continuum ($q>0$) is
shown.} \label{fig:cont}
\end{figure}

Let's look at the simplest case of a point-like interaction. A
frequency dependence of the self-energy arises already at the
second order. At this order, two diagrams in
Fig.~\ref{fig:selfenergy_2nd} are of interest to us. For a contact
interaction, diagram b) is just -1/2 of a) (which can be seen by
integrating over the fermionic momentum $\vec{p}$ first), so we
will lump them together. Two given fermions interact via
polarizing the medium consisting of other fermions. Hence, the
effective interaction at the second order is just proportional to
the polarization bubble
\[
\mathrm{Im}V^{R}(\omega ,q)=-U^{2}\mathrm{Im}\Pi ^{R}(\omega ,q).
\]
Let's focus on small angle-scattering first: $q\ll 2k_{F}$. It
turns out that in all three dimensions, the bubble has a similar
form (see Appendix \ref{sec:pi_anyD} for an explicit derivation of
this result)
\begin{equation}
-\mathrm{Im}\Pi ^{R}(\omega ,q)= \nu _{D}\frac{\omega }{v_{F}q}%
B_{D}\left( \frac{\omega }{v_{F}q}\right) ,  \label{bubble}
\end{equation}
where $\nu _{D}=a_{D}mk_{F}^{D-2}$ is the DoS in $D$ dimensions
[$a_{3}=(2\pi) ^{-2},$ $a_{2}=(2\pi)^{-1},$ $a_{1}=1/2\pi$] and
$B_{D}$
is a dimensionless function, whose main role is to impose a constraint $%
\omega \leq v_{F}q$ in 2D and 3D and $\omega =v_{F}q$ in 1D. \textbf{\ }Eq.~%
\textbf{(}\ref{bubble}\textbf{) }entails the physics of
\emph{Landau damping.} The constraint arises because collective
excitations--charge- and spin-density waves-- decay into
particle-hole pairs. Decay occurs only if bosonic momentum and
frequency ($q$ and $\omega $) are within the particle-hole
continuum (cf. Fig.~\ref{fig:cont}).
For $D>1,$ the boundary of the continuum for small $\omega $ and
$q$ is $\omega =v_{F}q$, hence the decay takes place if $\omega
<v_{F}q.$ The rest of Eq.~(\ref{bubble}) can be understood by
dimensional analysis. Indeed, $\Pi ^{R}$ is the retarded
density-density correlation function; hence, by the same argument
we applied to Im$V^{R},$ its imaginary part must be odd in $\omega
.$ For $q\ll k_{F},$ the only combination of units of frequency is
$v_{F}q,$ and the frequency enters as $\omega /v_{F}q.$ Finally, a
factor $\nu _{D}$ makes the overall units right. In 1D, the
difference is that the continuum shrinks to a single line $\omega
=v_{F}q,$ hence decay of collective excitations is possible only
on this line. In 3D, function $B_{3}$ is simply a $\theta -$
function
\[
\mathrm{Im}\Pi ^{R}(\omega ,q)=-\nu _{3}\frac{\omega }{v_{F}q}\theta \left(
q-\left| \omega /v_{F}\right| \right) .
\]

\begin{figure}[tbp]
\begin{center}
\epsfxsize=0.7 \columnwidth
%\epsffile{fig3lanie.eps}
\epsffile{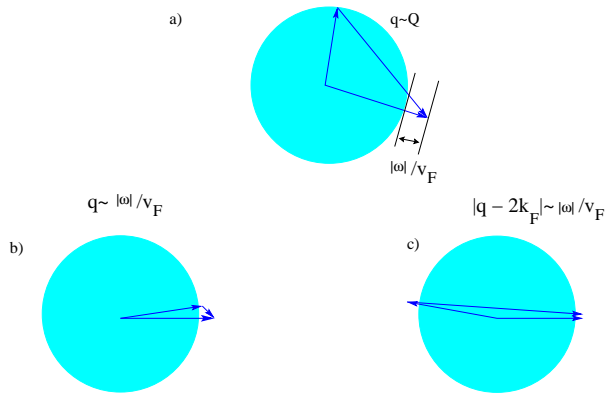}
\end{center}
\caption{Kinematics of scattering. a) ``Any-angle'' scattering.
Momentum transfer $q$ is of order of the intrinsic scale of the
interaction or $k_F$, whichever is smaller, and is independent of
the energy transfer, $\omega$, which is of order of the initial
energy $\protect\epsilon$. This process contributes regular terms
to the self-energy. b) Dynamic forward scattering: $q\sim |\protect\omega|%
/v_F$. c) Dynamic backscattering: $|q-2k_F|\sim
|\protect\omega|/v_F$. Processes b) and c) are responsible for the
non-analytic terms in the self-energy.} \label{fig:kinematics}
\end{figure}

\begin{figure}[tbp]
\begin{center}
\epsfxsize=1.0 \columnwidth \epsffile{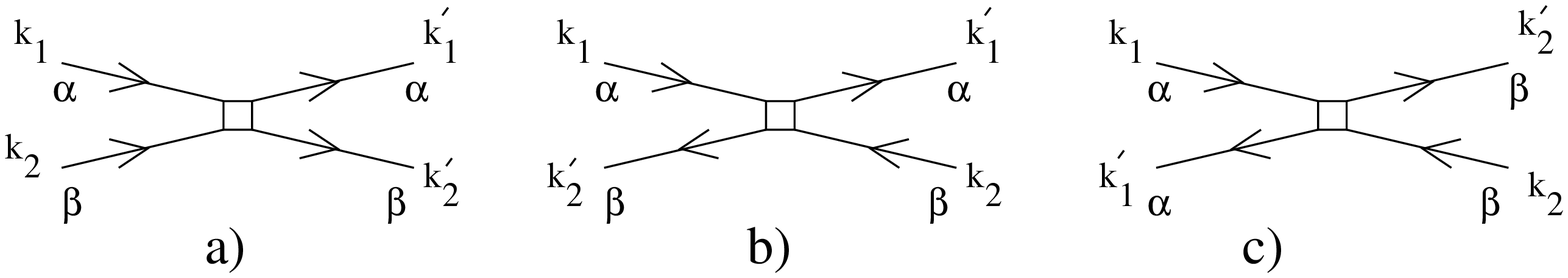}
\end{center}
\caption{ Scattering processes responsible for divergent and/or non-analytic
corrections to the self-energy in 2D. a) ``Forward scattering''--an analog
of the ``$g_{4}$''-process in 1D. All four fermionic momenta are close to
each other. b) Backscattering--an analog of the ``$g_{2}$''-process in 1D.
The net momentum before and after collision is small. Initial momenta are
close to final ones. Although the momentum transfer in such a process is
small, we still refer to this process as ``backscattering'' (see the
discussion in the main text). c) $2k_{F}-$ scattering. }
\label{fig:processes}
\end{figure}

Next-to-leading term in the expansion of Im$\Sigma ^{R}$ in
$\varepsilon $ comes from retaining the lower limit in the
momentum integral of Eq.~(\ref {imsigma_2}), upon which we get
\begin{eqnarray*}
-\mathrm{Im}\Sigma ^{R} &\sim &U^{2}mk_{F}\int_{0}^{\varepsilon }d\omega
\int_{\omega /v_{F}}^{Q\sim k_{F}}dqq^{2}\frac{1}{v_{F}q}\frac{\omega }{%
v_{F}q} \\
&\sim &U^{2}\frac{mk_{F}}{v_{F}^{2}}\int_{0}^{\varepsilon }d\omega \omega %
\left[ \underbrace{k_{F}}_{\text{FL }}-\underbrace{\frac{\omega }{v_{F}}}_{%
\text{beyond FL}}\right] \\
&\sim &a\varepsilon ^{2}-b\left| \varepsilon \right| ^{3}.
\end{eqnarray*}
The first term in the square brackets is the FL contribution that comes from
$q\sim Q.$ The second term is a correction to the FL coming from $q\sim
\omega /v_{F}.$ Thus, contrary to a naive expectation an expansion in $%
\varepsilon $ is \emph{non-analytic.} The fraction of  phase space
for small-angle scattering is small--most of the self-energy comes
from large-angle scattering events ($q\sim Q$); but we already
start to see the importance of the small-angle processes. Applying
Kramers-Kronig transformation to the non-analytic part ($\left|
\varepsilon \right| ^{3})$ in Im$\Sigma ^{R},$ we get a
corresponding non-analytic contribution to the real part as
\[
\left( \mathrm{Re}\Sigma ^{R}\right) _{\mathrm{non-an}}\propto \varepsilon
^{3}\ln \left| \varepsilon \right| .
\]
Correspondingly, specific heat which, by power counting, is obtained from Re$%
\Sigma ^{R}$ by replacing each $\varepsilon $ by $T$$,$ also acquires a
non-analytic term\footnote{%
One has to be careful with the argument, as a general relation
between $C(T)$ and the single-particle Green's function \cite{agd}
involves the self-energy on the mass shell. In 3D, the
contribution to $\Sigma$ from forward scattering, as defined in
Fig.~\ref{fig:processes}, vanishes on the mass shell; hence there
is no contribution to $C(T)$ \cite{chmm}. The non-analytic part of
$C(T)$ is related to the backscattering part of the self-energy
(scattering of fermions with small total momentum), which remains
finite on the mass shell. That forward scattering does not
contribute to non-analyticities in thermodynamics is a general
property of all dimensions, which can be understood on the basis
of gauge-invariance \cite{catelani}.}
\[
C(T)=\gamma _{3}T+\beta _{3}T^{3}\ln T.
\]
This is the familiar $T^{3}\ln T$ term, observed both in He$^{3}$
\cite{he3_3D} and metals \cite {stewart} (mostly, heavy-fermion
materials)
\footnote{%
The $T^{3}\ln T$-term in the specific heat coming from the
electron-electron interactions is often referred to in the
literature as to the ``spin-fluctuation'' or ``paramagnon''
contribution \cite{doniach,brinkman}. Whereas it is true that this
term is enhanced in the vicinity of a ferromagnetic (Stoner)
instability, it exists even far way from any critical point and
arises already at the second order in the interaction
\cite{amit}.}.

In 2D, the situation is more dramatic. The $q-$ integral diverges now
logarithmically in the infrared:
\begin{eqnarray*}
-\mathrm{Im}\Sigma ^{R}\left( \omega \right) &\sim &\frac{U^{2}}{v_{F}^{2}}%
m\int_{0}^{\varepsilon }d\omega \omega \int_{\sim \left| \omega \right|
/v_{F}}^{\sim k_{F}}\frac{dq}{q} \\
&\sim &\frac{U^{2}}{v_{F}}m\varepsilon ^{2}\ln \frac{E_{F}}{\left|
\varepsilon \right| }.
\end{eqnarray*}
Now, dynamic forward-scattering (with transfers $q\sim \omega
/v_{F})$ is not a perturbation anymore: on the contrary, the
$\varepsilon $ dependence of Im$\Sigma ^{R}$ is dominated by
forward scattering (the $\varepsilon ^{2}\ln \left| \varepsilon
\right|$-term is larger than the ``any-angle'' $\varepsilon
^{2}$-contribution ). Correspondingly, the real part acquires a
non-analytic term Re$\Sigma \propto \varepsilon \left| \varepsilon
\right|$, and the specific heat behaves as \footnote{again, only
processes with small total momentum contribute}
\[
C(T)=\gamma _{2}T-\beta _{2}T^{2}.
\]
The non-analytic $T^{2}$-term in the specific heat has been
observed in recent experiments on monolayers of He$^{3}$ adsorbed
on a solid substrate \cite
{he3_2D}\footnote{%
If a $T^{2}$ term in $C(T)$ does not fit your definition of
non-analyticity, you have to recall that the right quantity to
look at is the ratio $C(T)/T.$ Analytic behavior corresponds to
series $C(T)/T=\gamma +\delta T^{2}+\sigma T^{4}+\dots .$ whereas
we have a $T^{2}\ln T$ and $T$ terms as the leading order
corrections to the Sommerfeld constant $\gamma $ for $D=3$ and
$D=2,$ correspondingly.}.

\begin{figure}[tbp]
\begin{center}
\epsfxsize=0.8 \columnwidth
\epsffile{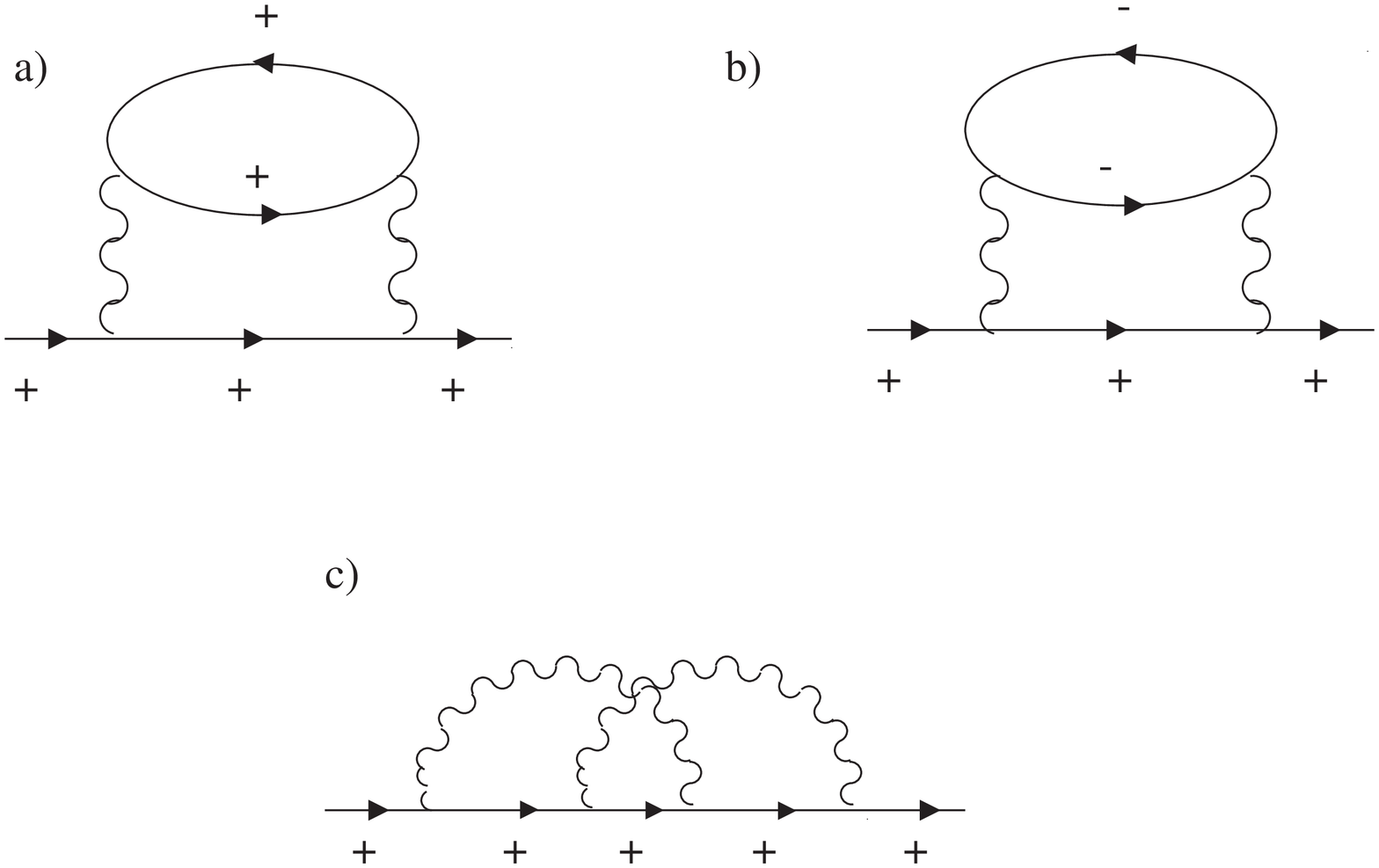}
\end{center}
\caption{ Self-energy in 1D. $\pm$ refer to right (left)-moving fermions.}
\label{fig:sigma_1D}
\end{figure}

Finally, in 1D the same power-counting argument leads to Im$\Sigma
^{R}\propto \left| \varepsilon \right| $ and Re$\Sigma ^{R}\propto $ $%
\varepsilon \ln \left| \varepsilon \right| $ \footnote{%
Special care is required in 1D as in the perturbation theory one
gets a strong divergence in the self-energy corresponding to
interactions of fermions of the same chirality
(Fig.~\ref{fig:sigma_1D}a,c). This point will be discussed in more
detail in Section \ref{sec:mass-shell} (along with a weaker but
nonetheless singularity in 2D). For now, let us focus on a regular
part of the self-energy corresponding to the interaction of
fermions of opposite chirality
(Fig.~\ref{fig:sigma_1D}b).}Correspondingly, the ``correction'' to
the specific heat behaves as $T\ln T$ and is larger than the
leading, $T-$ term. This is the ultimate case of dynamic forward
scattering, whose precursors we have already seen in higher
dimensions
\footnote{%
Bosonization predicts that $C(T)$ of a fermionic system is the
same as that of 1D bosons, which scales as $T$ for $D=1$
\cite{giamarchi_book}. This is true only for spinless fermions, in
which case bosonisation provides an asymptotically exact solution.
For  electrons with spins, the bosonized theory is of the
sine-Gordon type with the non-Gaussian (cos$\phi )$ term coming
from the backscattering of fermions of opposite spins. Even if
this term is marginally irrelevant and flows down to zero at the
lowest energies, at intermediate energies it results in a
multiplicative $\ln T$ factor in $C(T)$ and a $\ln \max \{q,T,H\}$
correction to the spin susceptibility (where $H$ is the magnetic
field, and units are such that $q,T,$ and $H$ have the
units of energy). The difference between the non-analyticities in $D>1$ and $%
D=1$ is that the former occurs already at the second order in the
interaction, whereas the latter start only at \emph{third }order. Naive
power-counting breaks down in 1D because the coefficient in front of $T\ln T$
term in $C(T)$ vanishes at the second order, and one has to go to third
order. In the sine-Gordon model, the third order in the interaction is quite
natural: indeed, one has to calculate the correlation function of the cos$%
\phi $ term, which already contains two coupling constants; the
third one occurs by expanding the exponent to leading (first)
order. For more details, see
\cite{DL72},\cite{nersesyan},\cite{ronojoy}.}.

Even if the bare interaction is point-like, the effective one
contains a long-range part at finite frequencies. Indeed, the
non-analytic parts of $\Sigma $ and $C(T)$ come from the region of
small $q,$ and hence large distances. Already to the second order
in $U$, the effective interaction $\tilde{U}=U^{2}\Pi (\omega ,q)$
is proportional to the \emph{dynamic } polarization bubble of the
electron gas, $\Pi \left( \omega ,q\right) $. In all dimensions,
Im$\Pi ^{R}$ is universal and singular in $q$ for $\left| \omega
\right| /v_{F}\ll q\ll k_{F}$
\[
\text{Im}\Pi ^{R}\left( \omega ,q\right) \sim \nu _{D}\frac{\omega }{v_{F}|q|%
}.
\]
Although the effective interaction is indeed screened at $q\rightarrow 0$
--and this is why the FL survives even if the bare interaction has a
long-range tail--it has a slowly decaying tail in the intermediate range of $%
q.$ In real space, $\tilde{U}(r)$ behaves as $\omega /r^{D-1}$ at distances $%
k_{F}^{-1}\ll r\ll v_{F}/|\omega |$.

Thus, we have the same singular behavior of the bubble in all
dimensions, and the results for the self-energy differ only
because the phase volume $q^{D}$ is more effective in suppressing
the singularity in higher dimensions than in lower ones.

There is one more special interval of $q$: $q\approx 2k_{F}$ ,
\emph{i.e.}, Kohn anomaly. Usually, the Kohn anomaly is associated
with the $2k_{F}$- non-analyticity of the \emph{static }bubble,
and its most familiar manifestation is the Friedel oscillation in
electron density produced by a static impurity (discussed later
on). Here, the static Kohn anomaly is of no interest for us as we
are dealing with dynamic processes. However, the dynamic bubble is
also singular near $2k_{F}$. For example, in 2D,
\[
\mathrm{Im}\Pi ^{R}\left( q\approx 2k_{F},\omega \right) \propto \frac{%
\omega }{\sqrt{k_{F}\left( 2k_{F}-q\right) }}\theta \left(
2k_{F}-q\right).
\]
Because of the one-sided singularity in $\mathrm{Im}\Pi ^{R}$ near
$q=2k_F$, the effective interaction oscillates and falls off as a
power of $r$. By power counting, if a static Friedel oscillation
falls off as $\sin 2k_{F}r/r^{D}$, then the dynamic one behaves as
\[
\tilde{U}\propto \frac{\omega \sin 2k_{F}r}{r^{(D-1)/2}}.
\]
Dynamic Kohn anomaly results in the same kind of non-analyticity
in the self-energy (and thermodynamics) as the forward scattering.
The
``dangerous'' range of $q$ now is $\left| q-2k_{F}\right| \sim \omega /v_{F}$%
--``dynamic backscattering''. It is remarkable that the non-analytic term in
the self-energy is sensitive only to strictly forward or backscattering
events, whereas processes with intermediate momentum transfers contribute
only to analytic part of the self-energy. To see this, we perform the
analysis of kinematics in the next section.

\bigskip

\subsection{1D kinematics in higher dimensions}

The similarity between non-FL behavior in 1D and non-analytic features in
higher dimensions occurs already at the level of kinematics. Namely, one can
make a rather strong statement: \emph{the non-analytic terms in the
self-energy in higher dimensions result from essentially 1D scattering
processes.} Let's come back to self-energy diagram \ref{fig:selfenergy_2nd}%
a. In general, integrations over fermionic momentum $\vec{p}$ and bosonic $%
\vec{q}$ are independent of each other: one can first integrate over ($\vec{p%
},\varepsilon ),$ forming a bubble, and then integrate over ($\vec{q},\omega $%
). Generically, $\vec{p}$ spans the entire Fermi surface. However, the
non-analytic features in $\Sigma $ come not from generic but very specific $%
\vec{p}$ which are close to either to $\vec{k}$ or to $-\vec{k}.$

\begin{figure}[tbp]
\begin{center}
\epsfxsize=0.7 \columnwidth
%\epsffile{fig2lanie.eps}
\epsffile{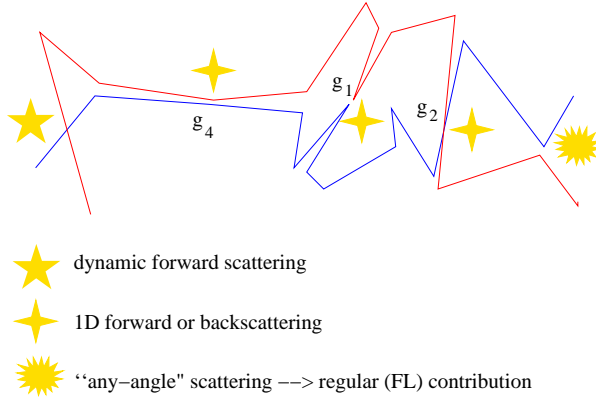}
\end{center}
\caption{Typical trajectories of two interacting fermions. Explosion:
``any-angle'' scattering at a third fermion (not shown) which leads to a
regular (FL) contribution. Five-corner star: dynamic forward scattering $%
q\sim |\protect\omega|/v_F$. This process contributes to
non-analyticity in 3D (to third order in the interaction) but not
in 2D. Four-corner star: 1D dynamic forward and backscattering
events, contributing to non-analyticities both in 3D and 2D.}
\label{fig:trajectory}
\end{figure}

Let's focus on the 2D case. The $\varepsilon ^{2}\ln |\varepsilon
|$ term results from the product of two $q^{-1}$ -singularities:
one is from the angular average of Im$G$ and the other one from
the dynamic, $\omega /v_{F}q$,
part of the bubble. In  Appendix \ref{sec:pi_anyD}, it is shown that the $%
\omega /v_{F}q$ singularity in the bubble comes from the region where $\vec{p%
}$ is almost perpendicular to $\vec{q}.$ Similarly, the angular averaging of
Im$G$ also pins the angle between $\vec{k}$ and $\vec{q} $ to almost $%
90^{\circ }.$
\begin{eqnarray*}
\mathrm{Im}G^{R}(\varepsilon -\omega ,\vec{k}-\vec{q}) &=&-\pi \delta \left(
\varepsilon -\omega -qv_{F}\cos \theta ^{\prime }\right) \rightarrow \\
\cos \theta ^{\prime } &=&\frac{\varepsilon -\omega }{v_{F}q}\sim \frac{%
\omega }{v_{F}q}\ll 1\rightarrow \theta ^{\prime }\approx \pi /2.
\end{eqnarray*}
As $\vec{p}$ and $\vec{k}$ are almost perpendicular to the same vector ($%
\vec{q}$), they are either almost parallel or anti-parallel to each other.
In terms of a symmetrized (``sunrise'') self-energy (cf. Fig.~\ref
{fig:selfenergy_2nd}), it means that either all three internal momenta are
parallel to the external one or one of the internal one is parallel to the
external whereas the other two are anti-parallel \footnote{%
In 3D, conditions $\vec{p}\perp \vec{q}$ and $\vec{k}\perp
\vec{q}$ mean only that $\vec{p}$ and $\vec{k}$ lie in the same
plane. However, it is still possible to show that for a closed
diagram, \emph{e.g., }thermodynamic potential, $\vec{p}$ and
$\vec{k}$ are either parallel or anti-parallel to each other.
Hence, the non-analytic term in $C(T)$ also comes from the 1D
processes. In addition, there are dynamic forward scattering
events (marked with a star in Fig.~\ref{fig:trajectory}) which,
although not being 1D in nature, do lead to a non-analyticity in
3D. Thus, the $T^{3}\ln T$ anomaly in $C(T)$ comes from both 1D
and non-1D processes \cite{chmm} . The difference is that the
former start already at the second order in the interaction
whereas the latter occur only at the third order. In 2D, the
entire $T-$ term in $C(T)$ comes from the 1D processes.}. Thus we
have three almost 1D processes:

\begin{itemize}
\item  all four momenta (two initial and two final) are almost
parallel to each other;

\item  the total momentum of the fermionic pair is near zero,
whereas the transferred momentum is small;

\item he total momentum of the fermionic pair is near zero,
whereas the transferred momentum is near $2k_F$.

\end{itemize}

These are precisely the same 1D processes we are going to deal
with in the next Section--the only difference is that in 2D,
trajectories do have some angular spread, which is of order
$\left| \omega \right| /E_{F}.$ The first one is known as
``$g_{4}"$ (meaning: all four momenta are in the same direction)
and the other one as ``$g_{2}"$ (meaning: two out of four momenta
are in the same direction). Both of these processes are of the
forward-scattering type as the transferred momentum is small. In
1D, these processes correspond to scattering of fermions of same
($g_{4}$) or opposite chirality ($g_{2}$). The last ($2k_F$)
process is known ``$g_{1}"$ in 1D.

It turns out  that of these two processes, the $g_{2}$- and
$2k_{F}$- ones, are directly responsible for the $\varepsilon
^{2}\ln \varepsilon $ behavior. The $g_{4}$-process leads to a
mass-shell singularity in the self-energy both in 1D and 2D,
discussed in the next section, but does not affect the
thermodynamics, so we will leave it for now.

What about $2k_{F}-$ scattering? Suppose electron $\vec{k}$ scatters into $%
-\vec{k}$ emitting an electron-hole pair of momentum $2\vec{k}.$ In general,
$2\vec{k}$ of the e-h pair may consist of any two fermionic momenta which
differ by $2\vec{k}:$ $\vec{p}$ and $\vec{p}+2\vec{k}.$ But since $\left| 2%
\vec{k}\right| \approx 2k_{F},$ the components of the e-h pair will be on
the Fermi surface only if $\vec{p}\approx -\vec{k}$ and $\vec{p}+2\vec{k}%
\approx \vec{k}.$ Only in this case  does the effective
interaction (bubble) have a non-analytic form at finite frequency.
Thus $2k_{F}$\-- scattering is also of the 1D nature for $D>1.$

What we have said above, can be summarized in the following
pictorial way. Suppose we follow the trajectories of two fermions,
as shown in Fig.~\ref {fig:trajectory}. There are several types of
scattering processes. First, there is ``any-angle'' scattering
which, in our particular example, occurs at a third fermion whose
trajectory is not shown. This scattering contributes regular, FL
terms both to the self-energy and thermodynamics. Second, there
are dynamic forward-scattering events, when $q\sim \left| \omega
\right| /v_{F}.$ These are \emph{not }1D processes, as fermionic
trajectories enter the interaction region at an arbitrary angle to
each other. In 3D, a third order in such processes results in the
non-analytic behavior of $C(T)$--this is the origin of the
``paramagnon'' anomaly in $C(T).$ In 2D, dynamic forward
scattering does not lead to non-analyticity. Finally, there are
processes, marked by``$g_1$'', ``$g_2$'', and ``$g_4$", when
electrons conspire to align their initial momenta so that they are
either parallel or antiparallel to each other. These processes
determine the non-analytic parts of $\Sigma $ and
thermodynamics in 2D (and also, formally, for $D<2.$) A crossover between $%
D>1 $ and $D=1$ occurs when all other processes but $g_{1},$ $g_{2},$ and $%
g_{4}$ are eliminated by a geometrical constraint.

We see that for non-analytic terms in the self-energy (and
thermodynamics), large-angle scattering does not matter.
Everything is determined by essentially 1D processes. As a result,
if the bare interaction has some $q$ dependence, only two Fourier
components matter: $U(0)$ and $U(2k_{F}).$ For example, in 2D
\begin{eqnarray*}
\text{Im}\Sigma ^{R}\left( \varepsilon \right) &\propto &\left[ U^{2}\left(
0\right) +U^{2}\left( 2k_{F}\right) -U(0)U(2k_{F})\right] \varepsilon
^{2}\ln \left| \varepsilon \right| ; \\
\text{Re}\Sigma ^{R}\left( \omega \right) &\propto &\left[ U^{2}\left(
0\right) +U^{2}\left( 2k_{F}\right) -U(0)U(2k_{F})\right] \varepsilon \left|
\varepsilon \right| ; \\
C(T)/T &=&\gamma ^{\ast }-a\left[ U^{2}\left( 0\right) +U^{2}\left(
2k_{F}\right) -U(0)U(2k_{F})\right] T; \\
\chi _{s}(Q,T) &=&\chi _{s}^{\ast }\left( 0\right) +bU^{2}\left(
2k_{F}\right) \max \left\{ v_{F}Q,T\right\} ;
\end{eqnarray*}
where $a$ and $b$ are coefficients. These perturbative results can be
generalized for the Fermi-liquid case, when the interaction is not
necessarily weak. Then the leading, analytic parts of $C(T)$ and $\chi _{s}$
are determined by the angular harmonics of the \emph{Landau interaction
function}
\[
\hat{F}\left( \vec{p},\vec{p}^{\prime }\right) =F_{s}\left( \theta \right)
\hat{I}+F_{a}\left( \theta \right) \vec{\sigma}\cdot \vec{\sigma}^{\prime },
\]
where $\theta $ is the angle between $\vec{p}$ and $\vec{p}^{\prime }.$ In
particular,
\begin{eqnarray*}
\gamma ^{\ast } &=&\gamma _{0}\left( 1+\langle \cos \theta F_{s}\rangle
\right) ; \\
\chi _{s}^{\ast }\left( 0\right) &=&\chi _{s}^{0}\frac{1+\langle \cos \theta
F_{s}\rangle }{1+\langle F_{a}\rangle },
\end{eqnarray*}
where $\gamma _{0}$ and $\chi _{s}^{0}$ are the corresponding quantities for
the Fermi gas. Because of the angular averaging, the FL part is rather
insensitive to the details of the interaction. As generically $F_{s}$ and $%
F_{a}$ are regular functions of $\theta ,$ the whole Fermi surface
contributes to the FL renormalizations. Vertices $U(0)$ and $U(2k_{F})$,
occurring in the perturbative expressions, are replaced by \emph{scattering
amplitudes} at angle $\theta =\pi $%
\[
\hat{A}\left( \vec{p},\vec{p}^{\prime }\right) =A_{s}\left( \theta \right)
\hat{I}+A_{a}\left( \theta \right) \vec{\sigma}\cdot \vec{\sigma}^{\prime },
\]
Beyond the perturbation theory \cite{zerosound},
\begin{eqnarray*}
C(T)/T &=&\gamma ^{\ast }-\bar{a}\left[ A_{s}^{2}\left( \pi \right)
+3A_{a}^{2}\left( \pi \right) \right] T; \\
\chi _{s}(Q,T) &=&\chi _{s}^{\ast }\left( 0\right) +\bar{b}A_{a}^{2}\left(
\pi \right) \max \left\{ v_{F}Q,T\right\} .
\end{eqnarray*}
Non-analytic parts are not subject to angular averaging and are
sensitive to a detailed behavior of $A_{s,a}$ near $\theta =\pi
$\footnote {The renormalization of the scattering amplitudes by
the Cooper channel of the interaction results in  additional $\ln
T$-dependences of $A_{c,s}(\pi)$}.

\subsection{Infrared catastrophe}

\label{sec:mass-shell}

\subsubsection{1D}

By now, it is well-known that the FL breaks down in 1D and an
attempt to apply the perturbation theory to 1D problem results in
singularities. Let's see what precisely goes wrong in 1D. I begin
with considering the interaction of fermions of opposite
chirality, as in diagram Fig.~\ref{fig:sigma_1D}b. Physically, a
right-moving fermion emits (and then re-absorbs) left-moving
quanta of density excitations (same for left-moving fermion
emitting/absorbing right-moving quanta). Now, instead of the
order-of-magnitude estimate (\ref{bubble}), which is good in all
dimensions but only for power-counting purposes, I am going to use
an exact expression for the bubble, Eq.~(\ref{pito_-}), formed by
left-moving fermions. On the Fermi surface ($k=k_F$, we have
\begin{eqnarray*}
-\mathrm{Im}\Sigma _{+-}^{R}\left( \varepsilon \right) &\sim
&U^{2}\nu _{1}\int_{0}^{\varepsilon }d\omega \int
dq\mathrm{Im}G_{+}^{R}\left(
\varepsilon -\omega ,k-q\right) \text{Im}\mathrm{\Pi }_{-}^{R} \\
&\sim &U^{2}\nu _{1}\int_{0}^{\varepsilon }d\omega \int dq\delta \left(
\varepsilon -\omega +v_{F}q\right) \left( \omega /v_{F}\right) \delta \left(
\omega +v_{F}q\right) \\
&\sim &g^{2}\left| \varepsilon \right| ,
\end{eqnarray*}
where $g\equiv U/v_{F}$ is the dimensionless coupling constant. The
corresponding real part behaves as $\varepsilon \ln \left| \varepsilon
\right| .$ What we got is bad, as $\mathrm{Im}\Sigma ^{R}$ scales with $%
\varepsilon $ in the same way as the energy of a free excitation
above the Fermi level and Re$\Sigma ^{R}$ increases faster than
$\varepsilon $ (which means that the effective mass depends on
$\varepsilon $ as $\ln |\varepsilon |) $, but not too bad because,
as long as $g\ll 1$, the breakdown of the quasi-particle picture
occurs only at exponentially small energy scales: $\varepsilon
\lesssim E_{F}\exp (-1/g^2).$  Now, let's look at scattering of
fermions of the same chirality. This time, I choose to be away
from the mass shell.
\begin{eqnarray}
-\mathrm{Im}\Sigma _{++}^{R} &\propto &\int_{0}^{\varepsilon }d\omega \int dq%
\underbrace{\delta \left( \varepsilon -\omega -v_{F}\left( k-q\right)
\right) }_{\mathrm{Im}G_{+}^{R}}\underbrace{\omega \delta \left( \omega
-v_{F}q\right) }_{\mathrm{Im}\Pi _{+}^{R}} \\
&=&\varepsilon ^{2}\delta \left( \varepsilon -v_{F}k\right) .
\end{eqnarray}
It is not difficult to see that the full (complex) self-energy is simply
\begin{equation}
\Sigma _{++}^{R}\propto -\frac{\varepsilon ^{2}}{\varepsilon -v_{F}k+i0^{+}}.
\label{sigma_pole}
\end{equation}
On the mass shell ($\varepsilon =v_{F}k)$ we have a
strong--delta-function--singularity. This anomaly was discovered
by Bychkov, Gor'kov, and Dzyaloshinskii back in the 60s
\cite{bychkov}, who called it the ``infrared catastrophe''.
Indeed, it is similar to an infrared catastrophe in QED, where an
electron can emit an infinite number of soft photons. Likewise,
since we have linearized the spectrum, a 1D fermion can emit an
infinite number of soft bosons: quanta of charge- and spin-density
excitations. The point is that in 1D there is a perfect match
between momentum and energy conservations for a process of
emission (or absorption)
of a boson with energy and momentum related by $\omega =v_{F}q:$%
\begin{eqnarray*}
k^{\prime } &=&k-q \\
\varepsilon ^{\prime } &=&\varepsilon -\omega =v_{F}k-v_{F}q=v_{F}\left(
k-q\right) .
\end{eqnarray*}
On the mass-shell, the energy and momentum conservations are equivalent.
Imagine that you want to find a probability of certain scattering process
using a Fermi Golden rule. Then you have a product of two $\delta -$
functions: one reflecting the momentum and other energy conservation. But if
the arguments of the delta-functions are the same, you have an essential
singularity: a square of the delta-function. As a result, the corresponding
probability diverges.

A pole in the self-energy [Eq.~(\ref{sigma_pole})] indicates the
non-perturbative and specifically 1D effect: spin-charge
separation. Indeed, substituting Eq.~(\ref{sigma_pole}) we get two
poles corresponding to excitations propagating with velocities
$v_{F}\left( 1\pm g\right) $ (recall that $g\ll 1).$ This peculiar
feature is confirmed by an exact solution (see Section
\ref{sec:DL}): already the $g_{4}-$interaction leads to a
spin-charge separation (but not to anomalous scaling). What we did
not get quite right is that the velocities of both--spin- and
charge-modes--are modified by the interactions. In fact,the exact
solution shows that the velocity of the spin-mode remains equal to
$v_{F},$ whereas the velocity of the charge mode is modified.

Obviously, there is no spin-charge separation for spinless electrons.
Indeed, in this case diagram Fig.~\ref{fig:sigma_1D}a does not have an
additional factor of two as compared to Fig.~\ref{fig:sigma_1D}c (but is
still of  opposite sign), so that the forward-scattering parts of these
two diagrams cancel each other. As a result, there is no infrared
catastrophe for spinless fermions.

\subsubsection{2D}

What we considered in the previous section sounds like an
essentially 1D effect. However, a similar effect exists also in 2D
(more generally, for $1\leq D\leq 2$).  This emphasizes once again
that the difference between $D=1$ and $D>1$ is not as dramatic as
it seems.

In 2D, the self-energy also diverges on the mass shell, if one linearizes
the electron's spectrum, albeit the divergence is weaker than in 1D--to
second order, it is logarithmic \footnote{%
In 3D, there is no mass-shell singularity to any order of the perturbation
theory.}. The origin of the divergence can be traced back to the form of the
polarization bubble at small momentum transfer, Eq.~(\ref{impi_anyD}).
Integrating over the angle in 2D, we get
\begin{equation}
\text{Im}\Pi ^{R}(\omega ,q)=-\left( \frac{m}{2\pi }\right) \frac{\omega }{%
\sqrt{\left( v_{F}q\right) ^{2}-\omega ^{2}}}\theta \left( v_{F}q-\left|
\omega \right| \right) .  \label{sqrt1}
\end{equation}
$\text{Im}\Pi ^{R}(\omega ,q)$ has a square-root singularity at
the boundary of the particle-hole continuum, \emph{i.e.,
}at\emph{\ }$\omega =v_{F}q$. (This is a threshold singularity of
the van Hove type: the band of soft electron-hole pairs is
terminated at $\omega =v_{F}q,$ but the spectral weight of the
pairs is peaked at the band edge). On the other hand,
expanding $\epsilon _{\mathbf{k}+\mathbf{q}}$ in $G^{R}(\varepsilon +\omega ,%
\mathbf{k}+$\textbf{$q$}$)$ as $\xi _{\mathbf{k+q}}=\xi _{k}+v_{F}q\cos
\theta $ and integrating over $\theta $, we obtain another square-root
singularity
\begin{equation}
\int d\theta \text{Im}G^{R}=-2\pi \left[ \left( v_{F}q\right) ^{2}-\left(
\varepsilon +\omega -\xi _{k}\right) ^{2}\right] ^{-1/2}.  \label{sqrt2}
\end{equation}
On the mass shell ($\omega =\xi _{k}$), the arguments of the square roots in
Eqs.~(\ref{sqrt1}) and (\ref{sqrt2}) coincide, and the integral over $q$
diverges logarithmically. The resulting contribution to Im$\Sigma ^{R}$
diverges on the mass shell ($\varepsilon =\xi _{k})$ \cite
{castellani,fukuyama,metzner},\cite{chm_long},\cite{zerosound}
\[
\text{Im}\Sigma _{g_{4}}^{R}\left( \varepsilon ,k\right) =-\frac{u^{2}}{8\pi
}\frac{\varepsilon ^{2}}{E_{F}}\ln \frac{E_{F}}{|\varepsilon -\xi _{k}|},
\label{c1b}
\]
where $\Delta \equiv \varepsilon -\xi _{k}$ and $u\equiv mU/2\pi
.$ The process responsible for the log-singularity is the
``$g_{4}"$ process in Fig.~\ref{fig:processes}. On the other hand,
$g_{2}$ and $g_{1}$ processes give a contribution which is finite
on the mass shell
\[
\text{Im}\Sigma _{g_{1}+g_{2}}^{R}\left( \varepsilon ,k\right) =-\frac{u^{2}%
}{4\pi }\frac{\varepsilon ^{2}}{E_{F}}\ln \frac{E_{F}}{|\varepsilon +\xi
_{k}|}.
\]
(The divergence at $\varepsilon =-\xi _{k}$ is spurious and is
removed by going beyond the log-accuracy
\cite{chm_long},\cite{zerosound}.) We see therefore that the
familiar form of the self-energy in 2D [$\varepsilon ^{2}\ln
\left| \varepsilon \right|$, see Ref.~\cite{2D}] is valid only on
the Fermi
surface $\left( \xi _{k}=0\right) .$ The logarithmic singularity in Im$%
\Sigma ^{R}$ on the mass shell  is eliminated by retaining the
finite curvature of single-particle spectrum (which amounts to
keeping the $q^{2}/2m$ term in $\xi _{\vec{k}+\vec{q}}).$ This
brings in a new scale $\varepsilon ^{2}/E_{F}$. The emerging
singularity in (\ref{c1b}) is regularized at $\left| \varepsilon
-\xi _{k}\right| \sim \varepsilon ^{2}/E_{F}$ and the $\varepsilon
^{2}\ln \left| \varepsilon \right| $ behavior is restored.
However, higher orders diverge as power-laws and finite curvature
does not help to regularize them. This means that--in contrast to
3D--the perturbation theory must be re-summed even for an
infinitesimally weak interaction. Once this is done, the
singularities are removed. Re-summation also helps to understand
the reason for the problems in the perturbation theory. In fact,
what we were trying to do was to take into account a
non-perturbative effect--an interaction with the zero-sound
mode--via a perturbation theory. Once all orders are re-summed,
the zero-sound mode splits off the continuum boundary--now it is a
propagating mode with velocity $c>v_{F}.$ This splitting is what
regularizes the divergences. The resulting state is essentially a
FL: the leading term in $\Sigma $ behaves as $\varepsilon ^{2}\ln
\left| \varepsilon \right| .$ However, some non-perturbative
features remain: for example, the spectral function exhibits a
second peak away from the mass shell corresponding to the emission
of the zero-sound waves by fermions. A two-peak structure of the
spectral function is reminiscent of the spin-charge separation,
although we do not really have a spin-charge separation here: in
contrast to the 1D case, the spin-density collective mode lies
within the continuum and is damped by the particle-hole pairs.

%%%%%%%%%%%%%%%%%%%%%%%%%%%%%%%%%%%%%%%%%%%%%%%%%%%%%%%%%
%%%LECTURE 2
%%%%%%%%%%%%%%%%%%%%%%%%%%%%%%%%%%%%%%%%%%%%%%%%%%%%%%%%

\section{Dzyaloshinskii-Larkin solution of the Tomonaga-Luttinger model}

\label{sec:DL}

\subsection{Hamiltonian, anomalous commutators, and conservation laws}

In the Tomonaga-Luttinger model \cite{tomonaga},\cite{luttinger}
one considers a system of 1D spin-1/2 fermions with a linearized
dispersion. Only forward scattering of left- and right-moving
fermions is taken into account ($g_{2}$ and $g_{4}-$ processes),
whereas backscattering is neglected. This last assumption means
that the interaction potential is of sufficiently long-range, so
that $U\left( 2k_{F}\right) \ll U\left( 0\right) . $ [We will come
back to this condition later.] Coupling between fermions of the
same chirality ($g_{4}$) is assumed to be different from coupling
between fermions of different chirality ($g_{2}).$ If the original
Hamiltonian contains only density-density interaction, then
$g_{2}=g_{4}.$ A difference between $g_{2}$ and $g_{4}$ leads to
an unphysical (within this model) current-current interaction. We
will keep $g_{2}\neq g_{4}$, however, at the intermediate steps of
the calculations as it helps to elucidate certain points. At the
end, one can make $g_{2}$ equal to $g_{4}$ without any penalty. In
addition, in some physical situations, $g_{2}\neq g_{4}$ .
\footnote{%
For example, Coulomb interaction between the electrons at the
edges of a finite-width Hall bar (in the Integer Quantum Hall
Effect regime) has this feature: electrons of the same chirality
are situated on the same edge, whereas electrons of different
chirality are on opposite edges; hence the matrix elements for the
$g_{2}-$ and $g_{4}$- interactions are different.} In what follows
I will follow the original paper by Dzyaloshinskii and Larkin (DL)
\cite{DL} and a paper by Metzner and di Castro \cite
{metzner_castro}, where the Ward identity used by Dzyaloshinskii
and Larkin is derived in a detailed way.

The Hamiltonian of the model is written as
\begin{eqnarray*}
H &=&H_{0}+H_{\mathrm{int}}; \\
H_{\mathrm{int}} &\equiv &H_{2}+H_{4},
\end{eqnarray*}
where
\[
H_{0}=v_{F}\sum_{k,\sigma }k\left( a_{+,\sigma }^{\dagger }(k)a_{+,\sigma
}(k)-a_{-,\sigma }^{\dagger }(k)a_{-,\sigma }(k)\right)
\]
is the Hamiltonian of free fermions ($\pm $ denote right/left moving
fermions and $\sigma $ is the spin projection) and

\begin{eqnarray*}
H_{2} &=&\frac{g_{2}}{L}\sum_{q}\sum_{\sigma ,\sigma ^{\prime }}\rho
_{+,\sigma }\left( q\right) \rho _{-,\sigma ^{\prime }}\left( -q\right) ; \\
H_{4} &=&\frac{g_{4}}{2L}\sum_{q}\sum_{\sigma ,\sigma ^{\prime }}\rho
_{+,\sigma }\left( q\right) \rho _{+,\sigma ^{\prime }}\left( -q\right)
+\rho _{-,\sigma }\left( q\right) \rho _{-,\sigma ^{\prime }}\left(
-q\right) ,
\end{eqnarray*}
with
\[
\rho _{\pm ,\sigma }=\sum_{k}a_{\pm ,\sigma }^{\dagger }\left( k+q\right)
a_{\pm ,\sigma }\left( k\right) .
\]
To avoid additional complications, I assume that the interaction
is spin-independent. To simplify the notations and to emphasize
the similarity between this model and QED, I will set $v_{F}$ to
unity in this section.

Introducing the chiral charge- and spin densities as
\begin{eqnarray*}
\rho _{\pm }^{c} &=&\rho _{\pm ,\uparrow }+\rho _{\pm ,\downarrow }; \\
\rho _{\pm }^{s} &=&\rho _{\pm ,\uparrow }-\rho _{\pm ,\downarrow },
\end{eqnarray*}
and total charge density and current as
\begin{eqnarray*}
\rho ^{c} &=&\rho _{+}^{c}+\rho _{-}^{c}; \\
j^{c} &=&\rho _{+}^{c}-\rho _{-}^{c},
\end{eqnarray*}
the interaction part of the Hamiltonian reduces to
\begin{equation}
H_{\mathrm{int}}=\sum_{q}\frac{1}{2}\left( g_{2}+g_{4}\right) \rho
^{c}\left( q\right) \rho ^{c}\left( -q\right) +\frac{1}{2}\left(
g_{4}-g_{2}\right) j^{c}\left( q\right) j^{c}\left( -q\right) .
\label{Hrrjj}
\end{equation}
As we have already said, for $g_{2}=g_{4},$ the interaction is of
a pure density-density type. Notice also that the spin density and
current drop out of the Hamiltonian--this is to be expected for a
spin-invariant interaction.
To make a link with QED, let us introduce Minkowski current $j^{\mu }$ with $%
\mu =0,1$ so that $j^{0}=\rho ^{c}$ (=$j_{0})$ and $j^{1}=j^{c}$
(=-$j_{1}). $ Then the interaction can be written as a 4-product
of Minkowski currents in a Lorentz-invariant form
\[
H_{\mathrm{int}}=\sum_{q}g_{\mu \nu }j_{\nu }j^{\nu },
\]
where
\begin{eqnarray}
g_{00} &=&\frac{1}{2}\left( g_{2}+g_{4}\right) ;  \nonumber \\
g_{11} &=&\frac{1}{2}\left( g_{4}-g_{2}\right) ;  \nonumber \\
g_{01} &=&g_{10}=0.  \label{gmunu}
\end{eqnarray}
In what follows, we will need the following anomalous commutators
\begin{eqnarray*}
\left[ \rho _{\pm ,\sigma }\left( q\right) ,H_{0}\right] &=&\pm q\rho _{\pm
,\sigma }\left( q\right) ; \\
\left[ \rho _{\pm ,\sigma }\left( q\right) ,H_{2}\right] &=&\pm \frac{g_{2}}{%
2\pi }q\rho _{\mp ,\sigma }\left( q\right) ; \\
\left[ \rho _{\pm ,\sigma }\left( q\right) ,H_{4}\right] &=&\pm \frac{g_{4}}{%
2\pi }q\rho _{\pm ,\sigma }\left( q\right).
\end{eqnarray*}
The derivation of these commutation relations can be found in a number of
standard sources \cite{mahan,giamarchi_book} and I will not present it here.
Adding up the commutators, we get
\begin{eqnarray*}
\left[ \rho _{\pm ,\sigma },H\right] &=&\left[ \rho _{\pm ,\sigma
},H_{0}+H_{2}+H_{4}\right] \\
&=&\pm q\rho _{\pm ,\sigma }\pm \frac{g_{2}}{2\pi }q\rho _{\mp }^{c}\pm
\frac{g_{4}}{2\pi }q\rho _{\pm }^{c}
\end{eqnarray*}
Adding up equations for spin-up and -down fermions, we obtain
\[
\left[ \rho _{\pm ,}^{c},H\right] =\pm q\rho _{\pm }^{c}\pm \frac{g_{2}}{\pi
}q\rho _{\mp }^{c}\pm \frac{g_{4}}{\pi }q\rho _{\pm }^{c}.
\]
Finally, adding up the $\pm $ components yields \beq i\partial
_{t}\rho ^{c}=\left[ \rho ^{c},H\right] =v_{c}qj^{c},
\label{cont}\eeq where
\[
v_{c}\equiv 1+\frac{g_{4}-g_{2}}{\pi }
\]
(recall that $v_{F}=1).$ Eq. (\ref{cont}) is a continuity
equation reflecting charge conservation. As if we did not have
enough new notations, here is another one
\[
Q^{\mu }=\left( \omega ,q\right)
\]
and
\[
\mathcal{Q}^{\mu }=\left( \omega ,v_{c}q\right) .
\]
In these notations and after a Fourier transform, the continuity
equation can be written as
\[
\mathcal{Q}_{\mu }j^{\mu }=0.
\]
The same relation for free particles reads
\[
Q_{\mu }j^{\mu }=0.
\]

\subsection{Reducible and irreducible vertices}

\begin{figure}[tbp]
\begin{center}
\epsfxsize=1.0 \columnwidth
\epsffile{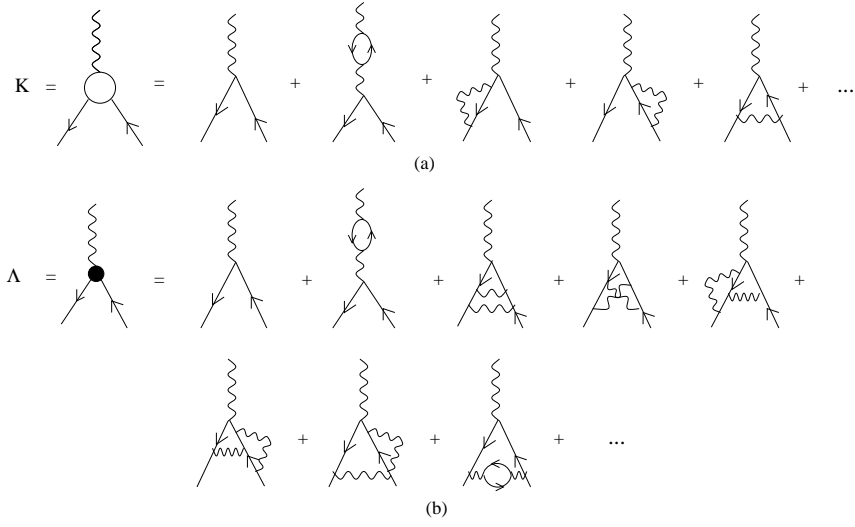}
\end{center}
\caption{a) Three-leg correlator $K$. b) Vertex part $\Lambda$.}
\label{fig:vertexk}
\end{figure}
Now, construct a mixed (fermion-boson) correlator
\begin{equation}
K_{\pm ,\sigma }^{\mu }\left( k,q|t,t_{1},t_{1}^{\prime }\right) =-\langle
Tj^{\mu }\left( q,t\right) a_{\pm ,\sigma }\left( k,t_{1}\right) a_{\pm
,\sigma }^{\dagger }\left( k+q,t_{1}\right)\rangle,  \label{kmu}
\end{equation}
where $\mu =0,1$ and
\begin{eqnarray*}
j^{0} &=&\rho _{+}^{c}+\rho _{-}^{c}; \\
j^{1} &=&\rho _{+}^{c}-\rho _{-}^{c}.
\end{eqnarray*}
$K^{\mu }$ is an analog of the three-leg vertex in QED , except that in QED
the ``boson'' is the $\mu -$ the component of the photon field
\[
\mathrm{QED:}\text{ }K^{\mu }=-\langle TA^{\mu }a\bar{a}\rangle .
\]
A diagrammatic representation of $K^{\mu }$ is a three-particle (one boson
and two fermions) diagram (cf. Fig. \ref{fig:vertexk}a).

The diagrams with self-energy insertions to solid lines simply renormalize
the Green's functions. Absorbing these renormalizations, we single out the
vertex part, re-writing $K^{\mu }$ as
\begin{equation}
K^{\mu }=G^{2}\Lambda ^{\mu }.  \label{kvsgamma}
\end{equation}
\begin{figure}[tbp]
\begin{center}
\epsfxsize=0.8 \columnwidth
\epsffile{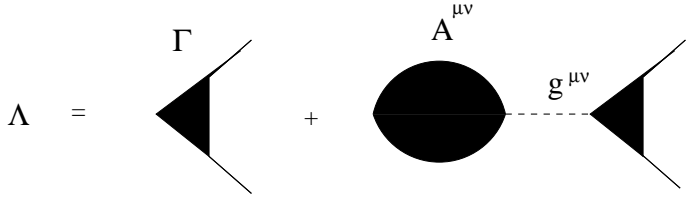}
\end{center}
\caption{Relation between vertices $\Lambda$ and $\Gamma$.}
\label{fig:dyson}
\end{figure}
Notice that there are as many vertex parts as there are bosonic degrees of
freedom. In (3+1) QED, $\Lambda ^{0}$ is a\emph{\ scalar} vertex and $%
\Lambda ^{\mu =1,2,3}$ are the components of the \emph{vector} vertex.
Diagrams representing $\Lambda ^{\mu }$ are shown in Fig. \ref{fig:vertexk}%
b. These series can be re-arranged further by separating the \emph{%
photon-irreducible } vertex part, $\Gamma ^{\mu }$. A
photon-irreducible part is obtained by separating the corrections
to the bosonic line, i.e., taking into account polarization.
Vertices $\Lambda ^{\mu }$ and $\Gamma ^{\mu }$ are related via a
kind of  Dyson equation, which is simpler than the usual Dyson in
a sense that there is no $\Lambda ^{\mu }$ on the right-hand-side.
Diagrammatically, this relation is represented by Fig.\ref
{fig:dyson} where a shaded bubble is an exact (renormalized)
current-current correlation function
\[
A^{\mu \nu }\left( q,t\right) =-\frac{i}{V}\langle j^{\mu }\left(
q,t\right) j^{\mu }\left( -q,0\right) \rangle .
\]
Algebraically, equation in Fig.\ref{fig:dyson} says
\begin{equation}
\Lambda _{\pm ,\sigma }^{\mu }=\Gamma _{\pm ,\sigma }^{\mu }+A^{\mu \nu
}g_{\mu \lambda }\Gamma _{\pm ,\sigma }^{\lambda }.  \label{Dyson_GL}
\end{equation}
(We remind the reader that indices $\pm ,\sigma $ simply specify
the fermionic flavor which is not mixed in our approximation of
forward-scattering and spin-independent forces, so all relations
are applicable to each individual flavor). The coupling constants
$g^{\mu \nu }$
relate currents to densities. According to Eqs. (\ref{Hrrjj}) and (\ref{gmunu}%
), densities couple to densities and currents to currents with no cross
terms. Opening the matrix product in Eq. (\ref{Dyson_GL}), we obtain
\begin{equation}
\Lambda _{i}^{\mu }=\Gamma _{i}^{\mu }+A^{\mu 0}g_{00}\Gamma ^{0}+A^{\mu
1}g_{11}\Gamma ^{1}.  \label{Dyson_GL_expl}
\end{equation}

\subsection{Ward identities}

A Ward identity for vertex $\Lambda ^{\mu }$ is obtained by applying $%
i\partial _{t}$ to $K^{\mu }$ in Eq.~(\ref{kmu}) and using the continuity
equation (\ref{kmu}).\footnote{%
When differentiating, recall that the $T-$ product can be represented by
step-functions in time which, upon differentiating, yield delta-functions}
Performing this operations and Fourier transforming in time, we obtain
\[
\mathcal{Q}_{\mu }K_{i}^{\mu }\left( K,Q\right) =G_{i}\left( K\right)
-G_{i}\left( K+Q\right) ,
\]
where $i$ denotes the branch
\[
i\equiv \pm ,\sigma .
\]
Recalling Eq.~(\ref{kvsgamma}), we see that the Ward identity
becomes
\begin{equation}
\mathcal{Q}_{\mu }\Lambda _{i}^{\mu }\left( K,Q\right) =G_{i}^{-1}\left(
K+Q\right) -G_{i}^{-1}\left( K\right) ,  \label{ward_gamma}
\end{equation}
which is identical to a corresponding identity in QED. For those who like
to see things not masked by fancy notations, here is Eq.~(\ref{ward_gamma})
in an explicit form
\begin{equation}
\omega \Lambda _{i}^{0}\left( \varepsilon ,k;\omega ,q\right) -v_{c}q\Lambda
_{i}^{1}\left( \varepsilon ,k;\omega ,q\right) =G_{i}^{-1}\left( \varepsilon
+\omega ,k+q\right) -G_{i}^{-1}\left( \varepsilon ,k\right) .
\label{ward_gamma_expl}
\end{equation}
Notice that Eqs.(\ref{ward_gamma},\ref{ward_gamma_expl}) contain \emph{%
renormalized }velocity $v_{c}.$ In what follows, we will actually need a
Ward identity not for $\Lambda ^{\mu }$ but for the photon-irreducible
vertex $\Gamma ^{\mu }.$ This one is obtained by deriving the continuity
equation for 4-current correlation function $A^{\mu \nu }.$ To this end, one
applies $i\partial _{t}$ to $A^{0\nu }$ and uses continuity equation (\ref{cont}%
), which yields \footnote{%
To get this result, recall the form of the anomalous density-density
commutator
\[
\lbrack j^{\mu }\left( q\right) ,j^{\nu }\left( -q\right) ]=\epsilon ^{\mu
\nu }\frac{2}{\pi }qL,
\]
where $\epsilon ^{00}=\epsilon ^{11}=0,$\ $\epsilon ^{01}=-\epsilon ^{10}=1.$
Now open the $T-$\ product in $A^{0\nu }$\ and apply $i\partial _{t}$%
\begin{eqnarray}
i\partial _{t}A^{0\nu }\left( q,t\right) &=&-\frac{i}{V}\left( i\partial
_{t}\right) \langle \theta \left( t\right) j^{0}\left( q,t\right) j^{\nu
}\left( -q,0\right) +\theta \left( -t\right) j^{\nu }\left( -q,0\right)
j^{0}\left( q,t\right) \rangle  \nonumber \\
&=&\frac{1}{V}\delta \left( t\right) [j^{0}\left( q,0\right) ,j^{\nu }\left(
-q,0\right) ]+v_{a}qA^{1\nu }=2\frac{q}{\pi }\delta _{\nu ,1}+v_{a}qA^{1\nu
}.  \label{Acont_expl}
\end{eqnarray}
In 4-notations, (\ref{Acont_expl}) is equivalent to (\ref{Acont}).}
\begin{equation}
Q_{\mu }A^{\mu \nu }=\frac{2}{\pi }q\delta _{\nu ,1}.  \label{Acont}
\end{equation}
{\small \ }

Now, we form a scalar product between $\mathcal{Q}_{\mu }$ and Eq. (\ref
{Dyson_GL_expl}), using continuity equation for $A^{\mu \nu }$ (\ref{Acont}%
). This brings us to
\[
\mathcal{Q}_{\mu }\Lambda _{i}^{\mu }=Q_{\mu }A^{\mu },
\]
where
\begin{eqnarray*}
\mathcal{Q}_{\mu }\Lambda _{i}^{\mu } &=&\mathcal{Q}_{\mu }\left( \Gamma
_{i}^{\mu }+A^{\mu 0}g_{00}\Gamma _{i}^{0}+A^{\mu 1}g_{11}\Gamma
_{i}^{1}\right) \\
&=&\mathcal{Q}_{\mu }\Gamma _{i}^{\mu }+\underbrace{\mathcal{Q}_{\mu }A^{\mu
0}}_{=0}g_{00}\Gamma _{i}^{0}+\underbrace{\mathcal{Q}_{\mu }A^{\mu 1}}%
_{=2q/\pi }g_{11}\Gamma _{i}^{1} \\
&=&\omega \Gamma _{i}^{0}-\underbrace{v_{c}}_{=1+\left( g_{4}-g_{2}\right)
/\pi }q\Gamma _{i}^{1}+\frac{2q}{\pi }\frac{1}{2}\frac{g_{4}-g_{2}}{\pi }%
\Gamma _{i}^{1} \\
&=&\omega \Gamma _{i}^{0}-q\Gamma _{i}^{1}=Q_{\mu }\Gamma ^{\mu }.
\end{eqnarray*}
Finally, the Ward identity for photon-irreducible vertex is
\begin{equation}
Q_{\mu }\Gamma ^{\mu }=G_{i}^{-1}\left( K+Q\right) -G_{i}^{-1}\left(
K\right) .  \label{Ward_Lambda}
\end{equation}
It is remarkable that the left-hand-side of
Eq.~(\ref{Ward_Lambda}) contains the \emph{bare }Fermi velocity
($=1)$ instead of the renormalized one. This is true even if we
allowed for spin-dependent interaction in the Hamiltonian.

It seems that we have not achieved much, as the conservation law
was simply cast into a different form. However, in our 1D problem
with a linearized spectrum a further progress can be made because
the current and density (for given chirality) are just the same
quantity (up to an overall factor of the Fermi velocity):
\[
\Gamma _{\pm ,\sigma }^{1}=\pm \Gamma _{\pm ,\sigma }^{0}
\]
Therefore, we have a closed relation between just one vertex and Green's
functions. Suppressing the 4-vector index $\mu ,$ we get the Ward identity
for the density vertex
\begin{equation}
\Gamma _{\pm ,\sigma }^{0}\left( K,Q\right) =\frac{G_{\pm ,\sigma
}^{-1}\left( K+Q\right) -G_{\pm ,\sigma }^{-1}\left( K\right) }{\omega \mp q}%
.  \label{ward_final}
\end{equation}
This is the identity that we need to proceed further with the
Dzyaloshinskii-Larkin solution of the Tomonaga-Luttinger problem. Notice
that (\ref{ward_final}) contains fully interacting Green's functions.

\subsection{Effective interaction}

\begin{figure}[tbp]
\begin{center}
\epsfxsize=1.0 \columnwidth
\epsffile{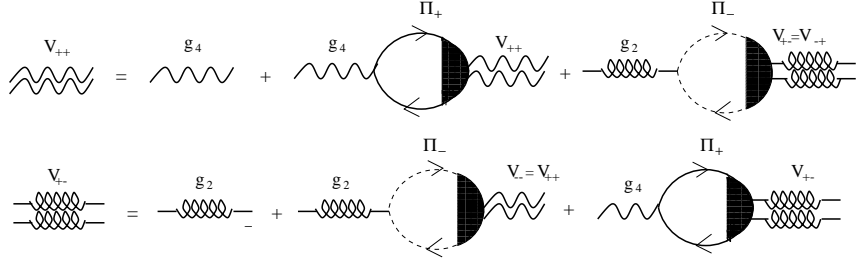}
\end{center}
\caption{Dyson equation for the effective interaction. Solid line: Green's
function of a right-moving fermion. Dashed line: Green's function of a
left-moving fermion. Single wavy line: bare interaction of fermions of the
same chirality; spiral line: same for the fermions of opposite chirality.
Double wavy and spiral lines represent the renormalized interactions.}
\label{fig:newpot}
\end{figure}
Effective interaction is obtained by collecting polarization corrections to
the bare one. Diagrammatically, this procedure is described by the Dyson
equation, represented in Fig.\ref{fig:newpot}. The interaction and
polarization bubble are matrices with components
\[
\hat{V}=\left(
\begin{array}{cc}
V_{++} & V_{+-} \\
V_{+-} & V_{++}
\end{array}
\right) ,\hat{V}_{0}=\left(
\begin{array}{cc}
g_{4} & g_{2} \\
g_{2} & g_{4}
\end{array}
\right) ,\hat{\Pi}=\left(
\begin{array}{cc}
\Pi _{+} & 0 \\
0 & \Pi _{-}
\end{array}
\right) ,
\]
where we used an obvious symmetry $V_{++}=V_{--},V_{+-}=V_{-+}.$ The Dyson
equation in the matrix form reads
\[
\hat{V}=\hat{V}_{0}+\hat{V}_{0}\hat{\Pi}\hat{V},
\]
or, in components,
\begin{eqnarray}
V_{++} &=&g_{4}+g_{4}\Pi _{+}V_{++}+g_{2}\Pi _{-}V_{+-};  \nonumber \\
V_{+-} &=&g_{2}+g_{2}\Pi _{-}V_{++}+g_{4}\Pi _{-}V_{+-}.  \label{rpa}
\end{eqnarray}
The bubbles in these equations are \emph{fully renormalized }ones, \emph{i.e., }%
they are built on exact Green's functions and contain a vertex
(hatched corner):
\[
\Pi _{\pm }\left( \omega ,q\right) =-2i\int \int \frac{dkd\varepsilon }{%
\left( 2\pi \right) ^{2}}G_{\pm }\left( \varepsilon +\omega ,k+q\right)
G_{\pm }\left( \varepsilon ,k\right) \Gamma _{\pm }^{0}\left( \varepsilon
,k;\omega ,q\right) .
\]
Now we use the Ward identity for $\Gamma _{\pm }^{0}$ (\ref{ward_final}) to
get \footnote{%
I skipped over a subtlety related to the infinitesimal imaginary parts $%
i0^{+}$in the denominator. Works the same way. If you are unhappy with
this, imagine that we work with Matsubara frequencies. Then there are no $%
i0^{+}$s whatsoever.}
\begin{equation}
\Pi _{\pm }\left( \omega ,q\right) =-2i\frac{1}{\omega \mp q}\int \int \frac{%
dkd\varepsilon }{\left( 2\pi \right) ^{2}}\left[ G_{\pm }\left( \varepsilon
,k\right) -G_{\pm }\left( \varepsilon +\omega ,k+q\right) \right] .
\label{bubble_1}
\end{equation}
 Eq.~(\ref{bubble_1}) looks exactly the same as a
\emph{free }bubble [cf. Eq.~(\ref{bubble_free})] except that it contains exact
rather than free Green's functions. Because we managed to transform the
product of two Green's functions into a difference, frequency integration in
Eq.~(\ref{bubble_1}) can be performed term by term yielding \emph{exact }%
momentum distribution functions $n_{\pm }\left( k\right) $ and $n_{\pm
}\left( k+q\right) :$%
\begin{equation}
\Pi _{\pm }\left( \omega ,q\right) =\frac{1}{\omega \mp q}\int \int \frac{dk%
}{\pi }\left[ n_{\pm }(k)-n_{\pm }(k+q)\right] .  \label{bubble_2}
\end{equation}
At the first glance, it seems that we have not achieved much so
far. Indeed, we traded one unknown quantity ($\Pi _{\pm }$) for
another ($n_{\pm }$). Both of them include the interaction to all
orders and without any further simplification we are stuck. In
fact, we have already made an important simplification: when
specifying the model, we assumed only forward scattering. This
means that the interaction is sufficiently long-range in real
space so that backscattering can be neglected. Equivalently, in
the momentum space it means that our interaction operates only in
a narrow window of width $q_{0}$ near the Fermi points, $\pm
k_{F}.$ Thus the states far away from the Fermi points are not
affected by the interaction. The momentum integral in (\ref
{bubble_2}) comes from regions far away from the Fermi surface
where unknown functions $n_{\pm }$ can be approximated by free
Fermi steps. This approximation is good as long as $q_{0}\ll
k_{F}.$ The solution is going to be exact only in a sense that
there will be no constraints on the amplitude of the interaction
(parameters $g_{2}$ and $g_{4})$ but not its
range.\footnote{%
In higher dimensions, we have a familiar problem of the Coulomb
potential. Because it's a power-law potential, one cannot separate
it into ``amplitude'' and ``range''. There is in fact a single
dimensionless parameter, $r_{s}$, which must be small for the
perturbation theory--Random Phase Approximation--to work. Once
$r_{s}\ll 1$, we have two things: the screened potential is
simultaneously weak \emph{and }long-ranged. The Tomonaga-Luttinger
model unties these two things: the interaction is assumed to be
long-ranged but not necessarily weak.}Now we understand better why
the title of the paper by Dzyaloshinskii and Larkin \cite{DL} is
``Correlation functions for a one-dimensional Fermi system with
\emph{long-range }interaction (Tomonaga
model)''\footnote{%
What seemed to be just a matter of mathematical convenience in the
70s, turns out to be quite a realistic case these days. If a wire
of width $a$ is located at distance $d$ to the metallic gate, the
Coulomb potential between
electrons in the wire is screened by their images in the gate. Typically, $%
d\gg a.$ A simple exercise in electrostatics shows that in this
case $U(0)$ is larger than $U(2k_{F})$ by large factor $\ln \left(
d/a\right) $ \cite{ruzin}.}.

With this simplification, the momentum integration proceeds in the same way
as for free fermions (see Appendix \ref{sec:free_bubble}) with the result
\emph{that the fully interacting bubbles are the same as free ones}
\begin{equation}
\Pi _{\pm }\left( \omega ,q\right) =\Pi _{\pm }^{0}\left( \omega ,q\right)
=\pm \frac{1}{\pi }\frac{q}{\omega -q+i0^{+}\mathrm{sgn}\omega }.
\label{int=free}
\end{equation}
This is a truly remarkable result which is a cornerstone for the DL solution
\footnote{%
In QED, this statement is known as Furry theorem (W. H. Furry, 1937)}.

Because our bubbles were effectively ``liberated'' from the interaction
effects, system (\ref{rpa}) is equivalent to what we would have obtained
from the Random Phase Approximation (RPA). It turns out that RPA is \emph{%
asymptotically }exact in 1D in the limit $q_{0}/k_{F}\rightarrow
0.$ Solving the 2 by 2 system, we obtain for the effective
interaction
\[
V_{++}\left( \omega ,q\right) =\left( \omega -q\right) \frac{g_{4}\left(
\omega +q\right) +\left( g_{4}^{2}-g_{2}^{2}\right) q/\pi }{\omega
^{2}-u^{2}q^{2}+i0^{+}},
\]
where \footnote{%
Notice that as long as $g_{4}\neq g_{2},$ the left-right symmetry is broken,
i.e., the potential is not symmetric with respect to $q\rightarrow -q.$}
\[
u=\sqrt{1+\frac{2g_{4}}{\pi }+\frac{g_{4}^{2}-g_{2}^{2}}{\pi }}.
\]
For $g_{4}=g_{2}\equiv g,$
\begin{equation}
V_{++}\left( \omega ,q\right) =g\frac{\omega ^{2}-q^{2}}{\omega
^{2}-u^{2}q^{2}+i0^{+}}.  \label{v_symm}
\end{equation}

\subsection{Dyson equation for the Green's function}

\begin{figure}[tbp]
\begin{center}
\epsfxsize=0.5 \columnwidth
\epsffile{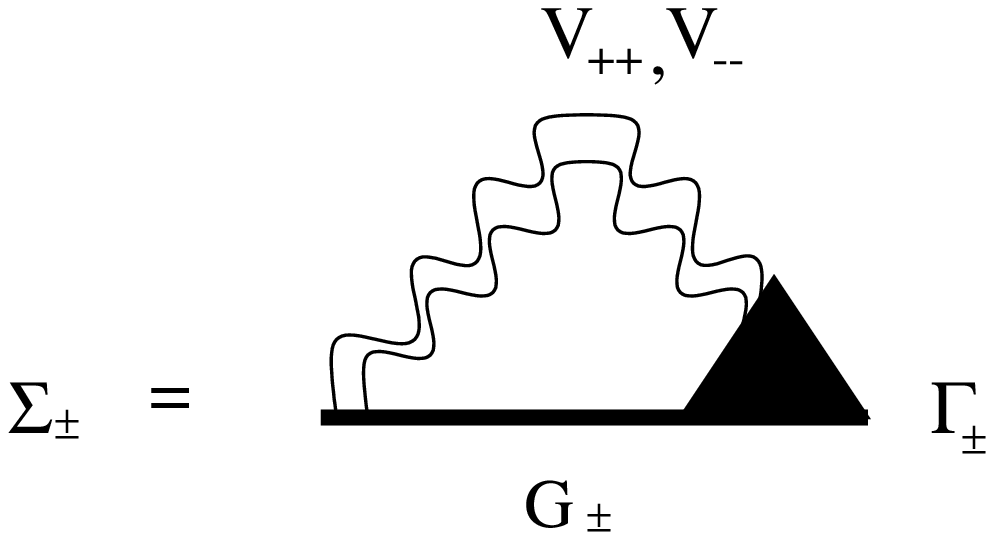}
\end{center}
\caption{Dyson equation for the self-energy.}
\label{fig:dysonself}
\end{figure}

The Dyson equation for right-moving fermions reads
\[
\Sigma _{+}\left( P\right) =i\int \frac{d^{2}Q}{\left( 2\pi \right) ^{2}}%
G_{+}\left( P-Q\right) V_{++}\left( Q\right) \Gamma _{+}^{0}\left(
P,Q\right) .
\]
Diagrammatically, this equation is shown in Fig. \ref{fig:dysonself}.
For linear dispersion,
\[
\Sigma _{\pm }\left( \varepsilon ,p\right) =\varepsilon \mp p-G_{\pm
}^{-1}\left( \varepsilon ,p\right)
\]
Substituting this relation back into the Dyson equations, we obtain
\[
\left( \varepsilon -p\right) G_{+}\left( \varepsilon ,p\right) =1+i\int \int
\frac{d\omega dq}{\left( 2\pi \right) ^{2}}G_{+}\left( \varepsilon ,p\right)
G_{+}\left( \varepsilon -\omega ,p-q\right) V_{++}\left( \omega ,q\right)
\Gamma _{+}^{0}\left( \varepsilon ,p;\omega ,q\right) .
\]
Using the Ward identity (\ref{ward_final}) , we get
\begin{eqnarray*}
\left( \varepsilon -p+\Sigma _{0}\right) G_{+}\left( \varepsilon ,p\right)
=1+i\int \int \frac{d\omega dq}{\left( 2\pi \right) ^{2}}G_{+}\left(
\varepsilon ,p\right) G_{+}\left( \varepsilon -\omega ,p-q\right) \\\times\frac{%
V_{++}\left( \omega ,q\right) }{\omega -q}
 \left[ G_{+}^{-1}\left( \varepsilon ,p\right) -G_{+}^{-1}\left(
\varepsilon -\omega ,p-q\right) \right]  \\
=1+i\int \int \frac{d\omega dq}{\left( 2\pi \right) ^{2}}G_{+}\left(
\varepsilon -\omega ,p-q\right) \frac{V_{++}\left( \omega ,q\right) }{\omega
-q}+G_{+}\left( \varepsilon ,p\right) \times \mathrm{const,}
\end{eqnarray*}
where
\[
\mathrm{const=}i\int \int \frac{d\omega dq}{\left( 2\pi \right) ^{2}}\frac{%
V_{++}\left( \omega ,q\right) }{\omega -q}
\].
A constant term can  always be absorbed into $\Sigma ,$ which simply results
in a shift of the chemical potential. We are free to choose this shift in
such a way that \textrm{const=0, }so that the Dyson equation reduces to
\begin{equation}
\left( \varepsilon -p\right) G_{+}\left( \varepsilon ,p\right) =1+i\int \int
\frac{d\omega dq}{\left( 2\pi \right) ^{2}}G_{+}\left( \varepsilon -\omega
,p-q\right) \frac{V_{++}\left( \omega ,q\right) }{\omega -q}.
\label{dyson_g}
\end{equation}
Notice that Eq.~(\ref{dyson_g}) is an integral equation with a difference
kernel, which can be reduced to a differential equation for $G.$ Before we
demonstrate how it is done, let's have a brief look at a case when there is
no coupling between left- and right-moving fermions: $g_{2}=0.$ In this
case,
\[
V_{++}\left( \omega ,q\right) =\pi \frac{\left( w-1\right) \left( \omega
-q\right) }{\omega -wq+i0^{+}},
\]
where
\[
w=1+g_{4}/\pi .
\]
Eq.~(\ref{dyson_g}) takes the form
\[
\left( \varepsilon -p\right) G_{+}\left( \varepsilon ,p\right) =1+i\left(
w-1\right) \int \int \frac{d\omega dq}{4\pi }\frac{G_{+}\left( \varepsilon
-\omega ,p-q\right) }{\omega -wq+i0^{+}}.
\]
This equation is satisfied by the following function
\begin{equation}
G_{+}\left( \varepsilon ,p\right) =\frac{1}{\sqrt{\varepsilon -p+i0^{+}}%
\sqrt{\varepsilon -wp+i0^{+}}}.  \label{branch}
\end{equation}
This is an example of a non-Fermi-liquid behavior: the pole of a
free $G$ splits into the product of two branch cuts, one peaked on
the mass shell of free
fermions ($\varepsilon =p)$ and another one at the renormalized mass shell ($%
\varepsilon =wp$). As left- and right movers are totally decoupled
in this problem, the same result would have been obtained for two
separate subsystems of left- and right movers. For example,
Eq.~(\ref{branch}) predicts that an edge state of an \emph{integer
}quantum Hall system is not a Fermi liquid, if spins are not yet
polarized by the magnetic field \cite{larkin_finn}. The same
procedure for a spinless system would give us a pole-like $G$ with
a
renormalized Fermi velocity. The non-Fermi-liquid behavior described by Eq.~(%
\ref{branch}) is rather subtle: it exists only if both
$\varepsilon $ and $p$ are finite. In the limiting case of $p=0$
(tunneling DoS) we are back to a free-fermion behavior $G\left(
\varepsilon ,0\right) =\varepsilon ^{-1}.$ Also, the $\varepsilon
-$ integral of Eq.~(\ref{dyson_g}) gives a step-like distribution
function in  momentum space. The spectral function, however, is
characteristically non-FL-like: instead of delta-function peak we
have a whole region $\left| p\right| <\left| \varepsilon \right|
<w\left| p\right| $ in which Im$G$ is finite. At the edges of this
interval Im$G$ has square-root singularities.

\subsection{Solution for the case $g_{2}=g_{4}$}

Substituting the effective interaction (\ref{v_symm}) into Dyson
equation (\ref{dyson_g}), we obtain
\[
\left( \varepsilon -p\right) G_{+}\left( \varepsilon ,p\right) =1+i\int
\frac{d\omega dq}{4\pi ^{2}}G\left( \varepsilon -\omega ,p-q\right) g\left(
q\right) \frac{\omega +q}{\omega ^{2}-u^{2}q^{2}},
\]
where
\[
u=\sqrt{1+2g/\pi }.
\]
Notice that the constant $g$ is replaced by a momentum-dependent
interaction, $g\left( q\right) .$ The reason is that without such a
replacement the integral diverges at the upper limit. Here, the assumption
of a cut-off in the interaction becomes important again. Transforming back
to real time and space
\[
G\left( x,t\right) =\int \int \frac{d\varepsilon dp}{\left( 2\pi \right) ^{2}%
}G\left( \varepsilon ,p\right) e^{i\left( px-\varepsilon t\right) },
\]
we obtain the Dyson equation in a differential form
\begin{equation}
\left( \frac{\partial }{\partial t}+\frac{\partial }{\partial x}\right)
G\left( x,t\right) =P\left( x,t\right) G\left( x,t\right) -i\delta \left(
x)\delta (t\right) ,  \label{dyson_diff}
\end{equation}
where
\begin{equation}
P\left( x,t\right) =\frac{1}{4\pi ^{2}}\int \int d\omega dqe^{i\left(
qx-\omega t\right) }g\left( q\right) \frac{\omega +q}{\omega
^{2}-u^{2}q^{2}+i0^{+}}.  \label{P}
\end{equation}
The integral for $P$ diverges if $g$ is constant. To ensure
convergence, we will approximate $g\left( q\right) =ge^{-\left|
q\right| /q_{0}}.$ An actual form of the cut-off function is not
important as long as we are interested in such times and spatial
intervals such that $x,t\gg q_{0}^{-1}.$ The integral over $\omega
$ is solved by closing the contour around the poles of the
denominator $\omega =\pm u\left| q\right|
\left( 1+i0^{+}\right) .$ For $t>0,$ we need to choose the one with Im$%
\omega <0.$ Doing so, we obtain
\[
\int \frac{d\omega }{2\pi }\dots =i\frac{\mathrm{sgn}q+u}{2u}e^{i\left(
qx-u\left| qt\right| \right) }.
\]
Solving the remaining $q-$ integral, we obtain for $P\left( x,t\right) $%
\[
P\left( x,t\right) =\frac{g}{4\pi u}\left( \frac{u+1}{x-ut+i/q_{0}}-\frac{u-1%
}{x+ut+i/q_{0}}\right) .
\]
For $t<0,$ one needs to change $q_{0}\rightarrow -q_{0}$ in the last formula.

The delta-function term can be viewed as a boundary condition
\begin{equation}
G\left( x,0+\right) -G\left( x,0-\right) =-i\delta \left( x\right) .
\label{bc}
\end{equation}
Once the function $P\left( x,t\right) $ is known, Eq.~(\ref{dyson_diff}) is
trivially solved in terms of new variables $r=x-t,$ $s=x+t.$ For example, for $t>0$%
\begin{equation}
G_{+}\left( r,t>0\right) =G_{0}\left( x,t\right) f_{>}\left( r\right) \exp
\left[ i\int_{r}^{s}ds^{\prime }P\left( r,s^{\prime }\right) \right] ,
\label{g_sol}
\end{equation}
where function $f_{>}(r)$ is determined by the analytic properties of $G$ as
a function of $\varepsilon .$ Substituting result for $P\left( x,t\right) $
into Eq.~(\ref{g_sol}), we get
\[
G_{+}\left( x,t>0\right) =\frac{1}{2\pi }G_{0}\left( x,t\right) f_{>}\left(
x-t\right) \left( \frac{x-t+i/q_{0}}{x-ut+i/q_{0}}\right) ^{\alpha
+1/2}\left( \frac{x-t-i/q_{0}}{x+ut-i/q_{0}}\right) ^{\alpha }.
\]
where
\begin{equation}
\alpha =\frac{\left( u-1\right) ^{2}}{8u}. \label{alpha_bulk}\eeq
Formula for $t<0$ is obtained by choosing another function $f_{<}$
and replacing $q_{0}\rightarrow -q_{0}.$ Functions $f_{>,<}$ are
determined from the analytic properties. First of all, recall that
\[
G_{0}\left( x,t\right) =\frac{1}{x-t+i\mathrm{sgn}t0^{+}}.
\]
We see that although $G_{0}$ is \emph{not }an analytic function of
$t$ for any $t,$ it is analytic for Re$t>0$ in the right lower
quadrant (Im$t<0)$ and for Re$t<0$ in the upper left quadrant
(Im$t>0$). The interaction cannot change analytic properties of a
Green's function hence we
should expect the same properties to hold for full $G.$ \footnote{%
Indeed, this property follows immediately from the Lehmann representation
for $G$
\begin{eqnarray*}
G\left( x,t\right) &=&-i\sum_{\nu }\left| M_{\nu 0}\right|
^{2}e^{ip_{\nu }x}e^{-i\left( E_{\nu }-E_{0}\right)
t},\;\mathrm{for}\;t>0;
\\
&=&i\sum_{\nu }\left| M_{\nu 0}\right| ^{2}e^{-ip_{\nu
}x}e^{i\left( E_{\nu }-E_{0}\right) t},\;\mathrm{for}\;t<0,
\end{eqnarray*}
where $M_{\nu 0}$ are the matrix elements between the ground state and state
$\nu $ with energy $E_{\nu }>E_{0}.$ The required property simply follows
from the condition for convergence of the sum.}

From the boundary condition (\ref{bc}), it follows that
\begin{eqnarray*}
f_{>}\left( x\right) &=&f_{<}\left( x\right) \\
\mathrm{and}\; f\left( 0\right) &=&0.
\end{eqnarray*}
Analyzing different factors in the formula for $G,$ we see that only the
term $\left( x-t\mp i/q_{0}\right) ^{\alpha }$ does not satisfy the required
analyticity property. This term is eliminated by choosing function $f\left(
x\right) $ as
\[
f\left( x\right) =\left( q_{0}^{2}x^{2}+1\right) ^{-\alpha }.
\]
Finally, the result for $G$ takes the form
\begin{eqnarray*}
G_{+}\left( x,t\right) &=&\frac{1}{2\pi }\frac{1}{x-t+i\mathrm{sgn}t0^{+}}%
\left( \frac{x-t+i\gamma }{x-ut+i{\gamma }}\right) ^{1/2} \\
&&\times \frac{1}{\left[ q_{0}^{2}\left( x-ut+i{\gamma }\right) \left( x+ut-i%
{\gamma }\right) \right] ^{\alpha }},
\end{eqnarray*}
where $\gamma =\mathrm{sgn}t/q_{0}.$ It seems somewhat redundant to keep two
different damping terms ($i\mathrm{sgn}t0^{+}$ and $\gamma )$ in the same
equation. However, these terms contain different physical scales. Indeed, $i%
\mathrm{sgn}t0^{+}$ enters a free Green's function and $0^{+}$ there has to
be understood as the limit of the inverse system size. On the other hand, $%
\gamma $ contains a cut-off of the interaction. Obviously, $|\gamma |\gg
1/L\rightarrow 0^{+}$ for a realistic situation. The difference between the
two cutoffs becomes important for the momentum distribution function and
tunneling DoS, discussed in the next Section.

\subsection{Physical properties}

\subsubsection{Momentum distribution}

Having an exact form of the Green's function, we can now calculate the
momentum distribution of, \emph{e.g., }right-moving fermions:
\begin{eqnarray*}
n_{+}\left( p\right)  &=&-i\int_{-\infty }^{\infty }dxe^{-ipx}G_{+}\left(
x,t\rightarrow 0^{+}\right)  \\
&=&-\frac{i}{2\pi }\int_{-\infty }^{\infty }dx\frac{e^{-ipx}}{x+i0^{+}}\frac{%
1}{\left[ q_{0}^{2}x^{2}+1\right] ^{\alpha }} \\
&=&-\frac{i}{2\pi }\int_{-\infty }^{\infty }dxe^{-ipx}\left[ \mathcal{P}%
\frac{1}{x}-i\pi \delta \left( x\right) \right] \frac{1}{\left[
q_{0}^{2}x^{2}+1\right] ^{\alpha }} \\
&=&\frac{1}{2}-\frac{1}{\pi }\text{sgn}p\int_{0}^{\infty }dx\frac{\sin
\left| p\right| x}{x}\frac{1}{\left[ q_{0}^{2}x^{2}+1\right] ^{\alpha }},
\end{eqnarray*}
We are interested in the behavior at $p\rightarrow 0$ (which means $\left|
p\right| \ll q_{0}).$ The final result for $n_{+}\left( p\right) $ depends
on whether $\alpha $ is larger or smaller than $1/2$ \cite{DL,guttfreund}.

\begin{itemize}
\item  For $\alpha <1/2$ (``weak interaction''), one cannot expand $\sin px$
in $x$ because the resulting integral diverges at $x=\infty .$ Instead,
rescale $px\rightarrow y$
\[
n_{+}\left( p\right) =\frac{1}{2}-\frac{1}{\pi }\int_{0}^{\infty }dy\frac{%
\sin y}{y}\frac{1}{\left[ \left( q_{0}/p\right) ^{2}y^{2}+1\right] ^{\alpha }%
}
\]
and neglect $1$ in the denominator. This gives
\begin{equation}
n_{+}\left( p\right) =\frac{1}{2}+C_{1}\left( \frac{\left| p\right| }{q_{0}}%
\right) ^{2\alpha }\text{sgn}p  \label{np_small}
\end{equation}
where
\[
C_{1}=\frac{\sin \pi \alpha }{\pi }\Gamma \left( -2\alpha \right) .
\]
Notice that $n_{+}\left( p\right) $ is finite ($=1/2$) at $p=0,$ although
its derivative is singular. We should be able to recover the Fermi-gas step
at $p=0$ by setting $\alpha =0$ in (\ref{np_small}). Indeed,
\[
\lim_{\alpha \rightarrow 0}C_{1}=\alpha \frac{1}{-2\alpha }=-\frac{1}{2}
\]
and
\[
n\left( p\right) =\frac{1-\text{sgn}p}{2},
\]
which is just the Fermi-gas result. Notice also that there is
nothing special about the limit $\alpha \rightarrow 0$ \footnote{%
contrary to some statements in the literature.}. Indeed,
constant $C_{1}$ has a regular expansion in $\alpha $%
\[
C_{1}=-\frac{1}{2}-\gamma \alpha+\dots ,
\]
where $\gamma =0.577 \dots$ and factor $\left( \left| p\right|
/q_{0}\right) ^{2\alpha }$ can be expanded for finite $p$ and
small $\alpha $ as
\[
\left( \left| p\right| /q_{0}\right) ^{2\alpha }=1+2\alpha \ln \left|
p\right| /q_{0}.
\]
To leading order in $\alpha $, we obtain
\[
n_{+}\left( p\right) =\frac{1}{2}-\text{sgn}p\frac{1}{2}\left[ 1+2\alpha \ln
\left| p\right| /q_{0}\right] =n_{0}\left( p\right) -\alpha \text{sgn}p\ln
\left| p\right| /q_{0},
\]
which is a perfectly regular in $\alpha $ (but logarithmically divergent at $%
p\rightarrow 0)$ behavior. Once again, it is not surprising: despite the
fact that the results for a 1D system differ dramatically from that for the
Fermi gas, they are still \emph{perturbative, i.e., analytic, }in the
coupling constant.

\item  For $\alpha >1/2$ (``strong interaction''), it is safe to expand $%
\sin px$ and the result is
\[
n_{+}\left( p\right) =\frac{1}{2}-C_{2}p/q_{0},
\]
where
\[
C_{1}=\frac{1}{2\sqrt{\pi }}\frac{\Gamma \left( \alpha -1/2\right) }{\Gamma
\left( \alpha \right) }.
\]
In this case, no remains of a jump at the Fermi point is present in $%
n_{+}\left( p\right) $ which is a regular, linear function near $p=0.$

\item  Finally, $\alpha =1/2$ is a special case, where expansion in $p$
results in a log-divergent integral. To log-accuracy
\[
n_{+}\left( p\right) =\frac{1}{2}-\frac{1}{\pi }\frac{p}{q_{0}}\ln \frac{%
q_{0}}{\left| p\right| }.
\]
\end{itemize}

In general, $n\left( p\right) $ is some hypergeometric function of $p/q_{0}$
which decays rapidly for $p\gg q_{0}$ and approaches $1$ for $p\ll -q_{0}$%
\textbf{. }\emph{A posteriori, }this justifies the replacement of exact $%
n\left( p\right) $ by its free form in the Dyson equation.

\subsubsection{Tunneling density of states}

Now we turn to the tunneling DoS

\[
N\left( \varepsilon \right) =-\frac{1}{\pi }\mathrm{Im}G^{R}\left(
\varepsilon ,x=0\right) .
\]
Recalling that \cite{agd}
\begin{eqnarray*}
G^{R}\left( \varepsilon \right) &=&G\left( \varepsilon \right) ,\mathrm{for}%
\text{ }\mathrm{\varepsilon >0;} \\
&=&G^{\ast }\left( \varepsilon \right) ,\mathrm{\ for}\text{ }\mathrm{%
\varepsilon <0,}
\end{eqnarray*}
we see that
\[
\mathrm{Im}G^{R}\left( \varepsilon ,0\right) =\text{sgn}\varepsilon \mathrm{%
Im}G\left( \varepsilon ,0\right)
\]
and
\begin{eqnarray*}
N\left( \varepsilon \right) &=&-\frac{1}{\pi }\text{sgn}\varepsilon \mathrm{%
Im}G\left( 0,\varepsilon \right) =-\frac{1}{\pi }\text{sgn}\varepsilon \left[
\int dte^{i\varepsilon t}G\left( 0,t\right) -\int dte^{-i\varepsilon
t}G^{\ast }\left( 0,t\right) \right] \\
&=&-\frac{1}{\pi }\text{sgn}\varepsilon \left[ \int dte^{i\varepsilon
t}\left\{ G\left( 0,t\right) -G^{\ast }\left( 0,-t\right) \right\} \right]
\end{eqnarray*}
For $t\rightarrow \infty ,$%
\[
G\left( 0,t\right) =\frac{\text{const}}{\left( -t\right) ^{1+2\alpha }}
\]
and
\[
G\left( 0,t\right) -G^{\ast }\left( -t\right)
\]
is an odd function of $t.$ Thus
\begin{eqnarray*}
N\left( \varepsilon \right) &=&-\frac{1}{\pi }\text{sgn}\varepsilon \mathrm{%
Im}G\left( 0,\varepsilon \right) =-\frac{1}{\pi }\text{sgn}\varepsilon \frac{%
1}{2i}\left[ \int dte^{i\varepsilon t}G\left( 0,t\right) -\int
dte^{-i\varepsilon t}G^{\ast }\left( 0,t\right) \right] \\
&=&-\frac{1}{\pi }\text{sgn}\varepsilon \left[ \int_{0}^{\infty }dt\sin
\varepsilon t\left\{ G\left( 0,t\right) -G^{\ast }\left( 0,-t\right)
\right\} \right] \propto \text{sgn}\varepsilon \int_{0}^{\infty }dt\frac{%
\sin \varepsilon t}{t^{1+2\alpha }}.
\end{eqnarray*}
The integral is obviously convergent for $\alpha <1/2.$ In this case,
\[
N_{s}\left( \varepsilon \right) \propto \left| \varepsilon \right|
^{2\alpha }, \label{nubulk}\] which means that the local tunneling
DoS is suppressed at the Fermi level. Actually, the exponent for
$\alpha
>1/2$ is the same, however, the prefactor is a different function
of $\alpha $ \cite{voit}.

The DoS in Eq.~(\ref{nubulk}) with exponent $2\alpha $, where
$\alpha$ is given by Eq.~(\ref{alpha_bulk}) corresponds to
tunneling into the ``bulk'' of a 1D system, \emph{i.e.}, when the
tunneling contact (with a tip of an STM or another carbon nanotube
crossing the first one) is far away from its ends. In the next
Section, we will analyze tunneling into an edge of a 1D conductor,
which is characterized by a different exponent, $\alpha'$.

%%%%%%%%%%%%%%%%%%%%%%%%%%%%%%%%%%%%%%%%%%%%%%%%%%%%%%%%%%
%%%%%%%%%%% LECTURE 3
%%%%%%%%%%%%%%%%%%%%%%%%%%%%%%%%%%%%%%%%%%%%%%%%%%%%%%%%
\section{Renormalization group for interacting fermions}
\label{sec:RG}The Tomonaga-Luttinger model can be solved exactly
as it was done in the previous Section-- only in the absence of
backscattering. Backscattering can be treated via the
Renormalization Group (RG) procedure. This treatment is standard
by now and discussed in a number of sources
\cite{solyom,emery,schulz_95,shankar,voit_review,tsvelik,fisher_glazman,vondelft,GNT,giamarchi_book}.
For the sake of completeness, I present here a short derivation of
the RG equations. A reader familiar with the procedure can skip
this Section and go directly to Sec. \ref{sec:ygm}, where these
equations will be used in the context of a single-impurity
problem.

An exact solution of the previous Section is parameterized by two
coupling constants, $g_{2}$ and $g_{4}$, which are equal to their
bare values. In the RG language, it means that these couplings do
not flow. Let's see if this is indeed the case. In what follows, I
will neglect the $g_{4}-$ processes, as their effect on the flow
of other couplings is trivial, and, for the sake of simplicity,
consider a spin-independent interaction. To second order, the
renormalization of the $g_{2}-$ coupling is accounted for by two
diagrams: diagrams a) and b) of Fig. \ref{fig:vertex1}.

\textbf{Diagram a) }is a correction to $g_{2}$ in the particle-particle
channel. The correction to $g_{2}$ is given by
\begin{equation*}
\left( g_{2}^{\left( 2\right) }\right)
_{a}=\frac{g_{2}^{2}}{\left( 2\pi \right) ^{2}}\int dq\int d\omega
G_{+}\left( i\varepsilon _{1}+i\omega ,k_{1}+q\right) G_{-}\left(
i\varepsilon _{2}-i\omega ,k_{2}-q\right).
\end{equation*}
Without a loss of generality, one can choose all momenta to be on the  Fermi
``surface'': $k_{1}=k_{2}=k_{3}=k_{4}=0.$ Choose $q>0$ (the other choice $q<0
$ will simply double the result)
\begin{eqnarray*}
\left( g_{2}^{\left( 2\right) }\right) _{a} &=&\frac{g_{2}^{2}}{\left( 2\pi
\right) ^{2}}\int dq\int d\omega G_{+}\left( i\varepsilon _{1}+i\omega
,q\right) G_{-}\left( i\varepsilon _{2}-i\omega ,-q\right)  \\
&=&\frac{g_{2}^{2}}{\left( 2\pi \right) ^{2}}\int_{0}^{\Lambda%
/2}dq\int d\omega \frac{1}{i\left( \varepsilon _{1}+\omega \right) -q}\frac{1%
}{i\left( \varepsilon _{2}-\omega \right) -q} \\
&=&\frac{2\pi ig_{2}^{2}}{\left( 2\pi \right) ^{2}}\int_{0}^{\Lambda
/2}dq\frac{1}{\varepsilon _{1}+\varepsilon _{2}+\omega +2iq}=\frac{g_{2}^{2}%
}{4\pi }\ln \frac{i\Lambda}{\varepsilon _{1}+\varepsilon _{2}}
\end{eqnarray*}
Adding the result up with the (identical) $q<0$ contribution, we find
\begin{equation*}
\left( g_{2}^{\left( 2\right) }\right) _{a}=\frac{g_{2}^{2}}{2\pi }\ln \frac{%
i\Lambda}{\varepsilon _{1}+\varepsilon _{2}}.
\end{equation*}
\textbf{Diagram b)  }is a correction to $g_{2}$ in the particle-hole channel:
\begin{eqnarray*}
\left( g_{2}^{\left( 2\right) }\right) _{b} &=&\frac{g_{2}^{2}}{\left( 2\pi
\right) ^{2}}\int dq\int d\omega G_{+}\left( i\varepsilon _{1}+i\omega
,q\right) G_{-}\left( i\varepsilon _{4}+i\omega ,q\right)  \\
&=&\frac{g_{2}^{2}}{\left( 2\pi \right) ^{2}}\int_{0}^{\Lambda%
/2}dq\int d\omega \frac{1}{i\left( \varepsilon _{1}+\omega \right) -q}\frac{1%
}{i\left( \varepsilon _{4}+\omega \right) +q} \\
&=&-\frac{2\pi i}{\left( 2\pi \right) ^{2}}g_{2}^{2}\int_{0}^{\mathbf{%
\Lambda }/2d}q\frac{1}{\varepsilon _{1}-\varepsilon _{4}+\omega +2iq}=-\frac{%
g_{2}^{2}}{4\pi }\ln \frac{i\Lambda}{\varepsilon _{1}-\varepsilon
_{4}}.
\end{eqnarray*}
As in the previous case, the final result is:
\begin{equation*}
\left( g_{2}^{\left( 2\right) }\right) _{b}=-\frac{g_{2}^{2}}{2\pi }\ln
\frac{i\Lambda}{\varepsilon _{1}-\varepsilon _{4}}.
\end{equation*}

\begin{figure}[tbp]
\begin{center}
\epsfxsize=0.8 \columnwidth \epsffile{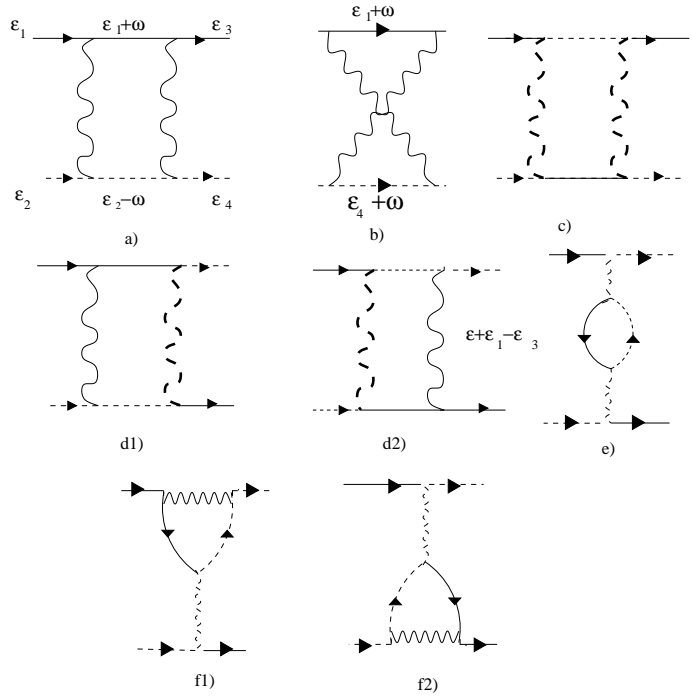}
\end{center}
\caption{ Second order diagrams for couplings $g_{2}$ (solid wavy line) and $%
g_{1}$ (dashed wavy line). Straight solid and dashed lines correspond to
Green's functions of right- and left moving fermions, correspondingly.}
\label{fig:vertex1}
\end{figure}

If we sum only the Cooper ladders, adding up more vertical interaction lines
to diagram a), the full vertex becomes
\begin{equation*}
\Gamma _{pp}=\frac{g_{2}}{1+g_{2}\ln \frac{i\Lambda}{\varepsilon
_{1}+\varepsilon _{2}}}.
\end{equation*}
(To keep track of the signs, one needs to recall that in Matsubara
frequencies each interaction line comes with the minus sign from
the expansion of the $S-$ matrix). The resulting vertex blows up
for attractive interaction ($g_{2}<0$) as $\varepsilon
_{1}+\varepsilon _{2}\rightarrow 0,$ which is nothing more than a
Cooper instability.

Likewise, untwisting the crossed lines in diagram b) and adding
more interaction lines, we get the particle-hole vertex
\begin{equation*}
\Gamma _{ph}=\frac{g_{2}}{1-g_{2}\ln \frac{i\Lambda}{\varepsilon
_{1}-\varepsilon _{4}}}.
\end{equation*}
This vertex has an instability for repulsive interaction ($g_{2}>0).$ In
fact, none of these instabilities occur. To see this, add up the results of
diagrams a) and b)
\begin{equation*}
\left( g_{2}^{\left( 2\right) }\right) _{a+b}=\frac{g_{2}^{2}}{2\pi }\left[
\ln \frac{i\Lambda}{\varepsilon _{1}+\varepsilon _{2}}-\ln \frac{i%
\Lambda}{\varepsilon _{1}-\varepsilon _{4}}\right] =\frac{g_{2}^{2}%
}{2\pi }\ln \frac{\varepsilon _{1}-\varepsilon _{4}}{\varepsilon
_{1}+\varepsilon _{2}}.
\end{equation*}
In the RG, one changes the cut-off and follow the corresponding evolution of
the couplings. As the cut-off dependence cancelled out in the result for $%
\left( g_{2}^{\left( 2\right) }\right) _{a+b},$ coupling $g_{2}$ remains
invariant under the RG flow.

Backscattering generates additional diagrams: diagrams c)-f) in
Fig.\ref{fig:vertex1}.

\textbf{Diagram c) }describes repeated backscattering in the
particle-particle channel, which is equivalent to forward scattering.
Therefore, this diagram gives a correction to $g_{2}-$ coupling. Using the
relation between $G_{\pm }$, \emph{i.e., }$G_{\pm }=-\left( G_{\mp }\right)
^{\ast },$ we find

\begin{eqnarray*}
\left( g_{2}^{\left( 2\right) }\right) _{c} &=&\frac{g_{1}^{2}}{\left( 2\pi
\right) ^{2}}\int dq\int d\omega G_{-}\left( i\varepsilon _{1}+i\omega
,q\right) G_{+}\left( i\varepsilon _{4}+i\omega ,q\right)  \\
&=&\frac{g_{1}^{2}}{\left( 2\pi \right) ^{2}}\left[ \int dq\int
d\omega G_{+}\left( i\varepsilon _{1}+i\omega ,q\right)
G_{-}\left( i\varepsilon _{4}+i\omega ,q\right) \right] ^{\ast }.
\end{eqnarray*}
The last integral is the same as for $\left( g_{2}^{\left( 2\right) }\right)
_{a}.$ Thus,
\begin{equation*}
\left( g_{2}^{\left( 2\right) }\right) _{c}=\frac{g_{1}^{2}}{g_{2}^{2}}\left[
dg_{2}^{\left( 1\right) }\right] ^{\ast }=\frac{g_{1}^{2}}{2\pi }\ln \frac{-i%
\Lambda}{\varepsilon _{1}+\varepsilon _{2}}.
\end{equation*}
The rest of the diagrams provide corrections to $g_{1}.$

\textbf{Diagram d1) \ }is the same as diagram a) except for the
prefactor being
equal to $g_{1}g_{2}:$%
\begin{equation*}
\left( g_{1}^{\left( 2\right) }\right) _{d1}=\frac{g_{1}g_{2}}{2\pi }\ln
\frac{i\Lambda}{\varepsilon _{1}+\varepsilon _{2}}
\end{equation*}

\textbf{Diagram d2) }is a complex-conjugate  of diagram d1). The
sum of diagrams d1) and d2) is equal to \textbf{\ }
\begin{eqnarray*}
\left( g_{1}^{\left( 2\right) }\right) _{d1+d2} &=&\frac{g_{1}g_{2}}{2\pi }%
\ln \frac{i\Lambda}{\varepsilon _{1}+\varepsilon _{2}}+\frac{%
g_{1}g_{2}}{2\pi }\ln \frac{-i\Lambda}{\varepsilon
_{1}+\varepsilon _{2}} \\
&=&\frac{g_{1}g_{2}}{\pi }\ln \frac{\Lambda}{\varepsilon
_{1}+\varepsilon _{2}}.
\end{eqnarray*}
\textbf{Diagrams e) }is a polarization\textbf{\ }correction\textbf{\ }to the
bare\textbf{\ }$g_{1}-$coupling:
\begin{equation*}
\left( g_{1}^{\left( 2\right) }\right) _{e}=\underbrace{-}_{\text{\textrm{%
fermionic loop}}}g_{1}^{2}\Pi _{2k_{F}}\left( \omega =\varepsilon
_{1}-\varepsilon _{2},q=0\right) .
\end{equation*}
Using Eq. (\ref{bubble_2kf})$,$ we obtain
\begin{equation*}
\left( g_{1}^{\left( 2\right) }\right) _{e}=\frac{N_{s}}{2\pi }g_{1}^{2}\ln
\frac{\Lambda}{\left| \varepsilon _{1}-\varepsilon _{2}\right| },
\end{equation*}
where $N_{s}$ is the degeneracy factor (=2 for spin 1/2 fermions,
occupying a single valley in the momentum space).

\textbf{Diagram f1) }is the\textbf{\ }same as the bubble
insertion, except for no minus sign, no degeneracy factor
($N_{s})$ factor, and the overall
coefficient is $g_{1}g_{2:}$%
\begin{equation*}
\left( g_{1}^{\left( 2\right) }\right) _{f1}=-\frac{1}{2\pi }g_{1}g_{2}\ln
\frac{\Lambda}{\left| \varepsilon _{1}-\varepsilon _{2}\right| }.
\end{equation*}
\textbf{Diagram f2) }is equal to f1). Their sum\textbf{\ \ }
\begin{equation*}
\left( g_{1}^{\left( 2\right) }\right) _{f1+f2}=-\frac{1}{\pi }g_{1}g_{2}\ln
\frac{\Lambda}{\left| \varepsilon _{1}-\varepsilon _{2}\right| }
\end{equation*}

Collecting all contributions together, we obtain
\begin{eqnarray*}
-\Gamma _{2} &=&-g_{2}+\left( g_{2}^{\left( 2\right) }\right) _{a}+\left(
g_{2}^{\left( 2\right) }\right) _{b}+\left( g_{2}^{\left( 2\right) }\right)
_{c}; \\
\Gamma _{2} &=&g_{2}\underbrace{-\frac{g_{2}^{2}}{2\pi }\ln \frac{i
\Lambda}{\varepsilon _{1}+\varepsilon _{2}}+\frac{g_{2}^{2}}{2\pi }\ln
\frac{i\Lambda}{\varepsilon _{1}-\varepsilon _{4}}}_{\text{\textrm{%
cancel out in the RG sense}}}-\frac{g_{1}^{2}}{2\pi }\ln \frac{-i
\Lambda }{\varepsilon _{1}+\varepsilon _{2}}; \\
-\Gamma _{1} &=&-g_{1}+\left( g_{2}^{\left( 2\right) }\right) _{d}+\left(
g_{2}^{\left( 2\right) }\right) _{e}+\left( g_{2}^{\left( 2\right) }\right)
_{f}; \\
\Gamma _{1} &=&g_{1}-\frac{g_{1}g_{2}}{\pi }\ln \frac{\Lambda}{%
\varepsilon _{1}+\varepsilon _{2}}-\frac{N_{s}}{2\pi }g_{1}^{2}\ln \frac{%
\Lambda}{\left| \varepsilon _{1}-\varepsilon _{2}\right| }+\frac{1%
}{\pi }g_{1}g_{2}\ln \frac{\Lambda}{\left| \varepsilon
_{1}-\varepsilon _{2}\right| }.
\end{eqnarray*}
Second and fourth terms in  $\Gamma _{1}$ also cancel out in the RG sense.
Changing the cut-off from $\Lambda$ to $\Lambda+d\
\Lambda$, we obtain two differential equations
\begin{eqnarray}
\frac{d\Gamma _{2}}{dl} &=&-\frac{\Gamma _{1}^{2}}{2\pi };  \label{rg1} \\
\frac{d\Gamma _{1}}{dl} &=&-N_{s}\frac{\Gamma _{1}^{2}}{2\pi },  \label{rg2}
\end{eqnarray}
where $l=\ln \Lambda.$ We see that a quantity
\begin{equation}
\bar{\Gamma}=\Gamma _{2}-\frac{1}{N_{s}}\Gamma _{1}=\text{const}=g_{2}-\frac{%
1}{N_{s}}g_{1}.  \label{invariant1}
\end{equation}
is invariant under RG flow, therefore its value can be obtained by
substituting the bare values of the coupling constants ($g_{2}$ and $g_{1})$
into  (\ref{invariant1}). The RG-invariant combination is then
\begin{equation}
\bar{\Gamma}=g_{2}-\frac{1}{N_{s}}g_{1}.  \label{invariant}
\end{equation}
For spinless electrons ($N_{s}=1)$,
\begin{equation*}
\bar{\Gamma}=\Gamma _{2}-\Gamma _{1}=U\left( 0\right) -U\left( 2k_{F}\right)
.
\end{equation*}
This last result can be understood just in terms of the Pauli principle.
Indeed, the anti-symmetrized vertex for spinless electrons is obtained by
switching the outgoing legs of the diagram ($p_{1},p_{2}\rightarrow
p_{3},p_{4}).$ To first order,
\begin{equation*}
\Gamma \left( p_{1},p_{2};p_{3},p_{4}\right) =U\left( p_{1}-p_{3}\right)
-U\left( p_{1}-p_{4}\right) .
\end{equation*}
Choosing $p_{3}=p_{1}-q$ and $p_{4}=p_{2}+q,$ we obtain [recall that $%
U\left( q\right) =U\left( -q\right) ]$%
\begin{equation*}
\Gamma \left( p_{1},p_{2}|q\right) =U\left( q\right) -U(p_{1}-p_{2}-q).
\end{equation*}
One of the incoming fermions is a right mover ($p_{1}=p_{F}$) and the other
one is a left mover ($p_{2}=-p_{F}$). As $q$ is small compared to $p_{F},$
we obtain
\begin{equation*}
\Gamma \left( p_{1},p_{2}|q\right) =U\left( 0\right) -U(2k_{F}).
\end{equation*}
In fact, for spinless electrons $g_{2}$ and $g_{1}$ processes are
indistinguishable\footnote{%
That does not mean that backscattering is unimportant! It comes
with a different scattering amplitude $U\left( 2k_{F}\right) .$ In
fact, it is only backscattering which guarantees that the Pauli
principle is satisfied, namely, for a contact interaction, when
$U\left( 0\right) =U\left( 2k_{F}\right) ,$ we must get back to a
Fermi gas as fermions are not allowed to occupy the same position
in space and hence they cannot interact via contact forced. Our
invariant combination $U\left( 0\right) -U\left( 2k_{F}\right) $
obviously satisfies this criterion. We will see that bosonization
does have a problem with respecting the Pauli principle, and it
takes some effort to recover it.} as we do not know whether the
right-moving electron in the final state is a right-mover of the
initial state, which experienced forward scattering, or the
left-mover of the initial state, which experienced backscattering.
A proper way to treat the case of spinless fermions is to include
backscattering into Dzyaloshinskii-Larkin scheme from the very
beginning, re-write the Hamiltonian in terms of \emph{forward
scattering }with invariant coupling $\bar{\Gamma},$ and proceed
with the solution. All the results will then be expressed in terms
of $\bar{\Gamma}$ rather than of $g_{2}.$

Solving the equation for $\Gamma _{1},$ gives on scale $\varepsilon $%
\begin{equation}
\Gamma _{1}=\frac{1}{\left( g_{1}\right) ^{-1}+\frac{N_{s}}{2\pi }\ln
\Lambda/\varepsilon }.  \label{g1_solution}
\end{equation}
At low energies, $\Gamma _{1}$ renormalizes to zero ($\Gamma _{1}^{\ast
}=\Gamma _{1}\left( l=\infty \right) =0)$, if the interaction is repulsive,
and blows up at $\varepsilon =\Lambda\exp \left( -1/|g_{1}|\right)
$, if the interaction is attractive. Coupling $\Gamma _{2}$ also flows to a
new value which can be read off from Eq.~(\ref{invariant})
\begin{equation*}
\Gamma _{2}^{\ast }=g_{2}-\frac{1}{N_{s}}g_{1}.
\end{equation*}
Roughly speaking, $g_{1}$ is not important for repulsive interaction as the
effective low-energy theory will look like a theory with forward scattering
only. This does not really mean, however, that one can consider a fixed
point as a new problem in which backscattering is absent, and apply our exact
solution to this problem. Instead, one should calculate observables, derive
the RG equations for flows, and use current values of coupling constants in
these RG equations. An example of this procedure will be given in the next
Section, where we will see that the flow of $\Gamma _{1}$ provides
additional renormalization of the transmission coefficient in an interacting
system.

Assigning different coupling constants to the interaction of fermions of
parallel ($g_{1||}$) and anti-parallel ($g_{1\perp }$) spins, one could see
that the coupling which diverges for attractive interaction is in fact $%
g_{1\perp }.$ This clarifies the nature of the gap that RG hints
at (in fact, a perturbative RG can at most just give a hint): it
is a spin gap. This becomes obvious in the bosonization technique,
as the instability occurs in the spin-sector of the theory. An
exact solution by Luther and Emery \cite{luther_emery} for a
special case of attractive interaction confirms this prediction.
%%%%%%%%%%%%%%%%%%%%%%%%%%%%LECTURE 4%%%%%%%%%%%%%%%%%%%%%%%%%%%%%%%%
\section{Single impurity in a 1D system: scattering theory for interacting fermions}
\label{sec:ygm} A single impurity or tunneling barrier placed  in
a 1D Fermi gas reduces the conductance from its universal
value--$e^2/h$ per spin orientation--to \beq {\cal
G}=N_s\frac{e^2}{h}|t_0|^2, \eeq where $t_0$ is the transmission
amplitude. The interaction renormalizes the bare transmission
amplitude. As a result, the conductance depends on the
characteristic energy scale (temperature or applied bias), which
is observed as a zero-bias anomaly in tunneling. This effect is
not really a unique property of 1D : in higher dimensions,
zero-bias anomalies in both dirty and clean (ballistic) regimes
\cite{aa,rag,reyzer_tunn,anton} as well as the interaction
correction to the conductivity \cite{aa,zna}, stem from the same
physics, namely, scattering of electrons from Friedel oscillations
produced by tunneling barriers or impurities. 1D is special in the
magnitude of the effect: the conductance varies significantly
already on the energy scale comparable to the Fermi energy,
whereas in higher dimensions the effect of the interaction is
either small at all energies or becomes significant only at low
energies (below some scale which is much smaller than $E_F$ as
long as the parameter $k_Fl$, where $l$ is the elastic mean free
path, is large. The 1D zero-bias anomaly is described quite simply
in a bosonized language \cite{kane_fisher}, which does not require
the interaction to be weak. We will use this description in
Sec.\ref{sec:bosonization}. However, in this Section I will choose
another description--via the scattering theory for fermions rather
than bosons--developed by Matveev, Yue, and Glazman \cite{YGM}.
Although this approach is perturbative in the interaction, it
elucidates the underlying mechanism of the zero-bias anomaly and
allows for an extension to higher-dimensional case (which was done
for the case of tunneling in Ref.\cite{rag} and transport in
Ref.\cite{zna}).
\subsection{First-order interaction correction to the transmission
coefficient} In this section we consider a 1D system of
\emph{spinless }fermions with a tunneling barrier located at $x=0$
\cite{YGM}. For the sake of simplicity, I assume that the barrier
is symmetric, so that transmission and reflection amplitude for
the waves coming from the left and right are the same. Also, I
assume that e-e interaction is present only to the right of the
barrier, whereas to the left we have a Fermi gas. Such a situation
models a setup when a tunneling contact separates a 1D interacting
system (quantum wire or carbon nanotube) and a ``good metal'',
where one can be neglect the interaction. We also assume that the
interaction potential $U\left( x\right) $ is sufficiently
short-ranged, so that $U\left( 0\right) $ is finite and one can
neglect over-the-barrier interaction. However, $U\left( 0\right)
\neq U\left( 2k_{F}\right) $ (otherwise, spinless
electrons do not interact at all\footnote{%
For a contact potential [which leads to $U\left( 0\right) =U\left(
2k_{F}\right) $], the four-fermion interaction for the spinless
case reduces to $\left[ \Psi ^{\dagger }\left( 0\right) \right]
^{2}\Psi ^{2}\left( 0\right) .$ By Pauli principle, $\left[ \Psi
^{\dagger }\left( 0\right) \right] ^{2}=\Psi ^{2}\left( 0\right)
=0,$ so that the interaction is absent.}).

The wave function of the free problem for a right-moving state is:
\begin{eqnarray}
\psi _{k}^{0}\left( x\right) &=&\frac{1}{\sqrt{L}}\left(
e^{ikx}+r_{0}e^{-ikx}\right) ,x<0;  \label{psi_k} \\
&=&\frac{1}{\sqrt{L}}t_{0}e^{ikx},x>0.  \nonumber
\end{eqnarray}
For a left-moving state:
\begin{eqnarray}
\psi _{-k}^{0}\left( x\right) &=&\frac{1}{\sqrt{L}}\left(
e^{-ikx}+r_{0}e^{ikx}\right) ,x>0;  \nonumber \\
&=&\frac{1}{\sqrt{L}}t_{0}e^{-ikx},x<0.  \label{psi_-k}
\end{eqnarray}
Here $k=\sqrt{2mE}>0.$ To begin with, we consider a high barrier:
$\left| t_{0}\right| \ll 1,r_{0}\approx -1.$ Then the free
wavefunction reduces to
\begin{eqnarray}
\psi _{k}^{0}\left( x\right) &=&\frac{2i}{\sqrt{L}}\sin kx,x<0\;\;\text{\textrm{%
(incoming from the left+reflected);}}  \label{ilr} \\
&=&\frac{1}{\sqrt{L}}t_{0}e^{ikx},x>0\;\;\text{\textrm{(transmitted left}}%
\rightarrow \text{\textrm{\ right);}}  \label{tlr}
\end{eqnarray}
\begin{eqnarray}
\psi _{-k}^{0}\left( x\right) &=&\frac{1}{\sqrt{L}}t_{0}e^{-ikx},x<0\;\;\text{%
\textrm{(transmitted right}}\rightarrow \text{\textrm{left);}}
\label{trl}
\\
&=&-\frac{2i}{\sqrt{L}}\sin kx,x>0\;\;\text{\textrm{(incoming from the
right+reflected).}}  \label{irl}
\end{eqnarray}
The barrier causes the Friedel oscillation in the electron density
on both sides of the barrier. The interaction is treated
perturbatively, via finding the corrections to the transmission
coefficient due to additional scattering at the potential produced
by the Friedel oscillation. Diagrammatically, the corrections to
the Green's function are described by the diagrams in Fig.\ref
{fig:HF}, where a) represents the Hartree and b) the exchange
(Fock) contributions, correspondingly. Compared to the textbook
case, though, the solid lines in these diagrams are the Green's
functions composed of the exact eigenstates in the presence of the
barrier (but no interaction). Because the barrier breaks
translational invariance, these Green's functions are not
translationally invariant as well. I emphasized this fact by
drawing the diagrams in  real space, as opposed to the momentum
-space representation. Notice also that the Hartree diagram is
usually discarded in textbooks because the bubble there
corresponds to the total charge density (density of electrons
minus that of ions), which is equal to zero in a translationally
invariant and neutral system. However, what we have in our case is
the \emph{local} density of electrons at some distance from the
barrier. Friedel oscillation is a relatively short-range
phenomenon (the period of the oscillation is comparable to the
electron wavelength), and it is possible to violate the charge
neutrality locally on such a scale. As a result, the Hartree
correction is not zero.

\begin{figure}[tbp]
\begin{center}
\epsfxsize=1.0 \columnwidth \epsffile{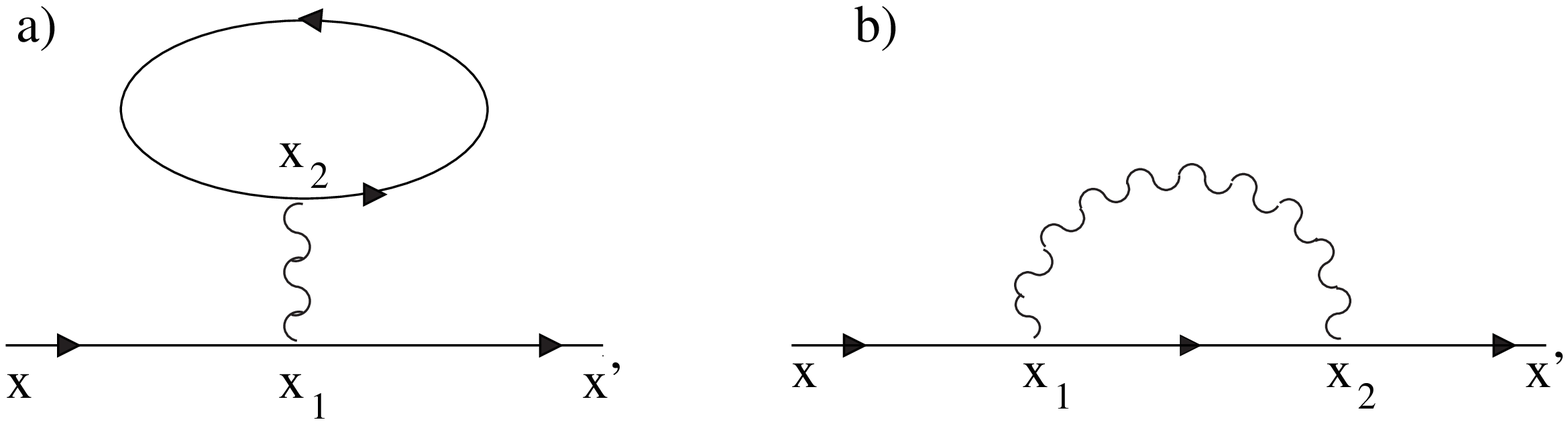}
\end{center}
\caption{ Correction to the Green's function: exact with respect
to the barrier and first order in the interaction.} \label{fig:HF}
\end{figure}

To first-order in the interaction, an equivalent way of solving
the problem is to find a correction to the wave-function, rather
than the Green's function, in the Hartree-Fock method. The
electron wave-function which includes both the barrier potential
and the electron-electron interaction is
\begin{eqnarray}
\psi _{k}(x)&=&\psi _{k}^{0}(x)+\int_{0}^{\infty }dx^{\prime }G_{0}^{>}(x%
\mathbf{,}x^{\prime },E)\notag\\
&&\times\int_{0}^{\infty }dx^{\prime \prime }[V_{H}(x^{\prime
\prime })\delta (x^{\prime }-x^{\prime \prime })+V_{ex}(x^{\prime
},x^{\prime \prime })]\psi _{k}^{0}\left( x^{\prime \prime
}\right) ,  \label{HF}
\end{eqnarray}
where $G_{0}^{>}$ is the Green's function of free electrons on the
right semi-line, $E$ is the full energy of an electron, $V_{H}$
and $V_{ex}$ are the Hartree and the exchange potentials. The
Hartree potential is
\begin{equation}
V_{H}(x)=\int dx^{\prime }U(x-x^{\prime })\delta n(x^{\prime }),
\end{equation}
where $\delta n(x)=n(x)-n_{0}$ is the deviation of the electron
density from its uniform value (in the absence of the potential)
and $U(x)$ is the interaction potential. Hartree interaction is a
direct interaction with the modulation of the electron density by
the Friedel oscillation. For a high barrier, which is essentially
equivalent to a hard-wall boundary condition, the electron density
is
\begin{eqnarray}
n(x) &=&4\int_{0}^{k_{F}}\frac{dk}{2\pi }\sin ^{2}(kx)=n_{0}\left( 1-\frac{%
\sin 2k_{F}x}{2k_{F}x}\right) \rightarrow \\
\delta n\left( x\right) &=&-\sin \left( 2k_{F}x\right) /2\pi x,
\end{eqnarray}
where $n_{0}=k_{F}/\pi $ is the density of electrons. Then,

\begin{equation}
V_{H}(x)=-\frac{1}{2\pi }\int_{0}^{\infty }dx^{\prime }U(x-x^{\prime })\frac{%
\sin 2k_{F}x^{\prime }}{x^{\prime }}.
\end{equation}
Notice that although the bare interaction is short-range, the
effective interaction has a slowly-decaying tail due to the
Friedel oscillation. (The integral goes over only for positive values of $x^{\prime }$
because electrons interact only there.)

The exchange potential is equal to
\begin{eqnarray}
V_{ex}(x,x)&=&-U(x-x^{\prime })\Big[
\int_{0}^{k_{F}}\frac{dk}{2\pi }\left[
\psi _{k}^{0}(x^{\prime })\right] ^{\ast }\psi _{k}^{0}(x)\notag\\&&+\int_{0}^{k_{F}}%
\frac{dk}{2\pi }\left[ \psi _{-k}^{0}(x^{\prime })\right] ^{\ast
}\psi _{-k}^{0}(x)\Big] .  \label{ex}
\end{eqnarray}
Since we assumed that electrons interact only if they are located
to the right of the barrier, the integral in (\ref{HF}) runs only
over $x,x^{\prime }>0$ and the Green's function is a Green's
function on a semi-line. The wave-function in (\ref{HF}) needs to
be evaluated at $x\rightarrow \infty ,$ which means that we will
only need an asymptotic form of the Green's function far away from the
barrier. This form is constructed by the method of images
\begin{equation}
G_{0}^{>}(x,x\mathbf{^{\prime }},E)=G_{0}(x,x^{\prime
},E)-G_{0}(x,-x^{\prime },E),
\end{equation}
where
\[
G_{0}\left( x,x^{\prime },E\right) =\frac{1}{iv_{k}}e^{ik\left|
x-x^{\prime }\right| }
\]
is the free Green's function on a line with $k=\sqrt{2mE}$ and
$v_{k}=k/m.$ Coordinate $x^{\prime }$ is confined to the barrier,
whereas $x\rightarrow \infty ,$ thus $x>x^{\prime }$ and
\[
G_{0}^{>}(x,x\mathbf{^{\prime }},E)=-\frac{2}{v_{k}}\sin
(kx^{\prime })e^{ikx}.
\]

\subsubsection{Hartree interaction}

Our goal is to present the correction to the wave-function for
electrons going from $x<0$ to $x>0$ in the form
\begin{equation}
\psi _{k}-\psi _{k}^{0}=\frac{1}{\sqrt{L}}\delta te^{ikx},
\label{wf}
\end{equation}
where $\delta t$ is the interaction correction to the transmission
coefficient. Substituting (\ref{wf}) into (\ref{HF}), we obtain
for the Hartree contribution to $t$

\[
\frac{\delta t^{H}}{t_{0}}=-\frac{2}{v_{F}}\int_{0}^{\infty
}dx\sin kxe^{ikx}V_{H}(x),
\]
where one can replace $v_{k}\rightarrow v_{F}$ in all
non-oscillatory factors. For a delta-function potential, $U\left(
x\right) =U\delta \left( x\right) $
\begin{equation}
V_{H}(x)=-\frac{U}{2\pi }\frac{\sin 2k_{F}x}{x}.
\end{equation}
However, the $\delta -$ function potential is not good enough for
us, because the Hartree and exchange contributions cancel each
other for this case.  Friedel oscillation arises due to
backscattering. With a little more effort, one can
show that $U$ in the last formula is replaced by $U\left( 2k_{F}\right) :$%
\footnote{%
Notice that the sign of the Hartree interaction is attractive near
the barrier (assuming the sign of the e-e interaction is repulsive
at $2k_{F}$): for $x\rightarrow 0,$ $V_{H}\left( x\right)
\rightarrow -U\left( 2k_{F}\right) k_{F}/\pi .$ The reason is that
the depletion of electron density near the barrier means that the
positive background is uncompensated. As a result, electrons are
\emph{attracted }to the barrier and transmission is \emph{enhanced
}by the Hartree interaction.}

\[
V_{H}(x)=-\frac{U\left( 2k_{F}\right) }{2\pi }\frac{\sin
2k_{F}x}{x}.
\]

Substituting this into $\delta t/t$ yields
\begin{eqnarray*}
\frac{\delta t_{H}}{t_{0}} &=&\frac{U(2k_{F})}{\pi
v_{F}}\int_{0}^{\infty
}dx\sin \left( kx\right) e^{ikx}\frac{\sin 2k_{F}x}{x} \\
&=&\frac{U(2k_{F})}{\pi v_{F}}\int_{0}^{\infty
}dx\frac{1}{2i}\left(
e^{2ikx}-\underbrace{1}\text{regular correction to Imt}  \right) \frac{\sin 2k_{F}x}{x} \\
&=&\frac{U(2k_{F})}{\pi v_{F}}\int_{0}^{\infty }dx\frac{1}{2i}e^{2ikx}\frac{%
\sin 2k_{F}x}{x}=\frac{U(2k_{F})}{2\pi v_{F}} \\
&&\times \int_{0}^{\infty }dx\left( \sin
2kx+\underbrace{i^{-1}\cos 2kx}
\text{yet another regular correction}  %
\right) \frac{\sin 2k_{F}x}{x} \\
&=&\frac{U(2k_{F})}{2\pi v_{F}}\int_{0}^{\infty }dx\sin 2kx\frac{\sin 2k_{F}x%
}{x} \\
&=&\frac{U(2k_{F})}{4\pi v_{F}}\ln \frac{k+k_{F}}{\left| k-k_{F}\right| }%
\approx \alpha _{2k_{F}}^{\prime }\ln \frac{k_{F}}{\left|
k-k_{F}\right| },
\end{eqnarray*}
where
\[
\alpha _{2k_{F}}^{\prime }=\frac{g_{1}}{4\pi v_{F}},
\]
and $g_{1}=U\left( 2k_{F}\right) .$ In deriving the final result,
all terms regular in the limit $k\rightarrow k_{F}$ were
discarded.

\subsubsection{Exchange}

Now both $x$ and $x^{\prime }>0$. We need to select the largest
wave-function, \emph{i.e.}, such that does not involve a small
transmitted component. Obviously, this is only possible for $k<0$
(second term in (\ref {ex})) and $\psi _{-k}^{0}$ , given by
(\ref{irl}). Substituting the free wave-functions into the equation
for the exchange interaction, we get
\begin{equation}
V_{ex}(x,x^{\prime })=-U(x-x^{\prime })\rho (x,x^{\prime }),
\end{equation}
where the 1D density-matrix is
\begin{eqnarray}
\rho (x,x^{\prime }) &=&4\int_{0}^{k_{F}}\frac{dk}{2\pi }\sin
(kx)\sin
(kx^{\prime }) \\
&=&2\int_{0}^{k_{F}}\frac{dk_{x}}{2\pi }\left[ \cos k(x-x^{\prime
})-\cos
k(x+x^{\prime })\right] \\
&=&\dots -\frac{\sin k_{F}(x+x^{\prime })}{\pi (x+x^{\prime })},
\end{eqnarray}
where $\dots $ stand for the term which depends on $x-x^{\prime
}.$ This term does not lead to the log-divergence in $\delta t$
and will be dropped
\footnote{%
Notice that the important part of the exchange potential is
\emph{repulsive} near the barrier. This means that electrons are
repelled from the barrier and transmission is suppressed.}. For
$x=x^{\prime },$ we get the correction to the density $\delta
n\left( x\right) ,$ as we should.

Correction to the transmission coefficient
\begin{equation}
\delta t_{ex}/t_{0}=-\frac{2}{\pi v_{F}}\int_{0}^{\infty
}dx^{\prime }\int_{0}^{\infty }dx^{\prime \prime }U\left(
x^{\prime }-x^{\prime \prime }\right) \sin kx^{\prime
}e^{ikx^{\prime \prime }}\frac{\sin k_{F}(x^{\prime }+x^{\prime
\prime })}{x^{\prime }+x^{\prime \prime }}.  \label{deltat_ex}
\end{equation}
After a little manipulation with trigonometric functions, which
involves dropping of the terms depending only on $x-x^{\prime },$
we arrive at
\begin{eqnarray}
\frac{\delta t_{ex}}{t_{0}} &=&-\frac{1}{4\pi ^{2}v_{F}}\int_{0}^{+\infty }%
\frac{dq}{q}U(q)\int_{0}^{\infty }\frac{dx_{+}}{x_{+}} \\
&&\times \{\sin 2(k-k_{F}+q)x_{+}-\sin 2(k-k_{F}-q)x_{+}\},
\end{eqnarray}
where
\begin{equation}
x_{+}=\frac{x^{\prime }+x^{\prime \prime }}{2}.
\end{equation}
Integral over $x_{+}$ \ provides a lower cut-off for the $q-$
integral

\begin{eqnarray}
&&\int_{0}^{+\infty }\frac{dx_{+}}{x_{+}}\{\sin
2(k-k_{F}+q)x_{+}-\sin
2(k-k_{F}-q)x_{+}\} \\
&=&\frac{\pi }{2}\text{sgn}(q+k-k_{F})+\frac{\pi }{2}\text{sgn}(q-k+k_{F}) \\
&=&\pi \theta (q-\left| k-k_{F}\right| ).
\end{eqnarray}
Now
\begin{equation}
\frac{\delta t_{ex}}{t_{0}}=-\frac{1}{4\pi v_{F}}\int_{\left|
k-k_{F}\right| }^{+\infty }\frac{dq}{q}U(q).  \label{a11}
\end{equation}
As $U\left( q\right) $ is regular at $q\rightarrow 0$ \footnote{%
If $U(q)$ has a strong dependence on $q$ for $q\rightarrow 0$
(which is the case for a bare Coulomb potential $U\left( q\right)
\propto \ln q),$ this dependence affects the resulting dependence
of the transmission coefficient on energy $\left| k-k_{F}\right|
,$ \emph{i.e., }on the temperature and/or bias. Instead of a
familiar power-law scaling of the tunneling conductance for the
short-range interaction, the conductance falls off with energy
faster than any power law for the bare Coulomb potential.}, one can take $%
U\left( q\right) $ out of the integral at $q=0$ (denoting $U\left(
0\right)
=g_{2})$%
\[
\frac{\delta t_{ex}}{t_{0}}\approx -\frac{1}{4\pi
v_{F}}g_{2}\int_{\left|
k-k_{F}\right| }^{q_{0}}\frac{dq}{q}=-\alpha _{0}^{\prime }\ln \frac{q_{0}}{%
\left| k-k_{F}\right| }.
\]
Combining the exchange and Hartree corrections together (in doing
so, we choose the smallest upper cut-off for the log which we
assume to be the inverse interaction range, $q_{0})$) we get
\begin{equation}
\delta t=-t_{0}\alpha ^{\prime }\ln \frac{q_{0}}{\left|
k-k_{F}\right| }, \label{delta_t}
\end{equation}
where
\begin{equation}
\alpha ^{\prime }=\alpha _{0}^{\prime }-\alpha _{2k_{F}}^{\prime }=\frac{%
g_{2}-g_{1}}{4\pi v_{F}};\text{\textrm{asymmetric geometry.}}
\label{alpha_s}
\end{equation}
It can be shown in a similar manner that if we had interacting
regions on \emph{both} sides of the barrier, the result for
$\alpha ^{\prime }$ would be double of that in Eq.~(\ref{alpha_s}).

\begin{equation}
\alpha ^{\prime }=\alpha _{0}^{\prime }-\alpha _{2k_{F}}^{\prime }=\frac{%
g_{2}-g_{1}}{2\pi v_{F}};\text{\textrm{symmetric geometry.}}
\label{alpha_spinless}
\end{equation}
The sign of the correction to $t$ depends on the sign of $%
g_{2}-g_{1}=U\left( 0\right) -U\left( 2k_{F}\right) .$ Notice that
transmission is \emph{enhanced, }if $U\left( 2k_{F}\right) >U(0).$
Usually, this behavior is associated with attraction. We see,
however, that even if the interaction is repulsive at all $q$ but
$U\left( 2k_{F}\right) >U(0),$ it works effectively as an
attraction. The case $U\left( 2k_{F}\right) >U(0)$ is not a very
realistic one, at least not in a situation when electrons
interact only among themselves. Other degrees of freedom, \emph{e.g., }%
phonons, must be involved to give a preference to $2k_{F}-$
scattering.

\subsection{ Renormalization group}

It is tempting to think that the first-order in interaction correction to $%
t_{0}$ in Eq.~(\ref{alpha_s}) is just an expansion of the scaling form $%
t\propto \left| k-k_{F}\right| ^{\alpha ^{\prime }}.$ A poor-man
RG indeed shows that this is the case. Near the Fermi level,
$k-k_{F}=\left( E-E_{F}\right) /v_{F}=\varepsilon /v_{F}$ so that
the first-order correction to $t$ is
\[
t_{1}=t_{0}\left( 1-\alpha ^{\prime }\ln \frac{W_{0}}{\left|
\varepsilon \right| }\right) ,
\]
where $W_{0}=q_{0}v_{F}$ is the effective bandwidth. The meaning
of this bandwidth is that the states at $\pm W_0$ from the Fermi
level (=0) are not affected by the interaction. For $\left| \varepsilon
\right| =W_{0}$, $t_{1}=t_{0}.$ Suppose that we want to reduce to
bandwidth $W_{0}\rightarrow W_{1}<W_{0}$ and find $t$ at
$|\varepsilon |=W_{1}$

\[
t_{1}=t_{0}\left( 1-\alpha ^{\prime }\ln
\frac{W_{0}}{W_{1}}\right) .
\]
It is of crucial importance here that coefficient $\alpha ^{\prime
}$ (which will become the tunneling exponent in the scaling form
we are about to get) is proportional to the \emph{RG-invariant}
combination $U\left( 0\right) -U\left( 2k_{F}\right) =g_{2}-g_{1}$
for spinless electrons. This means that $\alpha ^{\prime }$ is to
be treated as a constant under the RG flow. Repeating this
procedure using $t,$ found at the previous stage instead of a bare
$t_{0},$ $n$ times, we get
\[
t_{n+1}=t_{n}\left( 1-\alpha ^{\prime }\ln
\frac{W_{n}}{W_{n+1}}\right) .
\]
The renormalization process is to be stopped when the bandwidth
coincides with the physical energy $\left| \varepsilon \right| ,$
at which $t$ is measured. In the continuum limit
($t_{n+1}-t_{n}=dt;W_{n+1}=W_{n}-dW)$, this equation reduces to a
differential one
\begin{eqnarray*}
\frac{dt}{t} =\alpha ^{\prime }\frac{dW}{W}
\end{eqnarray*}
Integrating from $t\left( \varepsilon \right) $ to $t_{0}$ (and,
correspondingly, from $W=\left| \varepsilon \right| $ to
$W=W_{0})$, we obtain
\[
t\left( \varepsilon \right) =t_{0}\left( \left| \varepsilon
\right| /W_{0}\right) ^{\alpha ^{\prime }}.
\]

\subsection{Electrons with spins}

Now let's introduce the spin. The effect will be more interesting
than just multiplying the result for the tunneling conductance by
a factor of two (which is all what happens for non-interacting
electrons.)  To keep things general, I will assume an arbitrary
``spin'' (which may involve other degrees of freedom) degeneracy
$N_{s}$ and put $N_{s}=2$ at the end. In this section we will
exploit the result of Sec.\ref{sec:RG} stating the backscattering
amplitude flows under RG. This flow affects the renormalization of the
transmission coefficient at low energies.

Repeating the steps for the first-order correction to $t$ for the
case of electrons with spin is straightforward: one just has
 to recall that the Hartree correction is multiplied by $%
N_{s}$ (as the polarization bubble involves summation over all
isospin components, it is simply multiplied by a factor of
$N_{s}).$ On the contrary, the exchange interaction is possible
only between electrons of the same spin, so there are $N_{s}$
identical exchange potentials for every spin component. I am going
to discuss the strong barrier case first in the symmetric
geometry. Then, taking into account what we have just said about
the factor of $N_{s},$ we can replace the result for spinless
electrons (\ref{alpha_s}) by
\begin{equation}
\alpha ^{\prime }\rightarrow \alpha ^{\prime }=\alpha _{0}^{\prime
}-N_{s}\alpha _{2k_{F}}^{\prime }=\frac{g_{2}-N_{s}g_{1}}{4\pi v_{F}}\mathrm{%
.}  \label{alpha_spin}
\end{equation}
(and similarly for the symmetric geometry of the tunneling
experiment). Correspondingly, the correction to the transmission
coefficient (for a given spin projection) changes to
\[
t_{\sigma }=t_{0}\left( 1-\alpha ^{\prime }\ln W/\left|
\varepsilon \right| \right) .
\]
The tunneling conductance is found from the Landauer formula
\[
{\cal G}=\frac{e^{2}}{h}\sum_{\sigma =1}^{N_{s}}\left| t_{\sigma
}\right| ^{2},
\]
where, as the barrier is spin-invariant, the sum simply amounts to
multiplying the result for a given spin component by $N_{s}$. Now,
the result in Eq.~(\ref{alpha_spin}) seems to be interesting, as
the $2k_{F}$ contribution gets a boost. If $N_{s}U\left(
2k_{F}\right) >U\left( 0\right) $, we have in increase of the
barrier transparency. It does not seem too hard to satisfy this
condition. For
example, it is satisfied already for the delta-function potential \footnote{%
as now electrons have spins, they are allowed to be at the same
point in space and interact.} and $N_{s}=2.$ However,  as opposed
to the spinless case, $\alpha ^{\prime }$ is \emph{not }an
RG-invariant but flows under
renormalizations. Let's split $\alpha ^{\prime }$ into an RG-invariant part (%
\ref{invariant}) and the rest
\begin{eqnarray*}
\alpha ^{\prime } &=&\frac{1}{4\pi v_{F}}\left[ U\left( 0\right) -\frac{1}{%
N_{s}}U\left( 2k_{F}\right) \right] -\frac{1}{4\pi v_{F}}\frac{N_{s}^{2}-1}{%
N_{s}}U\left( 2k_{F}\right)  \\
&=&\alpha _{s}^{\prime }-\frac{1}{4\pi
v_{F}}\frac{N_{s}^{2}-1}{N_{s}}g_{1},
\end{eqnarray*}
where
\begin{equation}
\alpha _{s}^{\prime }=\frac{1}{4\pi v_{F}}\left( g_{2}-\frac{1}{N_{s}}%
g_{1}\right) =\frac{\bar{\Gamma}}{4\pi v_{F}}.
\label{edge}\end{equation} The condition for the tunneling
exponent to be negative is more restrictive that it seemed to be:
$g_{1}>N_{s}g_{2}.$ It is not hard to see that the RG equation for
$t_{\sigma }$ now changes to
\begin{equation}
\frac{dt}{dl}=-t\left( \alpha _{s}^{\prime }-\frac{1}{4\pi v_{F}}\frac{%
N_{s}^{2}-1}{N_{s}}\Gamma _{1}\left( l\right) \right) ,
\label{rgt_spin}
\end{equation}
where $\Gamma _{1}\left( l\right) $ is given by
\[
\Gamma _{1}=\frac{1}{\left( g_{1}\right) ^{-1}+\frac{N_{s}}{2\pi
}l}.
\]
Integrating (\ref{rgt_spin}), we find
\[
t_{\sigma }=t_{0}\left( 1+\frac{N_{s}g_{1}}{4\pi v_{F}}\ln
\frac{W}{\left| \varepsilon \right| }\right) ^{\beta _{s}}\left(
\left| \varepsilon \right| /W\right) ^{\alpha _{s}^{\prime }},
\]
where
\[
\beta _{s}=\frac{N_{s}^{2}-1}{N_{s}^{2}}.
\]
In particular, for $N_{s}=2$, we get
\[
t_{\sigma }=t_{0}\left( 1+\frac{g_{1}}{2\pi v_{F}}\ln
\frac{W}{\left| \varepsilon \right| }\right) ^{3/4}\left( \left|
\varepsilon \right| /W\right) ^{\alpha _{s}^{\prime }}.
\]
and conductance
\beq
{\cal G}={\cal G}_0\left( 1+\frac{g_{1}}{2\pi v_{F}}\ln
\frac{W}{\left| \varepsilon \right| }\right) ^{3/2}\left( \left|
\varepsilon \right| /W\right) ^{2\alpha _{s}^{\prime }},
\label{ygm_cond}\eeq
where ${\cal G}_0$ is the conductance for the free case.
Thus the flow of the backscattering amplitude results in a
multiplicative log-renormalization of the transmission
coefficient. One can check the first-order result is reproduced if
expand the RG result to the first log.

An interesting feature of this result is that it predicts \emph{three }%
possible types of behavior of the conductance as function of
energy.

\begin{enumerate}
\item  weak backscattering:
\[
\alpha ^{\prime }>0\rightarrow g_{1}<g_{2}/N_{s}.
\]
In this regime already the first-order correction corresponds to
suppression of the conductance, which decreases monotonically as
the energy goes down.

\item  intermediate backscattering:
\[
\alpha ^{\prime }>0\text{ but }\alpha _{s}^{\prime }<0\rightarrow
g_{2}/N_{s}<g_{1}<N_{s}g_{2}.
\]
In this regime, the first-order correction enhances the
transparency, but the RG result shows that the for $\varepsilon
\rightarrow 0,$ the transmission goes to zero. It means that at
higher energies, when the RG has not set in yet, the conductance
increases as the energy goes down, but at
lower energies the conductance decreases. The dependence of $%
{\cal G}\left( \varepsilon \right) $ on $\varepsilon $ is
non-monotonic--there is a maximum at the intermediate energies.

\item  strong backscattering:
\[
\alpha _{s}^{\prime }<0\rightarrow g_{1}>N_{s}g_{2}.
\]
In this regime, tunneling exponent $\alpha _{s}^{\prime }$ is
negative and the conductance increases as the energy goes down.
\end{enumerate}

\subsection{Comparison of bulk and edge tunneling exponents}

Tunneling into the bulk of a 1D system is described by the density
of states obtained, \emph{e.g.}, in the DL solution of the
Tomonaga-Luttinger model (no backscattering). The ``bulk''
tunneling exponent is equal to
\[
2\alpha =\frac{\left( u-1\right) ^{2}}{4u},
\]
where
\[
u=\sqrt{1+2g_{2}/\pi v_{F}}.
\]
In this Section, we considered  tunneling into the edge for weak
interaction and found that the conductance scales with exponent
$2\alpha'_s$ (\ref{edge}). To compare the two exponents, we need
to expand the DL exponent for weak interaction
\[
2\alpha =\frac{g_{2}^{2}}{4\pi ^{2}v_{F}}.
\]
For $g_{1}=0$, the edge exponent is
\[
2\alpha _{s}^{\prime }=\frac{1}{2\pi v_{F}}g_{2}.
\]
We see that for weak coupling tunneling into the edge is stronger
affected by the interaction than tunneling into the bulk: the
former effect starts at the first order in the interaction whereas
the latter starts at the second order. This difference has a
simple physical reason which is general for all dimensions. In a
translationally invariant system, the shape of the Green's
function is modified in a non-trivial way only starting at the
second order. For example, the imaginary part of the self-energy
(decay of quasi-particles) occur only at the second order. The
first-order corrections lead only to a shift in the chemical
potential and, if the potential is of a finite-range, to a
renormalization of the effective mass. If the translational
invariance is broken, non-trivial changes in the Green's function
occur already at the first order in the interaction. That
tunneling into the bulk and edge are characterized by different
exponents is also true in the strong-coupling case (cf.
Sec.\ref{sec:bosonization}). As the relation between the bulk and
edge exponents is known for an arbitrary coupling, one can
eliminate the unknown strength of interaction and express one
exponent via the other. Knowing one exponent from the experiment,
one can check if the observed value of the second exponent agrees
with the data. This cross-check was cleverly used in the
interpretation of the experiments on single-wall carbon nanotubes
\cite{bockrath,yao}.
%%%%%%%%%%%%%%%%%%%%%%%%%%%%%%%%%%%%%%%%%%%%%%%%%%%%%%%%%%%%%%%%%%%%%%%%%%
%%%%%%%%%%%%%%%%BOSONIZATION
%%%%%%%%%%%%%%%%%%%%%%%%%%%%%%%%%%%%%%%%%%%%%%%%%%%%%%%%%%%%%%%%%%%%%%%%%%

\section{Bosonization solution}
\label{sec:bosonization} Bosonization procedure in described in a
number of books and reviews \cite{solyom}-\cite{giamarchi_book}.
Without repeating all standard manipulations, I will only
emphasize the main steps in this Section, focusing on a couple of
subtle points not usually discussed in the literature. A reader
familiar with bosonization may safely skip the first part of this
Section, and go directly to Secs. \ref{sec:cf} and \ref{sec:dos},
where tunneling exponents are calculated. Some technical details
of the bosonization procedure are presented in Appendix
\ref{sec:bos_details}. As in Sec.\ref{sec:DL}, $v_F=1$ in this
Section, unless specified otherwise.
\subsection{Spinless fermions}

\subsubsection{Bosonized Hamiltonian}

We start from a Hamiltonian of interacting fermions without spin
\begin{equation*}
H=\frac{1}{L}\sum_{p,k,q}\xi _{k}a_{k}^{\dagger }a_{k}+\frac{1}{2L^{2}}%
\sum_{p,k,q}V_{q}a_{p-q}^{\dagger }a_{k+q}^{\dagger }a_{k}a_{p}.
\end{equation*}
The interacting part of the Hamiltonian can be re-written using
chiral densities
\begin{equation*}
\rho _{\pm }\left( q\right) =\sum_{p\gtrless 0}a_{p-q/2}^{\dagger
}a_{p+q/2}
\end{equation*}
as
\begin{equation*}
H_{\text{int}}=g_{2}\frac{1}{L}\sum_{q}\rho _{+}\left( q\right)
\rho _{-}\left( -q\right) +\frac{g_{4}}{2}\frac{1}{L}\sum_{q}\rho
_{+}\left( q\right) \rho _{+}\left( -q\right) +\rho _{-}\left(
q\right) \rho _{-}\left( -q\right) .
\end{equation*}
The interacting part is in the already bosonized form. For a
linearized dispersion
\begin{equation*}
\xi _{k}=\left| k\right| -k_{F},
\end{equation*}
it is also possible to express the free
part via densities. To check this, let's assume that $H_{0}$ can
indeed be written as
\begin{equation*}
H_{0}=\frac{1}{L}\sum_{q}A_{q}\left[ \rho _{+}\left( q\right) \rho
_{+}\left( -q\right) +\rho _{-}\left( q\right) \rho _{-}\left( -q\right) %
\right] ,
\end{equation*}
where $A_{q}$ is some unknown function. Commuting $\rho _{+}$ with
$H_{0},$ and making use of the anomalous commutator $\left[ \rho
_{+}\left( q\right) ,\rho _{+}\left( -q\right) \right] =qL/2\pi $,
we obtain
\begin{eqnarray*}
\left[ \rho _{+}\left( q\right) ,H_{0}\right]
&=&\frac{1}{L}\sum_{q^{\prime }}A_{q^{\prime }}[\rho _{+}\left(
q\right) ,\rho _{+}\left( q^{\prime
}\right) \rho _{+}\left( -q^{\prime }\right) ] \\
&=&A_{q}\rho _{+}\left( q\right) \frac{q}{\pi }.
\end{eqnarray*}
On the other hand, the same commutator can be calculated directly
in a model with the linearized spectrum, using only the fermionic
anticommutation relations \cite{levitov}. This gives
\begin{equation*}
\left[ \rho _{+}\left( q\right) ,H_{0}\right] =q\rho _{+}\left(
q\right) .
\end{equation*}
Comparing the two results, we see that
\begin{equation*}
A_{q}=\pi
\end{equation*}
and thus
\begin{equation*}
H_{0}=\pi \frac{1}{L}\sum_{q}\rho _{+}\left( q\right) \rho
_{+}\left( -q\right) +\rho _{-}\left( q\right) \rho _{-}\left(
-q\right) .
\end{equation*}

Combining the free and interacting parts of the Hamiltonian, we
obtain
\begin{equation*}
H=\pi \frac{1}{L}\left( 1+\frac{g_{4}}{2\pi }\right)
\sum_{q}\left\{ \rho _{+}\left( q\right) \rho _{+}\left( -q\right)
+\rho
_{-}\left( q\right) \rho _{-}\left( -q\right) \right\} + \\
\pi \frac{1}{L}%
g_{2}\sum_{q}\rho _{+}\left( q\right) \rho _{-}\left( -q\right) .
\end{equation*}
Notice that if only the $g_{4}-$ interaction is present, the
system remains free but the Fermi velocity changes.

It is convenient to expand the density operators over the normal
modes
\begin{eqnarray*}
\rho _{+}\left( x\right) &=&\sum_{q>0}\sqrt{\frac{\left| q\right| }{2\pi L}}%
\left( b_{q}e^{iqx}+b_{q}^{\dagger }e^{-iqx}\right); \\
\rho _{-}\left( x\right) &=&\sum_{q<0}\sqrt{\frac{\left| q\right| }{2\pi L}}%
\left( b_{q}e^{iqx}+b_{q}^{\dagger }e^{-iqx}\right) .
\end{eqnarray*}
One can readily make sure that density operators defined in this
way reproduce the correct commutation relations, given that
$\left[ b_{q},b_{q^{\prime }}^{\dagger }\right] =\delta
_{q,q^{\prime }}$ . In terms of these operators, the Hamiltonian
reduces to
\begin{equation*}
H=\pi \left( 1+\frac{g_{4}}{2\pi }\right)
\sum_{q>0}q\left\{ b_{q}^{\dagger }b_{q}+b_{-q}^{\dagger
}b_{-q}\right\} +\pi g_{2}\sum_{q}q\left( b_{q}^{\dagger
}b_{-q}^{\dagger }+b_{q}b_{-q}\right) .
\end{equation*}
Introducing new bosons via a Bogoliubov transformation
\begin{eqnarray*}
c_{q}^{\dagger } &=&\cosh \theta_q b_{q}+\sinh \theta_q b_{-q}^{\dagger }; \\
c_{-q}^{\dagger } &=&\cosh \theta_q b_{-q}^{\dagger }+\sinh \theta_q
b_{q}
\end{eqnarray*}
and choosing $\theta _{q}$ so that the Hamiltonian becomes
diagonal, \emph{i.e.},
\begin{equation*}
\tanh 2\theta_q =\frac{g_{2}/2\pi }{1+g_{4}/2\pi },
\end{equation*}
we obtain
\begin{equation*}
H=\frac{1}{L}\sum_{q}\omega _{q}c_{q}^{\dagger }c_{q},
\end{equation*}
where
\begin{equation*}
\omega _{q}=u\left| q\right|,
\end{equation*}
and
\footnote{%
Let's now introduce backscattering. Since for spinless fermions
backscattering is just an exchange process to forward scattering
of fermions of opposite chirality (think of a diagram where you
have right and left lines coming in and then interchange them at
the exit), the only effect of backscattering is to replace
$g_{2}\rightarrow g_{2}-g_{1}.$ (cf. discussion in
Sec.~\ref{sec:RG}). Instead of (\ref{u_spinless}), we then have
\begin{equation}
u=\left[ \left( 1+\frac{g_{4}}{2\pi }\right) ^{2}-\left( \frac{g_{2}-g_{1}}{%
2\pi }\right) ^{2}\right] ^{1/2}.
\end{equation}

Now, consider a delta-function interaction, when
$g_{1}=g_{2}=g_{4}.$ Pauli principle says that we should get back
to the Fermi gas in this case. However, $u$ still differs from
unity (Fermi velocity) and thus our result violates the Pauli
principle. See Appendix \ref{sec:app_back} for a resolution of the
paradox.}
\begin{equation}
u=\left[ \left( 1+\frac{g_{4}}{2\pi }\right) ^{2}-\left( \frac{g_{2}}{2\pi }%
\right) ^{2}\right] ^{1/2}.  \label{u_spinless}
\end{equation}

For the spinless case, backscattering can be absorbed into forward
scattering. The resulting expression for the renormalized velocity
for the case $g_{2}=g_{4}\neq g_{1}$ is (cf. Appendix
\ref{sec:app_back} )
\begin{equation*}
u=\sqrt{1+\frac{g_{2}-g_{1}}{2\pi }.}
\end{equation*}

\subsubsection{Bosonization of fermionic operators}

The $\Psi -$ operators of right/left movers can be represented as
\begin{equation}
\Psi _{\pm }\left( x\right) =\frac{1}{\sqrt{2\pi a}}e^{\pm 2\pi
i\int_{-\infty }^{x}\rho _{\pm }\left( x^{\prime }\right)
dx^{\prime }}, \label{mand1}
\end{equation}
where $a$ is the ultraviolet cut-off in real space. Using the
commutation relations for $\rho _{\pm },$ one can show that the
(anti) commutation relations
\begin{equation*}
\left\{ \Psi _{\pm }\left( x\right) ,\Psi _{\pm }^{\dagger }\left(
x^{\prime }\right) \right\} =\delta \left( x-x^{\prime }\right)
\end{equation*}
are satisfied.

The argument of the exponential can be re-written as two bosonic
fields
\begin{eqnarray}
\Psi _{\pm }\left( x\right) &=&\frac{1}{\sqrt{2\pi a}}e^{\pm i\sqrt{\pi }%
\varphi _{\pm }\left( x\right) } ; \label{mand2} \\
\varphi _{\pm }\left( x\right) &=&\varphi \left( x\right) \mp
\vartheta \left( x\right) .  \notag
\end{eqnarray}
Equating exponents in (\ref{mand1}) and (\ref{mand2}), we obtain
\begin{eqnarray}
\sqrt{\pi }\left[ \varphi \left( x\right) -\vartheta \left(
x\right) \right] &=&2\pi \int_{-\infty }^{x}dx^{\prime }\rho
_{+}\left( x^{\prime }\right)
\label{a1} \\
&=&2\pi \sum_{q>0}\sqrt{\frac{\left| q\right| }{2\pi
L}}\frac{1}{iq}\left(
b_{q}e^{iqx}-b_{q}^{\dagger }e^{-iqx}\right) ;  \notag \\
\sqrt{\pi }\left[ \varphi \left( x\right) +\vartheta \left(
x\right) \right] &=&2\pi \int_{-\infty }^{x}dx^{\prime }\rho
_{-}\left( x^{\prime }\right)
\label{a2} \\
&=&2\pi \sum_{q<0}\sqrt{\frac{\left| q\right| }{2\pi
L}}\frac{1}{iq}\left( b_{q}e^{iqx}-b_{q}^{\dagger }e^{-iqx}\right)
.  \notag
\end{eqnarray}
Solving for $\varphi \left( x\right) $ and $\vartheta \left(
x\right) $ gives
\begin{eqnarray}
\varphi \left( x\right) &=&-i\sum_{-\infty <q<\infty
}\frac{1}{\sqrt{2\left| q\right| L}}\text{sgn}q\left(
b_{q}e^{iqx}-b_{q}^{\dagger }e^{-iqx}\right) ;
\label{phi} \\
\vartheta \left( x\right) &=&i\sum_{-\infty <q<\infty }\frac{1}{\sqrt{%
2\left| q\right| L}}\left( e^{iqx}b_{q}-b_{q}^{\dagger
}e^{-iqx}\right). \label{theta}
\end{eqnarray}
Using (\ref{phi}) and (\ref{theta}), one can prove that $\varphi
\left( x\right) $ and $\partial _{x}\vartheta \left( x\right) $ satisfy canonical commutation relations
between coordinate and momentum (cf. Appenidx
\ref{sec:phi_theta})
\begin{equation}
\left[ \varphi \left( x\right) ,\partial _{x^{\prime }}\vartheta
\left( x^{\prime }\right) \right] =i\delta \left( x-x^{\prime
}\right) . \label{comm_main}
\end{equation}
Using Eqs. (\ref{a1}) and (\ref{a2}), we obtain the density and
current as the gradients of the bosonic fields
\begin{eqnarray*}
\varphi \left( x\right) &=&\sqrt{\pi }\int_{-\infty
}^{x}dx^{\prime }\left( \rho _{+}\left( x^{\prime }\right) +\rho
_{-}\left( x^{\prime }\right) \right) =\sqrt{\pi }\int_{-\infty
}^{x}dx^{\prime }\rho \left( x^{\prime
}\right) \rightarrow \\
\rho \left( x\right) &=&\frac{1}{\sqrt{\pi }}\partial
_{x}\varphi;\\
\vartheta \left( x\right) &=&-\sqrt{\pi }\int_{-\infty
}^{x}dx^{\prime }\left( \rho _{+}\left( x^{\prime }\right) +\rho
_{-}\left( x^{\prime }\right) \right) =-\sqrt{\pi }\int_{-\infty
}^{x}dx^{\prime }j\left(
x^{\prime }\right) \rightarrow \\
j\left( x\right) &=&-\frac{1}{\sqrt{\pi }}\partial _{x}\vartheta
\left( x\right).
\end{eqnarray*}
The continuity equation,
\begin{equation*}
\partial _{t}\rho +\partial _{x}j=0
\end{equation*}
relates the Heisenberg fields $\varphi \left( x,t\right) $ and
$\vartheta \left( x,t\right) $
\begin{equation*}
\partial _{t}\varphi =\partial _{x}\vartheta .
\end{equation*}
The current can be also found as
\begin{equation*}
j\left( x,t\right) =-\frac{1}{\sqrt{\pi }}\partial _{t}\varphi
\left( x,t\right) .
\end{equation*}
We will use this relation later. Expressing $H$ in the real space
via $\rho _{\pm }$
\begin{equation*}
H=\pi \int dx\left[ \rho _{+}^{2}+\rho _{-}^{2}\right]
+\frac{g_{4}}{2}\int dx\left[ \rho _{+}^{2}+\rho _{-}^{2}\right]
+g_{2}\int dx\rho _{+}\rho _{-}
\end{equation*}
and using the relations
\begin{equation*}
\rho _{\pm }=\frac{1}{2\sqrt{\pi }}\left( \partial _{x}\varphi \mp
\partial _{x}\vartheta \right),
\end{equation*}
we obtain a canonical form of $H$ in terms of the bosonic fields
\begin{eqnarray}
H &=&\frac{1}{2}\int dx\left[ \left( \partial _{x}\varphi \right)
^{2}+\left( \partial _{x}\vartheta \right) ^{2}\right] +\frac{g_{2}+g_{4}}{%
4\pi }\int dx\left( \partial _{x}\varphi \right) ^{2}+\frac{g_{2}-g_{4}}{%
4\pi }\int dx\left( \partial _{x}\vartheta \right) ^{2}\notag \\
&=&\frac{1}{2}\int dx\left[ \frac{u}{K}\left( \partial _{x}\varphi
\right) ^{2}+\text{ }uK\left( \partial _{x}\vartheta \right)
^{2}\right],   \label{H_bos}
\end{eqnarray}
where \footnote{%
As we have already seen in Sec.\ref{sec:DL}, a difference between
$g_{2}$ and $g_{4}$ leads to the current-current interaction in
the Hamiltonian. In the bosonized form, this interaction is the
$\left( g_{2}-g_{4}\right)
\left( \partial _{x}\vartheta \right) ^{2}$ term in the first line of Eq.~(%
\ref{H_bos}).}
\begin{eqnarray}
u =\sqrt{\left( 1+\frac{g_{4}}{2\pi }\right) ^{2}-\left(
\frac{g_{2}}{2\pi }\right) ^{2}};\;
K=\sqrt{\frac{1+\frac{g_{4}-g_{2}}{2\pi
}}{1+\frac{g_{4}+g_{2}}{2\pi }}}. \label{K_spinless}
\end{eqnarray}
For $g_{4}=g_{2}\equiv g,$ we have
\begin{eqnarray}
u=\sqrt{1+g/\pi }; \;K=\frac{1}{\sqrt{1+g/\pi }}.
\label{uK1}\end{eqnarray}
Notice that in this case $uK=1.$ This is important: in the next
Section, we will see that this product renormalizes the Drude
weight (and the persistent current). Neither of these
quantities are supposed to be affected by the interactions, as the Galilean
invariance remains intact. We see that it is indeed the case in
our model.

If backscattering is present (but $g_{2}=g_{4})$, the parameters change to (cf. Appendix \ref{sec:app_back})
\begin{eqnarray}
u &=&\sqrt{1+\frac{g_{2}-g_{1}}{\pi }};  \label{ug2g1} \\
K &=&\frac{1}{\sqrt{1+\left( g_{2}-g_{1}\right) /\pi }}.
\label{kg2g1}
\end{eqnarray}
Had we started with another microscopic model, \emph{e.g.}, with
fermions on a lattice but away from half-filling, the effective
low-energy theory would have also been described by Hamiltonian
(\ref{H_bos}) albeit with different--and, in general,
unknown--parameters $u$ and $K$. The term ``Luttinger liquid''
(LL) \cite{haldane} refers to a universal Hamiltonian of type
(\ref{H_bos}), which describes the low-energy properties of many
seemingly different systems. In that sense, the LL is a 1D analog
of higher-dimensional Fermi liquids, which describe the low-energy
properties of a large class of fermionic systems, while encoding
the quantitative differences in their high-energy properties by a
relatively small set of parameters.

\bigskip

\subsubsection{Attractive interaction}

What happens for the case of an attractive interaction, $g<0?$ Formally,
for $g<-\pi$ (or $g_2-g_1<\pi$), $u^{2}$ in Eqs.(\ref{uK1},\ref{ug2g1}) is negative, which seems
to suggest some kind of an instability. Actually, this is not the case \cite{DL},
as a 1D system of spinless fermions does not have any phase
transitions even at $T=0.$ All it means that the interacting
system is a liquid rather than a gas, \emph{i.e., }it does not
require external pressure to mantain its volume.
An equilibrium value of the density is fixed by  given ambient
pressure. To see this, restore the Fermi velocity $v_{F}=\pi n/m,$
where $n$ is the density
\begin{equation}
u^{2}=v_{F}^{2}\left( 1+\frac{g}{\pi v_{F}}\right) =\left( \frac{\pi n}{m}%
\right) ^{2}+\frac{ng}{m}
\end{equation}
and recall the thermodynamic relation
\begin{equation}
u^{2}=m^{-1}\partial P/\partial n,  \label{pressure}
\end{equation}
where $P$ is the pressure. Integrating (\ref{pressure}) with the
boundary condition $P\left( n=0\right) =0,$ we obtain the
constituency relation
\begin{equation*}
P=\left( \frac{\pi }{m}\right)
^{2}\frac{n^{3}}{3}+\frac{g}{2m}n^{2}.
\end{equation*}
For $g<0,$ there is a metastable region of negative pressure. This
means that if the ambient pressure is equal to zero, the
thermodynamically stable value of the density is given by the
non-zero root of the equation $P\left( n\right) =0$:
\begin{equation*}
n^{\ast }=\frac{3}{2\pi ^{2}}\left| g\right| m.
\end{equation*}
The square of the sound velocity at this density is positive:
\begin{equation*}
\left( u^{\ast }\right) ^{2}=\frac{3}{4\pi ^{2}}g^{2}.
\end{equation*}
The Fermi velocity at $n=n^{\ast }$ is
\begin{equation*}
v_{F}^{\ast }=\pi n^{\ast }/m=\frac{3}{2\pi ^{2}}\left| g\right|
\end{equation*}
and parameter $K$%
\begin{equation*}
K=\frac{v_{F}^{\ast }}{u^{\ast }}=\frac{\sqrt{3}}{\pi }
\end{equation*}
is a universal number, independent of the interaction.

\subsubsection{Lagrangian formulation}

In what follows, it will be more convenient to work in the
Lagrangian rather the Hamiltonian formulation (and also in complex
time). A switch from the Hamiltonian to Lagrangian formulation is
done via the usual canonical transformation
\begin{equation}
S=\int dx\int dt\left( \dot{q}p-\mathcal{H}\left( p,q\right)
\right) , \label{action}
\end{equation}
where $\mathcal{H}$ is the Hamiltonian density defined such that
\begin{equation*}
H=\int dx\mathcal{H}
\end{equation*}
and $q$ and $p$ are the canonical coordinate and momentum,
correspondingly. According to commutation relation
(\ref{comm_main}), \beq q =\varphi\; p =\partial _{x}\vartheta .
\eeq Performing a Wick rotation, $t\rightarrow -i\tau ,$ we reduce
the quantum-mechanical problem into a statistical-mechanics one
with the partition function
\begin{equation*}
Z=\int D\varphi \int D\vartheta e^{-S_{E}},
\end{equation*}
where the Euclidian action
\begin{equation*}
S_{E}=\int d\tau \int dx\left[\frac{1}{2}\frac{u}{K}
\left(\partial _{x}\varphi \right) ^{2}+\frac{1}{2}uK\left( \partial
_{x}\vartheta \right) ^{2}-i\partial _{\tau }\varphi \partial
_{x}\vartheta \right] .
\end{equation*}
In a Fourier-transformed form
\begin{equation*}
S_{E}=\int d^{2}k\left[ \frac{1}{2}\frac{u}{K}q^{2}\varphi_{\vec{k}}%
\varphi_{-\vec{k}} +\frac{1}{2}uK\vartheta_{\vec{k}}%
 \vartheta_{-\vec{k}} +iq\omega \varphi_{\vec{k}}\vartheta_{-\vec{k}}\right] ,
\end{equation*}
where $\vec{k}\equiv \left( q,\omega \right)$. If one needs only
an average composed of fields of one type ($\varphi $ or
$\vartheta ),$ then the other field can be integrated out.  This leads to two equivalent forms of the
action
\begin{subequations}
\begin{eqnarray}
S_{\varphi } &=&\frac{1}{2K}\int d^{2}k\left[ \frac{1}{u}\omega ^{2}+uq^{2}%
\right] \varphi_{\vec{k}} \varphi_{-\vec{k}}
\label{s_phi_1} \\
&=&\frac{1}{2K}\int dx\int d\tau \left[ \frac{1}{u}\left( \partial
_{\tau
}\varphi \right) ^{2}+\left( \partial _{x}\varphi \right) ^{2}\right]; \\
S_{\vartheta } &=&\frac{K}{2}\int d^{2}k\left[ \frac{1}{u}\omega ^{2}+uq^{2}%
\right] \vartheta_{\vec{k}} \vartheta
_{-\vec{k}}
\label{s_theta_1} \\
&=&\frac{K}{2}\int dx\int d\tau \left[ \frac{1}{u}\left( \partial
_{\tau }\vartheta \right) ^{2}+\left( \partial _{x}\vartheta
\right) ^{2}\right] .
\end{eqnarray}
\end{subequations}

In calculating certain correlation functions, \emph{e.g.}, the
fermionic Green's function, one also needs a cross-correlator
$\langle \varphi
\vartheta \rangle .$ This one is computed by keeping both $\varphi $ and $%
\vartheta $ in the action.

It is convenient to re-write the action in the matrix form
\begin{equation*}
S_{E}=\frac{1}{2}\int d^{2}k\hat{\eta}_{\vec{k}}^{\dagger }\hat{D}^{-1}\hat{%
\eta}_{\vec{k}},
\end{equation*}
where
\begin{equation*}
\hat{\eta}_{\vec{k}}=\left(
\begin{array}{c}
\varphi _{\vec{k}} \\
\vartheta _{\vec{k}}
\end{array}
\right)
\end{equation*}
and the inverse matrix of propagators
\begin{equation*}
\hat{D}^{-1}=\left(
\begin{array}{cc}
q^{2}uK & iq\omega \\
iq\omega & q^{2}\frac{u}{K}
\end{array}
\right) .
\end{equation*}
Inverting the matrix, we obtain
\begin{equation*}
\hat{D}=\frac{1}{u^{2}q^{2}+\omega ^{2}}\left(
\begin{array}{cc}
uK & -i\omega /q \\
-i\omega /q & \frac{u}{K}
\end{array}
\right) .
\end{equation*}
The space-time propagators can be found by performing the Fourier
transforms of $\hat{D}.$ For diagonal terms, one really does not
need to do it, as it is obvious from (\ref{s_phi_1}) and
(\ref{s_theta_1}) that these propagators just coincide with the
Green's function of a 2D Laplace's equations. Recalling that the
potential of a line charge is a log-function of the distance, we
obtain
\begin{eqnarray*}
\Phi \left( z\right) &=&\langle \varphi \left( z\right) \varphi
\left(
0\right) -\varphi ^{2}\left( 0\right) \rangle =\frac{K}{4\pi }\ln \frac{a^{2}%
}{x^{2}+\left( u\left| \tau \right| +a\right) ^{2}}, \\
\Theta \left( z\right) &=&\langle \vartheta \left( z\right)
\vartheta \left(
0\right) -\vartheta ^{2}\left( 0\right) \rangle =\frac{1}{4\pi K}\ln \frac{%
a^{2}}{x^{2}+\left( u\left| \tau \right| +a\right) ^{2}}
\end{eqnarray*}
where $a$ is ``lattice constant'', $z\equiv \left( x,\tau \right) ,$ and $%
x^{2}+u^{2}\tau ^{2}\gg a^{2}.$ \footnote{%
A non-symmetric appearance of the cut-off with respect to time and
space coordinates reflect an asymmetric way the sums over bosonic
momenta and frequencies were cut. We adopted a standard procedure
in which the sum of over $q$ is regularized by $\exp \left(
-\left| k\right| a\right) ,$ whereas
the frequency sum is unlimited. Other choices of regularization are possible.%
} These are the two correlation functions we will need the most.
In addition, there is also an off-diagonal propagator
\begin{eqnarray*}
\Xi \left( z\right) &=&\langle \varphi \left( z\right) \vartheta
\left( 0\right) -\varphi \left( 0\right) \vartheta \left( 0\right)
\rangle =\langle \vartheta \left( z\right) \varphi \left( 0\right)
-\vartheta \left( 0\right)
\varphi \left( 0\right) \rangle \\
&=&\int d^{2}k\left( e^{i\vec{k}\cdot \vec{z}}-1\right) \langle \varphi _{%
\vec{k}}\vartheta _{-\vec{k}}\rangle =-i\int d^{2}k\left(
e^{i\vec{k}\cdot \vec{z}}-1\right) \frac{\omega
/q}{u^{2}q^{2}+\omega ^{2}}.
\end{eqnarray*}
$\Xi \left( z\right) $ depends only on the ratio $x/u\tau $ and
thus does
not change the power-counting. To see this, introduce polar coordinates $%
q=k\cos \alpha /u,\omega =\sin \alpha ,x=(z/u)\cos \beta ,$ and
$\tau =z\sin \beta .$ Then
\begin{eqnarray*}
\Xi \left( x,\tau \right) &=&-i\int \frac{d^{2}k}{\left( 2\pi \right) ^{2}}%
\left( e^{i\left( qx-\omega \tau \right) }-1\right) \frac{\omega /q}{%
u^{2}q^{2}+\omega ^{2}} \\
&=&-i\frac{1}{\left( 2\pi \right) ^{2}}\int_{0}^{\infty }\frac{dk}{k}%
\int_{0}^{2\pi }d\alpha \left( e^{ik\cos \left( \alpha +\beta
\right) }-1\right) \tan \alpha .
\end{eqnarray*}
The resulting integral is a function of only $\beta =\tan
^{-1}\left( x/u\tau \right) $.

\subsubsection{Correlation functions}
\label{sec:cf}
Now we can calculate various correlation functions, including the
Green's function.

Non-time-ordered Green's function for right movers:
\begin{eqnarray}
G_{+}\left( x,\tau \right) &=&-\langle T_{\tau }^{B}\psi
_{+}\left( x,\tau
\right) \psi _{+}^{\dagger }\left( 0,0\right) \rangle  \label{gf_bos} \\
&=&\frac{1}{2\pi a}\langle T_{\tau }^{B}e^{i\sqrt{\pi }\left(
\varphi \left( 1\right) -\vartheta \left( 1\right) \right)
}e^{-i\sqrt{\pi }\left( \varphi \left( 0\right) -\vartheta \left(
0\right) \right) }\rangle ,  \notag
\end{eqnarray}
where $\left( 1\right) \equiv \left( x,\tau \right) $ and $\left(
0\right) \equiv \left( x=0,\tau =0\right) ,$ and where $T_{\tau
}^{B}$ is a bosonic
time-ordering operator. \footnote{%
Surely, it is not a conventional definition of the Green's
function, but it is easier to work with this one for now, and
restore the fermionic $T_{\tau } $ product at the end.} I will use
the well-known result, valid for an average of the product of the
exponentials of gaussian fields (see the books by Tsvelik
\cite{tsvelik} or Giamarchi \cite{giamarchi_book} for a
derivation)
\begin{equation}
\langle T_{\tau }\Pi _{j}e^{iA_{j}\gamma \left( z_{j}\right)
}\rangle =\delta _{\sum_{j}A_{j},0}\times
e^{-\sum_{k>j}A_{j}A_{k}\langle T_{\tau
}\gamma \left( z_{j}\right) \gamma \left( z_{k}\right) \rangle }e^{-\frac{1}{%
2}\sum_{k}A_{k}^{2}\langle \gamma ^{2}\left( z_{j}\right) \rangle
}. \label{identity}
\end{equation}
[Eq. (\ref{identity}) is essentially a field-theoretical analog of
the probability theory result for the average of $e^{iA\gamma },$
where $\gamma $ is a Gaussian random variable.] For example, in
the average
\begin{equation*}
Av\left( z\right) =\langle T_{\tau }e^{i\sqrt{\pi }\varphi \left(
z\right) }e^{-i\sqrt{\pi }\varphi \left( 0\right) }\rangle
\end{equation*}
$A_{1}=\sqrt{\pi }$, $A_{2}=-\sqrt{\pi }$ and
\begin{equation*}
Av\left( z\right) =e^{\pi \langle T_{\tau }\left[ \varphi \left(
z\right) \gamma \left( 0\right) -\varphi ^{2}\left( 0\right)
\right] \rangle }=\left( \frac{a^{2}}{x^{2}+\left( u\left| \tau
\right| +a\right) ^{2}}\right) ^{1/4K}.
\end{equation*}
Similarly, with the help of (\ref{identity}), Eq.~(\ref{gf_bos})
reduces to
\begin{eqnarray}
G_{+}\left( x,\tau \right) &=&\frac{1}{2\pi a}e^{\pi \langle
\varphi \left( 1\right) \varphi \left( 0\right) -\varphi
^{2}\left( 0\right) \rangle _{\tau }}e^{\pi \langle \vartheta
\left( 1\right) \vartheta \left( 0\right) -\vartheta ^{2}\left(
0\right) \rangle _{\tau }}e^{-2\langle \varphi \left( 1\right)
\vartheta \left( 0\right) -\varphi \left( 0\right) \vartheta
\left(
0\right) \rangle _{\tau }}  \notag \\
&=&\frac{1}{2\pi a}e^{\pi \Phi \left( x,\tau \right) }e^{\pi
\Theta \left(
x,\tau \right) }e^{-2\Xi \left( x,\tau \right) }  \label{gplus} \\
&=&\frac{1}{2\pi a}\left( \frac{a^{2}}{x^{2}+\left( u\left| \tau
\right| +a\right) ^{2}}\right) ^{\frac{K+K^{-1}}{4}}e^{if\left(
x/u\tau \right) }, \notag
\end{eqnarray}
where $\langle \dots \rangle _{\tau }$ stands for a time-ordered
product and
where it was used that in a translationally invariant and equilibrium system $%
\langle \varphi ^{2}\left( 0\right) \rangle =\langle \varphi
^{2}\left( 1\right) \rangle $ (same for $\vartheta ).$ Function
$f\left( x/u\tau \right) $ is a phase factor which does not effect
the power-counting.

\paragraph{Bulk tunneling DoS}

For $x=0,$%
\begin{equation*}
G\left( 0,\tau \right) \propto \tau ^{-\frac{K+K^{-1}}{2}}.
\end{equation*}
By power-counting,
\begin{equation}
\nu \left( \varepsilon \right) \propto \left| \varepsilon \right| ^{\frac{%
K+K^{-1}}{2}-1}=\left| \varepsilon \right| ^{\frac{\left( K-1\right) ^{2}}{2K%
}}. \label{nu_bulk}\end{equation} This is an analog of the DL
result for the spinless case.

\paragraph{Edge tunneling DoS}

In a tunneling experiment, one effectively measures the
\emph{local} DoS at the sample's surface. In a correlated electron
system, the boundary condition affects the wavefunction over a
long (exceeding the electron wavelength) distance from the
surface. Therefore, the surface DoS differs significantly from the
``bulk'' one. If a tunneling barrier is high, then--to leading
order in transmission-- the DoS can be found via imposing a
hard-wall boundary condition. The presence of the surface
(boundary) can be taken into account by imposing the boundary
conditions on the number current
\begin{equation}
j\left( x=0,\tau \right) =-\frac{1}{i\sqrt{\pi }}\partial _{\tau
}\varphi =0. \label{bcj}
\end{equation}
at $x=0.$ This means that $\varphi $ is \emph{pinned }at the boundary,\emph{%
\ i.e.}, it takes some time-independent value. In the
gradient-invariant theory, we can always choose this constant to
be zero. Thus,
\begin{equation*}
\varphi \left( 0,\tau \right) =0.
\end{equation*}
This suggests that the local correlator $\Phi (0,\tau )=0$, and
the long-time behavior of the Green's function in
Eq.~(\ref{gplus}) is determined only by the correlator of the
$\vartheta $ fields. Had the  boundary not affected this
correlator, we would have arrived at
\begin{equation}
G\left( x=x^{\prime }=0,\tau \right) \propto \exp \left( \pi
\Theta \left( x=0,\tau \right) \right) \propto \frac{1}{\left|
\tau \right| ^{1/2K}}. \tag{wrong}  \label{g_wrong}
\end{equation}
But then we have a problem, as Eq.~(\ref{g_wrong}) does not
reproduce the free-fermion behavior for $K=1.$ Consequently, the
DoS at the edge $\nu
_{e}\left( \varepsilon \right) \propto \left| \varepsilon \right| ^{\frac{1}{%
2K}-1}$ would have not reproduced the free behavior either. What
went wrong is that we pinned one field but forgot the other one is
canonical conjugate
to the first one. By the uncertainty principle, fixing the ``coordinate'' ($%
\varphi $) increases the uncertainty in the ``momentum''
($\vartheta $)--and vice versa. Thus, fluctuations of $\vartheta $
fields should increase. A rigorous solution to this problem is to
change the fermionic basis from the plane waves to the solutions
of the Schrodinger equation with the hard-wall boundary condition
and to bosonize in this basis. This was done by Eggert and Affleck
\cite{eggert_affleck} and Fabrizio and Gogolin \cite
{fabrizio_gogolin} , \cite{GNT}. Here I will give an heuristic
argument based on a simple image construction, which leads to the
same result.

Eq.~(\ref{bcj}) translates into the boundary conditions for the
bosonic propagators:
\beq
\Phi _{e}(x,x^{\prime },\tau )=0;\;\;
\partial _{x,x^{\prime }}\Theta _{e}\left( x,x^{\prime },\omega \right) =0,
\eeq
for $x,x^{\prime }=0$, where subindex $e$ denotes the correlators
in a semi-infinite system. Since $\Phi _{e}$ and $\Theta _{e}$
satisfy the Laplace's equation, we can view these propagators as
potentials produced by some fictitious charges. Then, $\Phi _{e}$
and $\Theta _{e}$ can be constructed from the propagators of an
infinite sample by the method of images:
\begin{eqnarray*}
\Phi _{e}(x,x^{\prime },\tau ) &=&\Phi (x-x^{\prime },\tau )-\Phi
(x+x^{\prime },\tau ); \\
\Theta _{e}\left( x,x^{\prime },\tau \right) &=&\Theta \left(
x-x^{\prime },\tau \right) +\Theta \left( x+x^{\prime },\tau
\right).
\end{eqnarray*}
For $x=x^{\prime },$

\begin{eqnarray}
\Phi _{e}(0,0,\tau )=0; \;\;
\Theta _{e}\left( 0,0,\tau \right) =2\Theta \left( 0,\tau
\right) .
\end{eqnarray}
Hence, pinning the $\varphi $ field enhances the rms fluctuations of the $%
\vartheta $ field by a factor of two. This leads us to
\begin{eqnarray*}
G_{+}\left( 0,0,\tau \right)  &\propto &\exp \left( \pi \Phi
_{e}\left( 0,0,\tau \right) \right) \exp \left( \pi \Theta
_{e}\left( 0,0,\tau \right)
\right)  \\
&=&\exp \left( 2\pi \Theta _{e}\left( 0,\tau \right) \right)
\propto \exp \left( \frac{2\pi }{2\pi K}\ln \frac{a}{\left| \tau
\right| }\right) \propto \left| \tau \right| ^{-1/K}.
\end{eqnarray*}
Consequently, the DOS becomes
\begin{equation}
\nu _{e}\left( \varepsilon \right) \propto \left| \varepsilon
\right| ^{K^{-1}-1}. \label{nu_edge}\end{equation} This result by
Kane and Fisher \cite{kane_fisher} initiated the new (and still
continuing) surge of interest to 1D systems (in terms of the
impurity scattering time, this result was obtained earlier in
Refs. \cite {mattis,gd}). For tunneling from a contact with
energy-independent DoS
(``Fermi liquid'') into a 1D system, the tunneling conductance scales as $%
\nu _{e}\left( \varepsilon \right) $
\begin{equation*}
{\cal G}(\varepsilon)\propto\nu_e(\varepsilon) \propto \left|
\varepsilon \right| ^{K^{-1}-1}.
\end{equation*}
Now we see that the free-fermion behavior is correctly reproduced
for $K=1.$

Expanding the tunneling exponent $K^{-1}-1$ with parameter $K$
from Eq.~(\ref {kg2g1}) for the weak-coupling case gives
\begin{equation*}
K^{-1}-1\approx \frac{g_{2}-g_{1}}{2\pi v_{F}}.
\end{equation*}
This is the same result as the weak-coupling tunneling exponent
(\ref {alpha_s}) obtained in Sec.\ref{sec:ygm} via the scattering
theory for interacting fermions.

Where do the ``bulk" and ``edge" forms of DoS match? Consider an
object $G(x=x',\varepsilon)$. At the boundary, the DoS is of the
``edge'' form (\ref{nu_edge}). Far away from the boundary, the
Green's function does not depend on $x$ and $\nu(\varepsilon)$
acquires a ``bulk" form (\ref{nu_bulk}). As a function of $x$,
$G(x=x',\varepsilon)$ varies on the scale $\simeq u/|\varepsilon|$
and the crossover between two limiting forms of $\nu$ occurs on
this scale. Choosing the energy in a tunneling experiment,
\emph{i.e.}, temperature or bias--whichever is larger, determines
how far from the boundary one should go in order to see a change
in the scaling behavior.
\subsection{Fermions with spin}

For fermions with spin, each component of the fermionic operator
is bosonized separately
\begin{equation*}
\psi _{\pm ,\sigma }=\frac{1}{\sqrt{2\pi a}}\exp \left[ \pm i\sqrt{\pi }%
\left( \varphi _{\sigma }\mp \vartheta _{\sigma }\right) \right] .
\end{equation*}
Index $\sigma $ of the bosonic field does not mean that bosons
acquired spin. We simply have more bosonic fields. Charge and spin
densities and currents are related to the derivatives of the
bosonic fields
\begin{eqnarray*}
\rho _{\pm ,\sigma } &=&\frac{1}{2\sqrt{\pi }}\left( \varphi
_{\sigma }^{\prime }\mp \vartheta _{\sigma }^{\prime }\right)
\;\rho _{\sigma }=\rho _{+,\sigma }+\rho _{-,\sigma
}=\frac{1}{\sqrt{\pi }}\varphi _{\sigma
}^{\prime }; \\
j_{\sigma } &=&\rho _{+,\sigma }-\rho _{-,\sigma }=\frac{1}{\sqrt{\pi }}%
\vartheta _{\sigma }^{\prime }; \\
\rho _{c} &=&\rho _{\uparrow }+\rho _{\downarrow }=\frac{1}{\sqrt{\pi }}%
\left( \varphi _{\uparrow }^{\prime }+\varphi _{\downarrow
}^{\prime
}\right) =\sqrt{\frac{2}{\pi }}\varphi _{c}^{\prime } \\
\rho _{s} &=&\rho _{\uparrow }-\rho _{\downarrow }=\frac{1}{\sqrt{\pi }}%
\left( \varphi _{\uparrow }^{\prime }-\varphi _{\downarrow
}^{\prime
}\right) =\sqrt{\frac{2}{\pi }}\varphi _{s}^{\prime }; \\
j_{c} &=&j_{\uparrow }+j_{\downarrow }=\frac{1}{\sqrt{\pi }}\left(
\vartheta
_{\uparrow }^{\prime }+\vartheta _{\downarrow }^{\prime }\right) =\sqrt{%
\frac{2}{\pi }}\vartheta _{c}^{\prime }; \\
j_{s} &=&j_{\uparrow }-j_{\downarrow }=\frac{1}{\sqrt{\pi }}\left(
\vartheta
_{\uparrow }^{\prime }-\vartheta _{\downarrow }^{\prime }\right) =\sqrt{%
\frac{2}{\pi }}\vartheta _{s}^{\prime },
\end{eqnarray*}
where $^{\prime }$ denotes $\partial _{x}$ and where the charge
and spin bosons are defined as
\beq
\varphi _{c,s}=\frac{\varphi _{\uparrow }\pm \varphi _{\downarrow }}{%
\sqrt{2}};\;
\vartheta _{c,s}=\frac{\vartheta _{\uparrow }\pm \vartheta _{\downarrow }%
}{\sqrt{2}}.
\eeq
I assume that the interaction is spin-invariant, i.e., couplings of $%
\uparrow \uparrow $ and $\uparrow \downarrow $ fermions are the
same. Substituting the relations between charge- and
spin-densities into the Hamiltonian, one arrives at the familiar
bosonized Hamiltonian which consists of  totally independent
charge and spin parts
\begin{eqnarray}
H &=&H_{c}+H_{s};  \notag \\
H_{c} &=&\frac{1}{2}\int dx\frac{u_{c}}{K_{c}}\left( \partial
_{x}\phi _{c}\right) ^{2}+u_{c}K_{c}\left( \partial _{x}\theta
_{c}\right) ^{2};
\notag \\
H_{s} &=&\frac{1}{2}\int dx\frac{u_{s}}{K_{s}}\left( \partial
_{x}\phi
_{s}\right) ^{2}+u_{s}K_{s}\left( \partial _{x}\theta _{s}\right) ^{2}\notag\\
&&+\frac{%
2g_{1}}{\left( 2\pi a\right) ^{2}}\int dx\cos \left( \sqrt{8\pi
}\phi _{\sigma }\right).\label{spin}
\end{eqnarray}
Parameters of the Gaussian parts are related to the microscopic
parameters of the original Hamiltonian

\begin{eqnarray*}
u_{c} &=&\left( 1+\frac{g_{1}}{2\pi }\right) ^{1/2}\left( 1+\frac{%
4g_{2}-g_{1}}{2\pi }\right) ^{1/2}; \;
K_{c} =\left( \frac{1+g_{1}/2\pi }{1+\left( 4g_{2}-g_{1}\right) /2\pi }%
\right) ^{1/2}; \\
u_{s} &=&\left( 1-\left( \frac{g_{1}}{2\pi }\right) ^{2}\right) ^{1/2}; \;
K_{s} =\left( \frac{1+g_{1}/2\pi }{1-g_{1}/2\pi }\right) ^{1/2}.
\end{eqnarray*}
Notice that $K_{c}<1$ for $g_{1}<2g_{2}$ (``repulsion'') and $K_{c}>1$ for $%
g_{1}>2g_{2}$ (``attraction''). The boundaries for ``repulsive''
and ``attractive'' behaviors coincide with those obtained when
studying tunneling of interacting electrons.
The velocity of the
charge part for $g_{1}=0$ coincides with that found in the DL
solution (Sec. \ref{sec:DL})
\begin{equation*}
u_{c}=\left( 1+\frac{2g_{2}}{\pi }\right) ^{1/2}.
\end{equation*}

%Some things in the bosonized Hamiltonian do not look right. For
%example, the product
%\begin{equation*}
%u_{c}K_{c}=1+\frac{g_{1}}{2\pi }\neq 1.
%\end{equation*}
%This means that the Drude weight in a Galilean-invariant system is
%renormalized by the interactions--and this cannot happen. However,
%due to the cos-term in $H_{s}$ in the hamiltonian is not Gaussian
%yet. For $K_{s}<1 $ (which means repulsive interaction in the spin
%channel), $g_{1}$ flows to zero. In this case, Galilean invariance
%is restored.

Scaling dimension of the backscattering term in the spin part can be read off
from the correlation function
\begin{equation*}
\frac{1}{a^{4}}\langle e^{i\sqrt{8\pi }\phi _{\sigma }\left( z\right) }e^{-i%
\sqrt{8\pi }\phi _{s}}\rangle =\frac{1}{a^{4}}\exp \left( \frac{8\pi K_{s}}{%
4\pi }\ln \frac{a^{2}}{z^{2}}\right) =\frac{1}{a^{4}}\left(
\frac{a}{\left| z\right| }\right) ^{4K_{s}}\propto a^{4\left(
K_{s}-1\right) }.
\end{equation*}
If we allowed for different coupling constants between electrons
of different spin orientations, then the coefficient in front of
the cos term would have been $g_{1\perp }.$ For $K_{s}>1,$ the
operator scales down to zero as $a\rightarrow 0$, whereas for
$K_{s}<1,$ it blows up signaling an instability: a spin-gap phase.

The RG-flow of the spin-part is described by the
Berezinskii-Kosterlitz-Thouless phase diagram. The fixed-point value of  $%
g_{1}^{\ast }=0$ for  $K_{s}^{\ast }>1.$ In the weak coupling
limit, the RG reduces to a single equation for  $g_{1},$ which we
have derived in the fermionic language in Sec.\ref{sec:RG}
\beq
\frac{dg_{1}}{dl}=-g_{1}^{2} \to
g_{1} =\frac{1}{\left( g_{1}^{0}\right) ^{-1}+l},
\eeq

\subsubsection{Tunneling density of states}
\label{sec:dos} The procedure of finding the scaling behavior for
the DoS reduces to a simple recipe.

\begin{itemize}
\item  Take the free Green's function and split it formally into
the spin and charge parts
\begin{equation*}
G\left( x,t\right) =\frac{1}{x-t}=\frac{1}{\left( x-t\right) ^{1/2}}\frac{1}{%
\left( x-t\right) ^{1/2}}.
\end{equation*}

\item  In an interacting system, the exponent of $1/2$ in the charge part is replaced by $%
\left( K_{c}+K_{c}^{-1}\right) /4$ and in the spin-part by $\left(
K_{s}+K_{s}^{-1}\right) /4.$ If the spin-rotational invariance is
preserved, then the spin exponent remains equal to $1/2.$

\item  For $x=0$,%
\begin{equation*}
G\left( t\right) \propto \frac{1}{t^{\left(
K_{c}+K_{c}^{-1}\right) /4+1/2}}
\end{equation*}
and the DoS behaves as
\begin{equation*}
\nu \left( \varepsilon \right) \propto \left| \varepsilon \right|
^{\left( K_{c}+K_{c}^{-1}\right) /4-1/2}=\left| \varepsilon
\right| ^{\frac{\left( K_{c}-1\right) ^{2}}{4K_{c}}}=\left| \varepsilon \right| ^{\frac{\left( u_{c}-1\right)
^{2}}{4u_{c}}}.
\end{equation*}

Comparing this result for $g_{1}=0$ with that by DL (Sec.
\ref{sec:DL}), we see that the bosonization solution gives the
same result as  the fermionic one.

\item  For tunneling into the edge, remove $K_{c},$ which comes
from the correlator $\langle \varphi \varphi \rangle $ pinned by
the boundary, and multiply $K_{c}^{-1}$, which comes from $\langle
\vartheta \vartheta \rangle $, by a factor of 2. This gives
\begin{equation*}
G_{e}\left( t\right) \propto \frac{1}{t^{K_{c}^{-1}/2+1/2}}
\end{equation*}
and
\begin{equation*}
{\cal G}\propto \nu _{e}\left( \varepsilon \right) \propto \left| \varepsilon \right| ^{%
\frac{1}{2}\left( K_{c}^{-1}-1\right) }.
\end{equation*}
\end{itemize}

Expanding $K_{c}$ back in the interaction
\begin{eqnarray*}
K_{c} &=&\left( \frac{1+g_{1}/2\pi }{1+\left( 4g_{2}-g_{1}\right) /2\pi }%
\right) ^{1/2}\approx 1-\frac{g_{2}-(1/2)g_{1}}{\pi },
\end{eqnarray*}
we obtain the weak-coupling limit for the tunneling exponent
\begin{equation*}
(1/2)\left(1/K_{c}-1\approx \frac{g_{2}-(1/2)g_{1}}{2\pi }\right).
\end{equation*}
This coincides with the result obtained in the fermionic language
(Sec.\ref {sec:ygm}). What was missed in a bosonization solution is a
multiplicative log-renormalization, present in Eq.~(\ref{ygm_cond}).
 This is because we evaluated
$G$ at the fixed point, where $g_{1}^{\ast }=0$, rather then
derived an independent RG equation for the flow of the
conductance.  This procedure should bring in the log-factors (cf.
Ref.\cite{gd} where these factors were obtained for the impurity
scattering time).
%%%%%%%%%%%%%%%%%%%%%%%%%%%%%%%%%%
%%%%%%%%%LECTURE 6
%%%%%%%%%%%%%%%%%%%%%%%%%%%%%%%%%

\section{Transport in quantum wires}
\label{transport}
\subsection{Conductivity and conductance}
\subsubsection{Galilean invariance}
Interactions between electrons cannot change the response to an electric
field in a Galilean-invariant system--the electric field couples only to the
center-of-mass whose motion is not affected by the inter-electron
interaction. This property is reproduced by the bosonized theory provided
that the product $uK=1$ ($=v_{F}$ in dimensional form.) To see this, combine
the Heisenberg equation of motion for density $\rho $ $\left( \text{spinless
fermions}\right) $ with the continuity equation:
\begin{equation}
\partial _{t}\rho =i[H,\rho ]=-\partial _{x}j.  \label{poisson}
\end{equation}
Calculating the commutator in Eq.~(\ref{poisson}) with the help of
Eq.~(\ref{comm_main}), we identify the current operator as
\begin{eqnarray*}
\partial _{t}\rho &=&\frac{uK}{\sqrt{\pi }}\partial _{x}^{2}\vartheta
=-\partial _{x}\left( -\frac{uK}{\sqrt{\pi }}\partial _{x}\vartheta \right)
\rightarrow \\
j &=&-\frac{uK}{\sqrt{\pi }}\partial _{x}\vartheta \text{ }.\text{ }
\end{eqnarray*}
The current is not affected by the interaction as long as $uK=1.$

\subsubsection{Kubo formula for conductivity}

The Kubo formula relates the conductivity, a response function to an
electric field at finite $\omega $ and $q,$ to the current-current
correlation function
\begin{equation}
\sigma \left( \omega ,q\right) =\frac{1}{i\omega }\left[ -\frac{e^{2}}{\pi }%
+\langle JJ\rangle _{q\omega }^{R}\right] ,  \label{kubo}
\end{equation}
where it was used that $n=k_{F}/\pi $ and $k_{F}/m=v_{F}=1$ in our
units.

Electric current for electrons $(e>0)$
\[
J=-ej=\frac{e}{\sqrt{\pi }}\partial _{x}\vartheta .
\]
In complex time,
\begin{eqnarray}
\langle JJ\rangle _{x,\tau }^{R} &=&\left( \frac{e}{\sqrt{\pi }}\right)
^{2}\left( -\partial _{x}^{2}\right) \langle \vartheta \vartheta \rangle
_{x,\tau }\rightarrow  \nonumber \\
\langle JJ\rangle _{q,\omega _{m}}^{R} &=&\frac{e^{2}}{\pi }q^{2}\langle
\vartheta \vartheta \rangle _{q,\omega _{m}}  \nonumber \\
&=&\frac{e^{2}}{\pi }-\frac{e^{2}}{\pi }\frac{\omega
_{m}^{2}}{\omega _{m}^{2}+u^{2}q^{2}}=\frac{e^{2}}{\pi }+\langle
\tilde{J}\tilde{J}\rangle _{q,\omega _{m}}.  \label{jj}
\end{eqnarray}
The first term in (\ref{jj}) cancels the diamagnetic response in
(\ref{kubo}). Continuing analytically to real frequencies, we find
\begin{eqnarray}
\sigma \left( \omega ,q\right) &=&\frac{1}{i\omega }\langle \tilde{J}\tilde{J%
}\rangle _{q,\omega _{m\rightarrow -i\omega +\delta }}=-\frac{e^{2}}{\pi }%
\frac{1}{i\omega }\frac{-\omega ^{2}}{-\left( \omega +i\delta \right)
_{m}^{2}+u^{2}q^{2}}  \label{sigma_inv} \\
&=&i\frac{e^{2}}{\pi }\frac{\omega }{\omega ^{2}-u^{2}q^{2}+i\text{sgn}%
\omega \delta }.  \nonumber
\end{eqnarray}
Consequently, the dissipative conductivity is equal to
\begin{eqnarray}
\text{Re}\sigma \left( \omega ,q\right) &=&-\frac{e^{2}}{\pi }\omega \text{Im%
}\frac{1}{\omega ^{2}-u^{2}q^{2}+i\text{sgn}\omega \delta }  \nonumber \\
&=&\frac{e^{2}}{2}\left[ \delta \left( \omega -uq\right) +\delta \left(
\omega +uq\right) \right] .  \label{sigma_omega_q}
\end{eqnarray}
\subsubsection{\protect\bigskip Drude conductivity}
In a macroscopic system, one is accustomed to take the limit
$q\rightarrow 0$ first: this corresponds to applying a spatially
uniform but time-dependent electric field \cite{mahan}. (For the
lack of a better name, I will refer to the conductivity obtained
in this way as to the \emph{Drude conductivity}). The Drude
conductivity in our case is the same as for the Fermi gas as the
charge velocity drops out from the result
\[
\text{Re}\sigma \left( \omega ,0\right) =e^{2}\delta \left( \omega \right)
\]
or, restoring the units,
\[
\text{Re}\sigma \left( \omega ,0\right) =\frac{e^{2}v_{F}}{\hbar }\delta
\left( \omega \right) .
\]
All it means that when a static electric field is applied to a continuous
system of either free or interacting electrons, the center-of-mass moves
with an acceleration and there is no linear response, as there is no
``friction'' that can balance the electric force.

For electrons with spins, the electrical current is related only to the
charge component of the $\vartheta -$ field:
\[
J_{c}=-ej_{c}=e\sqrt{\frac{2}{\pi }}\partial _{x}\vartheta _{c},
\]
where again $u_{c}K_{c}=1.$ Because of the $\sqrt{2}$ factor in the current,
the conductivity is by a factor of two different from that in the spinless
case
\[
\text{Re}\sigma \left( \omega ,0\right) =\frac{2e^{2}v_{F}}{\hbar }\delta
\left( \omega \right) .
\]
(Notice, however, that at fixed density $v_{F}$ is by a factor of
2 smaller for electrons with spin, so that the conductivity is the
same.)

\subsubsection{Landauer conductivity}

Let's consider now the opposite order of limits, corresponding to
a situation when a static electric field is applied over a part of
the infinite wire. (Again, for the lack of a better name, I will
refer to this conductivity as to the \emph{Landauer
conductivity}.) The electric field might as well be non-uniform;
the only constraint we are going to use is that the integral
\[
\int dxE\left( x\right) ,
\]
equal to the applied voltage, is finite. The induced current (which in 1D
coincides with the current density) is given by
\begin{eqnarray*}
J\left( t,x\right) &=&\int dx^{\prime }\int dt^{\prime }\sigma \left(
t-t^{\prime };x,x^{\prime }\right) E\left( t^{\prime },x^{\prime }\right) \\
&=&\int dx^{\prime }\int \frac{d\omega }{2\pi }e^{-i\omega t}\sigma (\omega
;x,x^{\prime })E\left( \omega ,x^{\prime }\right) .
\end{eqnarray*}
In linear response, the conductivity is defined in the absence of the field.
As such, it is still a property of a translationally invariant system and
depends thus only on $x-x^{\prime }.$ This allows one to switch to Fourier
transforms
\begin{equation}
J\left( t,x\right) =\int dx^{\prime }\int \frac{d\omega }{2\pi }\int \frac{dq%
}{2\pi }e^{iq\left( x-x^{\prime }\right) }e^{-i\omega t}\sigma \left( \omega
,q\right) E\left( \omega ,x^{\prime }\right) .  \label{cos}
\end{equation}
Now use the fact that the applied field is static: $E\left( x,\omega \right)
=2\pi \delta \left( \omega \right) E_{0}\left( x\right) $ (upon which the $t$%
-dependence of the current disappears, as it should be in the steady state)
\begin{equation}
J(x)=\int dx^{\prime }\int \frac{dq}{2\pi }e^{iq\left( x-x^{\prime }\right)
}\sigma \left( 0,q\right) E_{0}\left( x^{\prime }\right) .  \label{current}
\end{equation}
From (\ref{sigma_omega_q}),
\begin{equation}
\sigma \left( 0,q\right) =\frac{1}{u}e^{2}\delta \left( q\right)
=Ke^{2}\delta \left( q\right) ,  \label{sigma_q_spinless}
\end{equation}
where $uK=1$ was used again. Substituting (\ref{sigma_q_spinless}) into (\ref
{current}), we see that the $x-$ dependence of the current also disappears
\[
J=\frac{Ke^{2}}{2\pi }\int dx^{\prime }E_{0}\left( x^{\prime }\right) =\frac{%
Ke^{2}}{2\pi }V.
\]
Conductance $\mathcal{G}=J/V$ is given by
\[
\mathcal{G}=\frac{Ke^{2}}{2\pi },
\]
or, restoring the units,
\begin{equation}
\mathcal{G}=K\frac{e^{2}}{h}.  \label{withKnospin}
\end{equation}
For electrons with spin, a similar consideration gives
\begin{equation}
\mathcal{G}=K_{c}\frac{2e^{2}}{h}.  \label{withK}
\end{equation}
We see that the conductance is renormalized by the interactions from it
universal value given by the Landauer formula for an ideal wire \cite
{kane_fisher}.

\subsection{Dissipation in a contactless measurement}

What kind of an experiment Eqs. (\ref{withKnospin}) and (\ref{withK})
correspond to?

Suppose that we connect a wire of length $L$ to an external resistor and
place the whole circuit into a resonator \cite{levinson}. Now, we apply an \emph{ac }%
electric field $E\left( x,t\right) $ of frequency $\omega _{0}$
and parallel to a segment of the wire of length $L_{E}\ll L,$ and
measure the losses in the resonator. The external resistor takes
care of energy dissipation: as the wire is ballistic (also in a
sense that electrons travel through the wire without emitting
phonons), the Joule heat can be generated only outside the wire.
Dissipated energy, averaged over many periods of the field, is
given by
\[
\dot{Q}=-\int dx\langle J\left( x,t\right) E\left( x,t\right) \rangle .
\]
For a monochromatic field, $E\left( x,t\right) =E_{0}\left( x\right) \cos
\omega _{0}t$ and after averaging over many periods of oscillations, we
obtain
\[
\dot{Q}=-\int dx\int dx^{\prime }\text{Re}\sigma \left( \omega
_{0};x,x^{\prime }\right) E_{0}\left( x\right) E_{0}\left( x^{\prime
}\right) .
\]
Now, choose the frequency in such a way that
\begin{equation}
L_{E}\ll \frac{u}{\omega _{0}}\ll L,  \label{condition}
\end{equation}
where $u$ is the velocity of the charge mode in the wire. Because the
wavelength of the charge excitations at frequency $\omega _{0}$--acoustic
plasmons-- is much shorter than the distance to contacts ($L$), the
conductivity is essentially the same as for an infinite wire and depends
only on $x-x^{\prime }.$ Performing partial Fourier transform in Eq. (\ref
{sigma_omega_q}), we find \emph{\ }
\begin{equation}
\text{Re}\sigma \left( \omega ,x\right) =\frac{e^{2}}{2\pi u}\cos \left(
\omega x/u\right) =\frac{e^{2}}{2\pi }K\cos \left( \omega x/u\right) ,
\end{equation}
so that
\[
\dot{Q}=-\frac{1}{2}\frac{e^{2}}{2\pi }K\int dx\int dx^{\prime }\cos \left[
\omega _{0}\left( x-x^{\prime }\right) /u\right] E_{0}\left( x\right)
E_{0}\left( x^{\prime }\right) .
\]
On the other hand, because $\left| x,x^{\prime }\right| \leq L_{E}$ $\ll
u/\omega _{0},$ the cosine can be replaced by unity, and
\[
\dot{Q}=-\frac{e^{2}}{2\pi }KV^{2}\equiv -\mathcal{G}V^{2},
\]
or
\[
\mathcal{G=}\frac{e^{2}}{2\pi }K.
\]
Therefore, dissipation in a contactless measurements under the
conditions specified by Eq. (\ref{condition}) corresponds to a
renormalized conductance. To the best of my knowledge, this
experiment has not been performed. A typical (two-probe) transport
measurement is done by applying the current and measuring the
voltage drop between the reservoirs. In this case, the measured
conductance does \emph{not }correspond to Eqs.(\ref
{withKnospin},\ref{withK}) but is rather given simply by $e^{2}/h$
per spin projection--\emph{regardless of the interaction in the
wire }\cite {maslov_stone},\cite{ponomarenko},\cite{safi_schulz}.
This result is discussed in the next Section.

\subsection{Conductance of a wire attached to reservoirs}

\bigskip The reason why the two-terminal conductance is not renormalized by
the interactions within the wire is very simple. For the Fermi-gas case, the
conductance of $e^{2}/h$ per channel is actually not the conductance of wire
itself--a disorder-free wire by itself does not provide any resistance to the
current. In a four-probe measurement, when the voltage and current are
applied to and measured in different contacts, the conductance of a disorder-free
wire is, in fact, infinite. However, in a two-probe measurement, the voltage
and current contacts are the same. Finite resistance comes only from
scattering of electrons from the boundary regions, connecting wide
reservoirs to the narrow wire \cite{imry,landauer}, as shown in Fig.\ref
{fig:wire}a). The universal value of $e^{2}/h$ is approached in the limit of
an adiabatic (smooth on the scale of the electron wavelength) connection
between the reservoirs and the wire \cite{glazman_wire} \footnote{%
Accidentally, the actual constraint on the adiabaticity of the connection is
rather soft--it is enough to require the radius of curvature of the
transition region be just comparable to, rather than much larger than, the
electron wavelength \cite{glazman_wire}.}. As the resistance comes from the
regions \emph{exterior }to the wire, the interaction \emph{within} the wire
is not going to modify the $e^{2}/h-$ result. Another way to think about it
is to notice that the renormalized conductance (\ref{withKnospin},\ref{withK}%
) can be interpreted as a manifestation of the \emph{fractional charge }$%
e^{\ast }=\sqrt{K}(\sqrt{K_{c}})$, associated with the excitations
in a 1D system. However, the current coming from, \emph{e.g.,
}the\emph{\ }left reservoir is carried by integer charges, and as
all these charges  get eventually transmitted through the wire,
the current collected in the right reservoir is carried again by
integer charges. Fractional charges is a
transient phenomena which, in principle, can be observed in an \emph{ac }%
conductance or noise measurements but not in a \emph{dc
}experiment. In the rest of this Section, these arguments will be
substantiated with some simple calculations.
\begin{figure}[tbp]
\begin{center}
\epsfxsize=0.7\columnwidth\epsffile{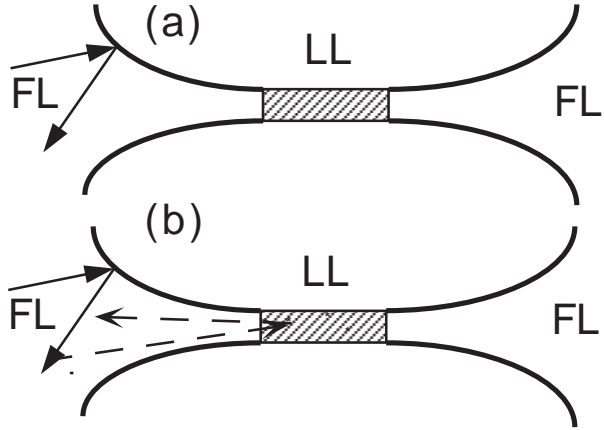}
\end{center}
\caption{ a) A Luttinger-liquid (LL) wire attached to Fermi-liquid (FL)
reservoirs. b) same for a single impurity within the wire.}
\label{fig:wire}
\end{figure}
\subsubsection{Inhomogeneous Luttinger-liquid model}
\label{sec:inhll}
An actual system consists of two Fermi-liquid reservoirs connected via a
Luttinger-liquid (LL) wire and , due to the presence of the reservoirs, is
not one-dimensional. In the \emph{inhomogeneous Luttinger-liquid model, }the
actual system is replaced by an effective 1D system, which is an infinite LL
with a position-dependent interaction parameter $K(x)$ (cf. Fig.~\ref{fig:1Dmodel}%
). The actual reservoirs are higher ($D=2$ or $3)$ systems, where
the effect of the interaction can be disregarded. Consequently,
the reservoirs are modelled by one-dimensional free conductors
with $K_{\text{L}}=1.$ In between, $K\left( x\right) $ goes
through some variation. Similarly, the
charge velocity is equal to the Fermi one in the reservoirs and varies with $%
x $ in the middle of the system. The potential difference applied
to the system produces some distribution of the electric field
along the wire. The shape of this distribution is irrelevant in
the {\em dc} linear response.
\subsubsection{Elastic-string analogy}

\begin{figure}[tbp]
\begin{center}
\epsfxsize=0.7\columnwidth \epsffile{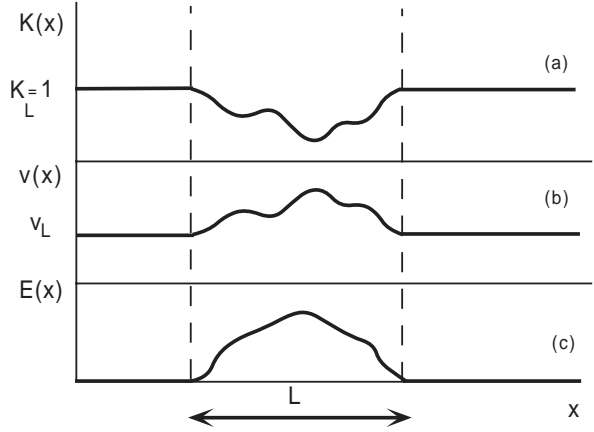}
\end{center}
\caption{ Inhomogeneous Luttinger-liquid model.}
\label{fig:1Dmodel}
\end{figure}

The (real-time) bosonic action for a spinless LL is
\begin{equation}
S=\frac{1}{2}\int d^{2}x\frac{1}{K(x)}\left\{ u(x)(\partial _{x}\varphi
)^{2}-\frac{1}{u(x)}(\partial _{t}\varphi )^{2}\right\} .
\label{(EQ:action_r)}
\end{equation}
The density of the electrons (minus the background density) and the (number)
current are given by
\begin{equation}
\rho =\partial _{x}\varphi /\sqrt{\pi },\qquad j=-\partial _{t}\varphi /%
\sqrt{\pi }.  \label{(EQ:rho_j)}
\end{equation}
The interaction with an external electromagnetic field $A_{\mu }$ is
described by
\begin{equation}
S_{int}=\frac{e}{2\sqrt{\pi }}\int d^{2}x\left\{ A_{0}\partial _{x}\varphi
-A_{1}\partial _{t}\varphi \right\} ,  \label{EQ:s_int}
\end{equation}
so that the equation of motion for $\varphi $ is
\begin{equation}
\partial _{t}\left( \frac{1}{Ku}\partial _{t}\varphi \right) -\partial
_{x}\left( \frac{u}{K}\partial _{x}\varphi \right) =\frac{e}{\sqrt{\pi }%
\hbar }E(x,t),  \label{EQ:eq_m}
\end{equation}
where $E=-\partial _{x}A_{0}+\partial _{t}A_{1}$ is the electric field. We
assume that the electric field is switched on at $t=0$, so that $E(x,t)=0$
for $t<0$ and $E(x,t)=E(x)$ for $t\geq 0$. The problem reduces now to
determining the profile of an infinite elastic string under the external
force. In this language, $\varphi (x,t)$ is the transverse displacement of
the string at point $x$ and at time $t$, while the number current $%
j=-\partial _{t}\varphi /\sqrt{\pi }$ is proportional to the transverse
velocity of the string.

To develop some intuition into the solution of Eq.~(\ref{EQ:eq_m}), we first
solve it in the homogeneous case, when $K=$const, $u=$const, and $E(x)=$%
const for $|x|\leq L/2,$ and is equal to zero otherwise. In this case, the
solution of Eq.~(\ref{EQ:eq_m}) for $t>L/u$, is
\[
\varphi (x,t)=\frac{KeV}{2\sqrt{\pi }}\times \left\{
\begin{array}{cl}
t-\frac{x^{2}+L^{2}/4}{Lu}\;\mathrm{for}\;|x|\leq L/2; &  \\
t-|x|/u\;\mathrm{for}\;L/2\leq |x|\leq ut-L/2 &  \\
\frac{u}{2L}\big(t-\frac{|x|-L/2}{u}\big)^{2}\;\mathrm{for}\;ut-L/2\leq
|x|\leq ut+L/2\cr0,\;\mathrm{for}\;|x|\geq ut+L/2, &
\end{array}
\right.
\]
where $V=EL$ is the total voltage drop. This solution is depicted in Fig.~%
\ref{fig:string}a.

\begin{figure}[tbp]
\begin{center}
\epsfxsize=1.0\columnwidth \epsffile{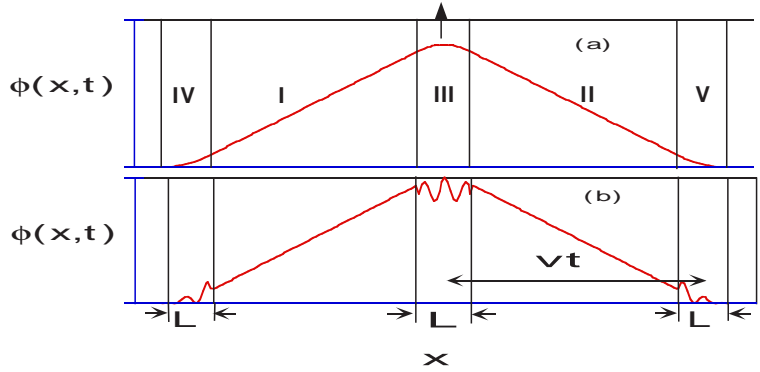}
\end{center}
\caption{ a)~Solution of the wave equation in the homogeneous case for $%
t=5L/u$. b)~Schematic solution in the inhomogeneous case for $t\gg L/u$.}
\label{fig:string}
\end{figure}
The profile of the string consists of two segments (I and II in
Fig.~\ref {fig:string}a) whose widths, equal to $(ut-L)$, grow
with time, and of three segments (III, IV and V in
Fig.~\ref{fig:string}a) whose widths are constant
in time and equal to $L$. In segments I and II, the profile of the string $%
\varphi (x,t)$ is linear in $x$, and therefore, being the solution
of the wave equation, also in $t$; in segments III-V, the profile
is parabolic. Outside segments IV and V, the string is not
perturbed yet, and $\varphi (x,t)=0$. As time goes on, the larger
and larger part of the profile becomes linear. For late times, the
pulse produced by the force spreads outwards with velocity $u$,
involving the yet unperturbed parts of the string in motion;
simultaneously, in all but narrow segments in the middle and at
the
leading edges of the pulse, the string moves upwards with the $t$- and $x$%
-independent ``velocity''\ $\partial _{t}\varphi =KeV/2\sqrt{\pi }$. In
terms of the original transport problem, it means that the charge current $%
J=-ej$ is constant outside the wire (but not too close to the edges of the
regions of where the electron density is not yet perturbed by the electric
field) and given by $J=Ke^{2}V/h$. Therefore, the conductance (per spin
orientation) is $\mathcal{G}=Ke^{2}/h$.

We now turn to the inhomogeneous case. As in the previous case, the profile
consists of several characteristic segments (cf. Fig.~\ref{fig:string}b). In
segments III-V, the profile is affected by the inhomogeneities in $K(x)$, $%
u(x)$, and $E(x)$ and depends on the particular choice of the $x$%
-dependences in all these quantities. In segments I and II however, the
profile, being the solution of the free wave equation, is again linear in $x$
(and in $t$). Requiring the slopes of the string be equal and opposite in
segments I and II (which is consistent with the condition of the current
conservation), the solution in these regions can be written as $\varphi
(x,t)=A(t-|x|/u_{\mathrm{L}})$. The constant $A$ can be found by integrating
Eq.~(\ref{EQ:eq_m}) between two symmetric points $\pm a$ , chosen outside
the wire
\begin{equation}
-\int_{-a}^{+a}dx\partial _{x}\Big(\frac{u}{K}\partial _{x}\varphi \Big)=%
\frac{e}{\sqrt{\pi }}\int_{-a}^{+a}dxE(x)=\frac{eV}{\sqrt{\pi }}.
\end{equation}
Outside the wire, $K(x)=K_{\mathrm{L}}$ and $u(x)=u_{\mathrm{L}}$, thus $%
A=K_{\mathrm{L}}eV/2\sqrt{\pi }\hbar $. Calculating the current, we get $%
\mathcal{G}=K_{\mathrm{L}}e^{2}/h$ and, recalling that $K_{\mathrm{L}}=1$,
we finally arrive at $\mathcal{G}=e^{2}/h$. \textit{Thus, the conductance is
not renormalized by the interactions in the wire.}

\subsubsection{Kubo formula for a wire attached to reservoirs}

The Kubo formula for a translationally non-invariant system can be written
as
\begin{eqnarray*}
\sigma \left( \omega ;x,x^{\prime }\right) &=&-\frac{e^{2}}{i\pi \omega }%
\delta \left( x-x^{\prime }\right) \\
&+&\frac{e^{2}}{i\pi \omega }\left\{ \int
d\left( \tau -\tau ^{\prime }\right) e^{i\omega _{m}\tau }\langle T_{\tau
}\partial _{\tau }\varphi \left( x,\tau \right) \partial _{\tau ^{\prime
}}\varphi \left( x^{\prime },\tau ^{\prime }\right) \rangle \right\}
|_{i\omega _{m}\rightarrow \omega +i\delta }.
\end{eqnarray*}
The diamagnetic contribution is cancelled by a delta-function term, which is
obtained when integrating by parts in the time-ordered product \cite
{shankar90,giamarchi_book}. Having this in mind, I will re-write the
conductivity via the Fourier transform of the $\varphi \varphi $- correlator
without the $T-$ product
\[
\sigma \left( \omega ;x,x^{\prime }\right) =i\frac{e^{2}}{\pi \omega }\omega
_{m}^{2}\Phi _{\omega _{m}}\left( x,x^{\prime }\right) |_{i\omega
_{m}\rightarrow \omega +i\delta }.
\]
For a translationally invariant case, this reduces back to Eq.~(\ref
{sigma_inv}). Now, $K\left( x\right) $ and $u\left( x\right) $ depend on
position. The propagator of the $\varphi $ fields satisfy the wave equation
(or a Laplace's equation as we are dealing with the imaginary time)
\begin{equation}
\left[ \frac{\omega _{m}^{2}}{u\left( x\right) K\left( x\right) }-\partial
_{x}\left( \frac{u\left( x\right) }{K\left( x\right) }\partial _{x}\right) %
\right] \Phi _{\omega _{m}}\left( x,x^{\prime }\right) =\delta \left(
x-x^{\prime }\right) .  \label{laplace_inh}
\end{equation}
In a model of step-like variation of $K\left( x\right) $ and $u\left(
x\right) $ ($K=K_{W}$ and $u=u_{W}$ within the wire and $K=K_{L}=1$ and $%
u=u_{L}=1$ outside the wire), Eq. (\ref{laplace_inh}) is complemented by the
following boundary conditions: 1) $\Phi _{\omega _{m}}\left( x,x^{\prime
}\right) $ is continuous at $x=\pm L/2,$ 2) $u\left( x\right) K\left(
x\right) \partial _{x}\Phi _{\omega _{m}}\left( x,x^{\prime }\right) $ is
continuous at $x=\pm L/2$; but 3) undergoes a jump of unit height at $%
x=x^{\prime }.$ Solution of this problem is totally equivalent to
finding a potential of a point charge located somewhere in a
sandwich-like system, consisting of three insulators with
different dielectric constants. Two of these layers are
semi-infinite and the third one (in the middle) is of finite
thickness. ``Potential'' $\Phi _{\omega _{m}}\left( x,x^{\prime
}\right) $ can be found in a general form for arbitrary
$x,x^{\prime }.$ In the expression for the current
\[
J\left( t,x\right) =\int dx^{\prime }\int \frac{d\omega }{2\pi }\sigma
\left( \omega ;x,x^{\prime }\right) E\left( \omega ,x^{\prime }\right) ,
\]
$x^{\prime }$ is within the wire; hence we need to know $\Phi
_{\omega _{m}}\left( x,x^{\prime }\right) $ only for $-L/2\leq
x^{\prime }\leq L/2.$ In a steady-state regime, one is free to
measure the current through any cross-section; let's choose $x$
also within the wire. As we are interested in the limit $\omega
\rightarrow 0,$ when the plasmon wavelength is larger
than the wire length, we can put $x=x^{\prime }.$ In the interval $%
-L/2<x=x^{\prime }<L/2$ the solution of the Laplace's equation is
\begin{equation}
\Phi _{\omega _{m}}\left( x,x\right) =\frac{K_{W}}{2\left| \omega
_{m}\right| }+\frac{K_{W}}{2\left| \omega _{m}\right| }\frac{\kappa
_{-}^{2}e^{-L/L_{\omega }}+\kappa _{+}\kappa _{-}\cosh \left( 2x/L_{\omega
}\right) }{e^{L/L_{\omega }}\kappa _{+}^{2}-e^{-L/L_{\omega }}\kappa _{-}^{2}%
},  \label{potential}
\end{equation}
where $L_{\omega }=u_{W}/\left| \omega _{m}\right| ,$ $u_{W}$ is the charge
velocity within the wire, and
\[
\kappa _{\pm }=K_{W}^{-1}-K_{L}^{-1}.
\]
Letting $\omega _{m}$ in (\ref{potential}) to zero (and thus $L_{\omega }$
to $\infty ),$ we find that
\[
\Phi _{\omega }\left( x,x\right) =\frac{K_{L}}{2i\omega }=\frac{1}{2i\omega }%
,
\]
as by our assumption $K_{L}=1.$ This result is true for any $x,x^{\prime }$
within the wire for $\omega \rightarrow 0$%
\[
\Phi _{\omega }\left( x,x^{\prime }\right) =\frac{1}{2i\omega },\text{ for }%
-L/2<x,x^{\prime }<L/2.
\]
The Luttinger-liquid parameters of the wire drop out from the answer. The
\emph{dc }conductivity reduces to its free value
\[
\sigma \left( \omega \rightarrow 0;x,x^{\prime }\right) =\frac{e^{2}}{2\pi },
\]
and, consequently, the conductance
\[
\mathcal{G}=\frac{e^{2}}{h}
\]
is not renormalized by the interaction. The same consideration for electrons
with spins gives
\beq
\mathcal{G}=\frac{2e^{2}}{h}\label{univ}
\eeq
\subsubsection{Experiment}
\label{sec:exp}
Most of the experiments on quantum wires indeed show that the conductance is
quantizated in units of $2e^2/h$ at relatively high temperatures.
\footnote{However, it has been observed recently
that the conductance of carbon nanotubes is  quantized in units of $e^2/h$
--as opposed to $4e^2/h$, predicted
by the non-interacting theory for this case \cite{marcus}.}
\footnote{A special case of a non-universal conductance quantization
is very long wires grown by cleaved-edge overgrowth technique \cite{yacoby} can
be attributed to a non-trivial coupling between the wire and 2D reservoirs
\cite{yacoby2}, characteristic for these systems.}
 At lower temperatures, the conductance decreases beyond
the universal value and also the quantization plateaux exhibit
some structure as a function of the gate voltage \cite{tarucha}.
This can be interpreted as the effect of residual disorder: as was
discussed in Secs. \ref{sec:ygm},\ref{sec:bosonization}
transmission decreases at lower energy scales. An effect of single
impurity in a quantum wire will be largely insensitive to the
presence of reservoirs: as long as the transmission coefficient
for an impurity is much smaller than one, the largest voltage
drops occurs near the impurity rather than at the contacts to the
wire. One can show that the scaling of the conductance with energy
is determined by the interaction parameter $K$ {\it inside} the
wire \cite{maslov_dirty}, in a contrast to the disorder-free case
when only $K$ outside the wire matters. Also, the mesoscopic
conductance fluctuations increase as the temperature goes down
(the theory predicts that this effect is enhanced by the
interaction \cite{renn}). As one is dealing here with a crossover
regime from scattering at a single impurity to that at many
impurities, a quantitative analysis of the temperature dependences
is difficult; another complication arises from the finite length
of the wire which cuts off scalings with temperature and voltage.
In addition, at higher temperatures the first quantization plateau
exhibits a well-defined step at about $0.7\times 2e^2/h$
\cite{thomas96,thomas98,kristensen00,kane,reilly01,marcus07}. This
``0.7'' feature is not likely to result from spurious impurity
scattering but rather reveals some interesting physics beyond what
has been discussed so far in this review. Although the ``0.7''
feature deserves a review on its own, I will come back to this
subject briefly in the next Section.
\subsection{Spin component of the conductance}
\label{sec:spin_incoh} As we have shown in the previous sections,
the Luttinger-liquid models predicts that conductance of a
disorder-free wire is given by $e^{2}/h$ per channel at any
temperature. Also, thanks to spin-charge separation, spin degrees
of freedom do not play an essential role in charge transport
except for giving an overall factor of two to the conductance.
These two results hold as long as the Luttinger-liquid model is a
good description for interacting electrons in the wire. When does
this model break down? If the interaction is strong, electrons
form almost a periodic 1D structure: quasi-Wigner crystal. The
exchange energy of almost localized electrons is exponentially
small and, correspondingly, the spin velocity is small too:
$u_{s}\ll u_{c}.$ The Luttinger-liquid model should work for
energies (temperatures) much smaller than the smallest of the two
(spin and charge) bandwidth $T\ll u_{s}k_{F}\ll u_{c}k_{F},$ when
both spin- and charge degrees of freedom are coherent.  The \emph{spin- incoherent regime}, {\emph i.e.},  $%
u_{s}k_{F}\ll T\ll u_{c}k_{F}$, has attracted considerable interest recently
\cite{matveev,zvonarev,balents}, and was shown to spoil the conductance
quantization in integer multiples of $2e^{2}/h$ \cite{matveev} at
temperatures larger than the spin bandwidth ($u_{s}k_{F}$). In what follows,
I present a short summary of Ref.\cite{matveev}.

In a quasi-Wigner-crystal regime,  a reasonable starting point for
describing the spin sector is the Heisenberg model
\begin{equation*}
H_{s}=J_{\text{ex}}\sum_{l}\vec{S}_{l}\cdot \vec{S}_{l+1},
\end{equation*}
where spins are localized at ``lattice sites'' corresponding to
positions of electrons. Because the Lieb-Mattis theorem
\cite{lieb} forbids ferromagnetic ordering in 1D, the sign of the
exchange interaction must be antiferromagnetic: $J_{\text{ex}}>0.$
Assuming that electrons are well
localized at distances $a=1/n$ from each other, $J_{%
\text{ex}}$ can be estimated in the WKB approximation: $J_{\text{ex}%
}\sim E_{F}\exp \left( -c/\sqrt{a_{B}n}\right) ,$ where $c\sim 1$
and $a_{B}$ is the Bohr radius. A spin-1/2 chain is then mapped
onto a Hubbard 1/2-filled model of spinless fermions via the
Jordan-Wigner transformation
\begin{eqnarray*}
S_{z}\left( l\right)  &=&a_{l}^{\dagger }a_{l}-1/2; \\
S_{x}\left( l\right) +iS_{y}\left( l\right)  &=&a_{l}^{\dagger }\exp \left(
i\pi \sum_{J_{\text{ex}}=1}^{l-1}a_{J_{\text{ex}}}^{\dagger }a_{J_{\text{ex}%
}}\right)
\end{eqnarray*}
with the result
\begin{equation*}
H_{s}=-\frac{J_{\text{ex}}}{2}\sum \left[ c_{l+1}^{\dagger }c_{l}+H.c.\right]
+J_{\text{ex}}\sum :c_{l}^{\dagger }c_{l}-1/2::c_{l+1}^{\dagger }c_{l}-1/2:.
\end{equation*}
The spinless Hubbard model can be bosonized
\begin{equation}
H_{s}=\frac{1}{2}\int dx\frac{u}{K}\left( \partial _{x}\phi \right)
^{2}+uK\left( \partial _{x}\theta \right) ^{2}+\frac{aJ_{\text{ex}}}{\left(
2\pi a\right) ^{2}}\cos \left( \sqrt{16\pi }\phi \right) .  \label{spin1/2}
\end{equation}
A comparison to the Bethe Ansatz solution of the Hubbard model at
half-filling enables one to identify the parameters of the
spinless LL with the microscopic parameters of the spin-1/2 chain.
In particular, for an isotropic spin chain ($J_x=J_y=J_z$) \beq u
=\frac{\pi }{2}J_{\text{ex}}a;\; K =1/2. \eeq
Comparing the spin-part of the Hamiltonian of the original LL model, Eq.~(\ref{spin}%
) with  that of the spinless LL, Eq.~(\ref{spin1/2}), one notices
that they the same upon the following mapping \beq\phi
=\frac{1}{\sqrt{2}}\phi _{s};\;\text{ }\theta =\sqrt{2}\theta
_{s};\; \frac{u_{s}}{K_{s}} =\frac{u}{2K};\;\text{
}u_{s}K_{s}=2uK, \eeq or \beq u_{s}=u=\frac{\pi
}{2}J_{\text{ex}}a;\; K_{s}=2K=1. \eeq As $J_{\rm ex}$ is
exponentially small, so is the spin bandwidth. Therefore, the
Luttinger-liquid description is valid only at very low
temperatures.

 A translationally invariant LL still possesses
spin-charge separation. However, this is no longer true for a wire
connected to non-interacting leads. To understand this point,
let's come back to the inhomogeneous LL model (cf. Sec.
\ref{sec:inhll}), where the electron density changes from a
higher value in the leads to a lower value within the wire. Because $J_{%
\text{ex}}$ depends on the local density, it is modulated along
the wire, and its minimum value is at the middle of the wire. In
the leads, we have a non-interacting system, where
$J_{\text{ex}}\sim E_{F}\gg T.$ However, in the middle of the wire
spins are incoherent, if $J_{\text{ex}}^{\min }\ll T.$ Thus a spin
part of the electron incoming from the lead at energy $T$ above
the Fermi level cannot propagate  freely through the wire because
the spin band narrows down: it works as if there is a barrier for
spin excitations in the wire. Although charge plasmons propagate
freely, backscattering of spin plasmons leads to additional
dissipation, and thus to additional resistance. The total
resistance of the wire consists now of two parts
\begin{equation*}
R=R_{c}+R_{s}.
\end{equation*}
The charge part, $R_{c},$ is due to propagation of charge plasmons.  Since
the charge part is still described by the LL model, our previous result for
universal conductance, Eq.~(\ref{univ}), still holds and $R_{c}=\mathcal{G}%
^{-1}=h/2e^{2}.$ For $T\ll J_{\text{ex}}^{\min},$ only athermal spin plasmons, with
energies exceeding the width of the spin band, contribute to $R_{s}.$ The
number of such plasmons is exponentially small, hence
\begin{equation*}
R_{s}\propto \exp \left( -J_{\text{ex}}^{\min}/T\right) ,
\end{equation*}
and total conductance $\mathcal{G=}\left( R_{c}+R_{s}\right)^{-1} $ is
exponentially close to $2e^{2}/h.$ At high temperatures $\left( T\gg J_{%
\text{ex}}^{\min}\right)$, almost all spin plasmons are reflected by the wire.
Then $R_{s}\sim R_{c}\sim h/e^{2}$, and the conductance differ substantially
from its universal value.  This qualitative picture can be confirmed in a
particular simple case of the $XY-$ model for the spin-chain. In this case, $%
R_{s}$ can be calculated explicitly  with the result
\cite{matveev}
\begin{equation*}
R_{s}=\frac{h}{2e^{2}}\frac{1}{\exp \left( J_{\text{ex}}^{\min}/T\right) +1}
\end{equation*}
and, consequently, the conductance is equal to
\begin{equation*}
\mathcal{G}=\frac{2e^{2}/h}{1+\left[\exp \left(
J_{\text{ex}}^{\min}/T\right) +1\right] ^{-1}}.
\end{equation*}
For $T\rightarrow 0,$ ${\cal G}$ approaches the universal value of $2e^{2}/h.$ For $%
T\gg J_{\text{ex}}^{\min}$,${\cal G}$ approaches another $T$
-independent limit, equal to $(2/3) 2e^{2}/h.$ The actual number
in the high-temperature limit of the conductance is
model-dependent (it is different, for example, for an isotropic
spin-chain), but the main result, \emph{i.e.}, the
non-universality of conductance quantization at higher
temperatures, survives.

As it was mentioned in Sec. \ref{sec:exp}, the experiment shows
that there is a shoulder in the conductance preceding the first
quantization plateau  at a fractional value of about 0.7$\times
2e^{2}/h.$ Surprisingly, this ``0.7 feature'' is more pronounced
at \emph{higher }temperatures, and the $T$- dependence of this
feature was reported to be of an activated type \cite
{kristensen00}. The magnetic field transforms the ``0.7 feature''
into a fully developed quantization plateau at $e^2/h$, which is
to be expected in a fully polarized, and thus spinless, regime.
The sensitivity to the magnetic field hints at the spin origin of
the effect, and a significant theoretical effort was invested in
understanding how spins can explain the observed phenomena.
Although the effect, described in this Section, does have all
qualitative characteristics of the observed ``0.7 feature'', it is
not clear at the moment whether this feature indeed corresponds to
the spin-incoherent regime. Other explanations of the ``0.7
feature'' have been suggested (most prominently, the Kondo physics
is believed to be involved \cite{meir,marcus07}), but a further
discussion of this point goes beyond the scope of these notes.

\subsection{Thermal conductance: Fabry-Perrot resonances of plasmons}
\label{sec:thermal} There is an important difference between
charge and thermal (electronic) conductances \cite{fazio}. As we
have just shown, the charge conductance is equal to $e^{2}/h$
regardless of interaction in the wire. This means that the
transmission coefficient of electrons is equal to unity. The
effect of the temperature on the charge conductance is the same as
for a non-interacting, perfectly transmitting wire:  at finite
temperature, not only the lowest but also higher subbands of
transverse quantization  are populated, and the quantization
plateaux are smeared. However, this effect is exponentially small
for temperatures smaller than either  the Fermi energy, $E_{F},$
or the difference between the Fermi energy and the threshold of
the next subband of transverse quantization, $\Delta$; whichever
is smaller.

Thermal current is carried not by electrons but bosonic
excitations: acoustic plasmons. In contrast to electrons, plasmons
get reflected at the boundary between the reservoirs and the wire
due to the mismatch of charge velocities (this reflection happens
even for an adiabatically smooth transition). From the plasmon's
point-of-view, a wire coupled to reservoirs represents a
Fabry-Perrot interferometer. Interference of plasmon waves
scattered off the opposite ends of the wire results in an
oscillatory dependence of the transmission coefficient on the
frequency with a period given by the travel time of plasmons
through the wire
\[
2\pi \omega _{L}=L/u_{W}.
\]
As long as $\lambda _{F}\ll L,$ this period is long: $\omega _{L}\ll E_{F}$ $%
.$ The difference between charge and heat transport is that the chemical
potential of plasmons is equal to zero, and thus the characteristic scale
for frequency is set by $T$. Therefore, the thermal conductance varies with
the temperature on a scale $T\simeq \omega _{L}.$

Suppose that a small temperature difference $\delta T$ is
maintained between the reservoirs, connected by a quantum wire. As
the Hamiltonian of an interacting system is diagonalized of terms
of plasmons, plasmons modes are decoupled and contribute to the
energy flux independently. Then the thermally averaged energy
current, \emph{i.e.}, the thermal current can be found via a
Landauer-like argument
\[
J_{T}=\int_{0}^{\infty }\frac{d\omega }{2\pi }\omega \left| t\left( \omega
\right) \right| ^{2}\left( n_{L}\left( \frac{\omega }{T+\delta T}\right)
-n_{R}\left( \frac{\omega }{T}\right) \right) ,
\]
where $n_{L,R}\left( \omega /T\right) $ are the Bose distribution functions
in the reservoirs. Expanding in small $\delta T,$ we obtain the thermal
conductance
\begin{equation}
\mathcal{G}_{T}=\frac{J_{T}}{\delta T}=\frac{1}{8\pi T^{2}}\int_{0}^{\infty
}d\omega \frac{\omega ^{2}}{\sinh ^{2}\omega /2T}\left| t\left( \omega
\right) \right| ^{2}.  \label{cond_heat}
\end{equation}
For a free system, $\left| t\left( \omega \right) \right| ^{2}=1$ and $%
\mathcal{G}_{T}=\pi T/6.$ The charge and thermal conductances of a free
system obey the Wiedemann-Franz law
\begin{equation}
\frac{\mathcal{G}_{T}}{T\mathcal{G}}=L_{0}=\frac{\pi ^{2}}{3e^{2}},
\label{WF}
\end{equation}
where $L_{0}$ is the Lorentz number. This means that charge and energy are
carried by the same excitations. This is not so in a Luttinger liquid.

For an interacting system, relation (\ref{WF}) holds in the limit of $%
T\rightarrow 0.$ The characteristic scale for frequencies in integral (\ref
{cond_heat}) is determined by $T.$ For $T\ll \omega _{L}$, one can
substitute $\omega =0$ into $\left| t\left( \omega \right) \right| ^{2}.$
Regardless of the interaction strength, $\left| t\left( 0\right) \right|
^{2}=1:$ a  Fabry-Perrot interferometer becomes transparent in the long
wavelength limit. For $T\gtrsim \omega _{L},$ the result for $\mathcal{G}_{T}
$ depends on how the charge velocity varies along the wire, and is thus
non-universal. On the other hand, the charge conductance is universal.
Therefore, their  ratio is non-universal and the Wiedemann-Franz law is
violated.

In a step-like model of Sec.~\ref{sec:inhll}, the transmission
coefficient of plasmons is equal to
\[
\left| t\left( \omega \right) \right| ^{2}=\frac{1}{1+\frac{\left(
K^{2}-1\right) ^{2}}{4K^{2}}\sin ^{2}\frac{\omega }{\omega _{L}}}.
\]
Obviously, $\left| t\left( 0\right) \right| ^{2}=1$ regardless of
$K,$ an agreement with what was said above.   For $T\ll \omega
_{L},$  the Lorentz number is close to the universal value of $\pi
^{2}/3e^{2}.$  For $T\gg \omega _{L}$, the oscillations of $\left|
t\left( \omega \right) \right| ^{2} $ become very fast, so that
$\left| t\left( \omega \right) \right| ^{2}$ can be replaced by
its averaged value
\[
\langle \left| t\left( \omega \right) \right| ^{2}\rangle =\int_{0}^{\omega
_{L}}\frac{d\omega }{\omega _{L}}\left| t\left( \omega \right) \right| ^{2}=%
\frac{2K}{K^{2}+1}.
\]
The thermal conductance increases linearly with $T,$ so that
$L_{0}$ approaches a constant but a non-universal value
\begin{equation} L|_{T\gg \omega _{L}}=\frac{2K}{K^{2}+1}L_{0}<L_{0}.
\label{WFLL}
\end{equation} As the Lorentz number varies with temperature in
between two limits specified by Eqs.(\ref{WF}) and (\ref{WFLL}),
the Wiedemann-Franz law is violated.

\section{Acknowledgments}
Late R. Landauer and H. J. Schulz helped me--either directly or
indirectly, through their writings--to form my way of thinking
about interactions and transport, and my gratitude goes to their
memories. I benefited tremendously from discussions with B. L.
Altshuler, C. Biagini, C. P{\`e}pin, A. V. Chubukov, V. V.
Cheianov, E. Fradkin, F. Essler, S. Gangadharaiah, Y. Gefen, L. I.
Glazman, P. M. Goldbart, I. V. Gornyi, Y. B. Levinson, D. Loss, K.
A. Matveev, B. N. Narozhny, A. A. Nersesyan, D. G. Polyakov, C.
P{\'e}pin, M. Yu. Reizer, I. Safi, R. Saha, B. I. Shklovskii, M.
Stone, O. A. Starykh, S. Tarucha, S.-W. Tsai, A. Yacoby, O. M.
Yevtushenko, and M. B. Zvonarev on various subjects discussed in
these notes. I am also grateful to the organizers of the 2004
Summer School in Les Houches--H. Bouchiat, Y. Gefen, S.
Gu{\'e}ron, and G. Montambaux--for assembling a very interesting
program and providing a stimulating environment, as well as to the
all participants of the School for their questions, attention, and
patience. S. Gangadharaiah, L. Merrill, and R. Saha kindly helped
me with producing the images.  S. Gangadharaiah, P. Kumar, C.
P{\'e}pin, R. Saha, and S.-W. Tsai proofread parts of the
manuscript (which does not make them responsible for the typos). I
acknowledge the hospitality of the Abdus Salam International
Centre for Theoretical Physics (Trieste, Italy), where a part of
this manuscript was written. Financial support from NSF
DMR-0308377 is greatly appreciated.
%%%%%%%%%%%%%%%%%%%%%%%%%%%%%%%%%%%%%%%%%%%%%%%%%%%%%%%%%%%%%%%%%%%%%%%%%%%
%%%%%%%%%%%%%%%END OF MAIN TEXT
%%%%%%%%%%%%%%%%%%%%%%%%%%%%%%%%%%%%%%%%%%%%%%%%%%%%%%%%%%%%%%%%%%%%%%%%%%%%%
\appendix
\section{Polarization bubble for small $q$ in arbitrary dimensionality}
\label{sec:pi_anyD} The polarization bubble in Matsubara frequencies and at $%
T=0$ is given by
\begin{eqnarray*}
\Pi \left( i\omega ,q\right)  &=&\frac{N_{s}}{\left( 2\pi \right) ^{D+1}}%
\int \int d^{D}pd\varepsilon G\left( i\varepsilon +i\omega ,\vec{p}+\vec{q}%
\right) G\left( i\varepsilon ,p\right)  \\
&=&\frac{N_{s}}{\left( 2\pi \right) ^{D+1}}\int \int
d^{D}pd\varepsilon \frac{1}{i\omega -\xi _{\vec{p}+\vec{q}}+\xi
_{\vec{p}}}\left[ G\left( i\varepsilon +i\omega
,\vec{p}+\vec{q}\right) -G\left( i\varepsilon
,p\right) \right]  \\
&=&\frac{N_{s}}{\left( 2\pi \right) ^{D+1}}\int d^{D}p\frac{f\left( |\vec{p}+%
\vec{q}|\right) -f\left( p\right) }{i\omega -\xi _{\vec{p}+\vec{q}}+\xi _{%
\vec{p}}},
\end{eqnarray*}
where $f$ is the Fermi function. Expanding in $\vec{q},$ and
switching from the integration over $d^{D}p$ to $d\xi ,$ we obtain
\[
\Pi \left( i\omega ,q\right) =-N_{s}\nu _{D}\left( 1-\int \frac{d\Omega }{%
\Omega _{D}}\frac{i\omega }{i\omega -v_{F}q\cos \theta }\right) .
\]
where  $\Omega _{D}=4\pi $ (3D), $=2\pi \left( 2D\right) ,$ $=2$ (1D) and $%
\nu _{D}$ is the DoS in $D$ dimensions per one of the $N_{s}$
isospin components.  For D=1, the integral over $\Omega $ is
understood as a sum of terms with $\cos \theta =\pm 1.$ It is
obvious already this form that  the small $q-$form of the bubble
depends on the combination $\omega /v_{F}q$ for any $D.$  The
final result depends on the dimensionality. Performing analytic
continuation to real frequencies $i\omega \rightarrow \omega
+i0^{+},$ we obtain
\[
\Pi ^{R}\left( \omega ,q\right) =-N_{s}\nu _{D}\left( 1-\int \frac{d\Omega }{%
\Omega _{D}}\frac{\omega }{\omega -v_{F}q\cos \theta +i0}\right) .
\]
Taking the imaginary part
\begin{equation}
\mathrm{Im}\Pi ^{R}\left( \omega ,q\right) =-\pi N_{s}\nu
_{D}\omega \int \frac{d\Omega }{\Omega _{D}}\delta \left( \omega
-v_{F}q\cos \theta \right) . \label{impi_anyD}
\end{equation}
From here
\[
\cos \theta =\omega /v_{F}q,
\]
which means that $\theta \approx \pi /2$ for $\omega \ll v_{F}q.$
Thus, the
fermionic momentum $\vec{p}$ is almost perpendicular to the bosonic one, $%
\vec{q},$ in this limit.

\section{Polarization bubble in 1D}
\label{sec:free_bubble}
\subsection{Small $q$}
Free time-ordered (causal) Green's function in 1D is equal to
\[
G_{\pm }^{0}\left( \varepsilon ,k\right) =\frac{1}{\varepsilon
-\xi _{k}^{\pm }+i0^{+}\mathrm{sgn}\xi _{k}^{\pm }},
\]
where
\[
\xi _{k}^{\pm }=\pm v_{F}\left( k\mp k_{F}\right) ,
\]
and $\pm $ signs correspond to right/left moving fermions. We will
be
measuring the momenta from the corresponding Fermi points. For +branch :$%
k-k_{F}\rightarrow k$ and for -branch: $k+k_{F}\rightarrow k.$
Consequently,
\[
G_{\pm }^{0}\left( \varepsilon ,k\right) =\frac{1}{\varepsilon \mp
v_{F}k+i0^{+}\mathrm{sgn}k}.
\]
I assume the $N_{s}-$ fold degeneracy ($N_{s}=2$ for electrons with spin, $%
N_{s}=1$ for spinless electrons), so that
\[
\Pi _{\pm }\left( \omega ,q\right) =-\frac{i}{\left( 2\pi \right) ^{2}}%
N_{s}\int d\varepsilon \int dkG_{\pm }^{0}\left( \varepsilon
+\omega ,k+q\right) G_{\pm }^{0}\left( \varepsilon ,k\right) .
\]
Calculate, \emph{e.g.}, $\Pi _{+}:$%
\begin{eqnarray}
\Pi _{+}\left( \omega ,q\right) &=&-\frac{i}{\left( 2\pi \right) ^{2}}%
N_{s}\int d\varepsilon \int dk\frac{1}{\varepsilon +\omega -v_{F}(k+q)+i0^{+}%
\mathrm{sgn(}k+q)}\frac{1}{\varepsilon
-v_{F}k+i0^{+}\mathrm{sgn}k}
\nonumber \\
&=&-\frac{i}{\left( 2\pi \right) ^{2}}N_{s}\int d\varepsilon \int dk\frac{1}{%
\omega -v_{F}q+i0^{+}\mathrm{sgn(}k+q)-i0^{+}\mathrm{sgn}k}
\label{bubble_free} \\
&&\times \left[ G_{+}^{0}\left( \varepsilon ,k\right)
-G_{+}^{0}\left( \varepsilon +\omega ,k+q\right) \right] .
\end{eqnarray}
The integral of the Green's over the frequency gives a Fermi
distribution function \cite{agd}
\[
n_{+}\left( k\right) =-i\int \frac{d\varepsilon }{2\pi
}G_{+}^{0}\left( \varepsilon ,k\right) .
\]
For free fermions,
\[
n_{+}\left( k\right) =\theta \left( -k\right)
\]
Now,
\[
\Pi _{+}^{0}\left( \omega ,q\right) =\frac{N_{s}}{2\pi }\int dk\frac{1}{%
\omega -v_{F}q+i0^{+}\mathrm{sgn(}k+q)-i0^{+}\mathrm{sgn}k}\left[
\theta \left( -k\right) -\theta \left( -k-q\right) \right] .
\]
The integral is not equal to zero only if the arguments of the
$\theta $ -functions are of the opposite signs. Consider different
situations.

1) $k>0$; $k+q<0\rightarrow 0<k<-q\rightarrow q<0.$ In this case,
\[
\Pi _{+}^{0}\left( \omega ,q\right) =\frac{N_{s}}{2\pi
}\frac{q}{\omega -v_{F}q-i0^{+}},\mathrm{\ }q<0;
\]

2) $k<0,$ $k+q>0\rightarrow -q<k<0\rightarrow q>0$%
\[
\Pi _{+}^{0}\left( \omega ,q\right) =\frac{N_{s}}{2\pi
}\frac{q}{\omega -v_{F}q+i0^{+}},q>0.
\]
Combining the results for $q>0$ and $q<0$ together,
\begin{equation}
\Pi _{+}^{0}\left( \omega ,q\right) =\frac{N_{s}}{2\pi
}\frac{q}{\omega -v_{F}q+i0^{+}\mathrm{sgn}q}.  \label{pito_+}
\end{equation}
Similarly,
\begin{equation}
\Pi _{-}^{0}\left( \omega ,q\right) =-\frac{N_{s}}{2\pi
}\frac{q}{\omega -v_{F}q+i0^{+}\mathrm{sgn}\omega }.
\label{pito_-}
\end{equation}
The total bubble
\begin{equation}
\Pi ^{0}\left( \omega ,q\right) =\Pi _{+}^{0}\left( \omega
,q\right) +\Pi _{-}^{0}\left( \omega ,q\right) =\frac{N_{s}}{\pi
}\frac{v_{F}q^{2}}{\omega ^{2}-v_{F}^{2}q^{2}+i0^{+}}.
\label{pito_full}
\end{equation}

In what follows, we will also need the retarded and advanced form
of the bubble. These forms can easily be obtained by repeating the
calculation above in Matsubara frequencies and analytically
continuing $i\omega _{m}\rightarrow \omega +i0$ . Even simpler,
one can use the general relation between time-ordered and retarded
propagators \cite{agd} (which works equally well for fermionic and
bosonic quantities)
\begin{eqnarray*}
\Pi _{\pm }^{R}\left( \omega ,q\right) &=&\Pi _{\pm }\left( \omega
,q\right)
,\text{ }\mathrm{for}\text{ }\mathrm{\omega >0} \\
&=&\Pi _{\pm }^{\ast }\left( \omega ,q\right) ,\text{ }\mathrm{for}\text{ }%
\mathrm{\omega <0.}
\end{eqnarray*}
Using Eqs.~(\ref{pito_+}) and (\ref{pito_-}) we obtain
\begin{equation}
\Pi _{\pm }^{R}\left( \omega ,q\right) =\pm \frac{N_{s}}{2\pi }\frac{q}{%
\omega -v_{F}q+i0^{+}}  \label{pi_r_pm}
\end{equation}
and
\begin{eqnarray}
\Pi^{R}\left( \omega ,q\right) &=&\frac{N_{s}}{\pi
}\frac{v_{F}q^{2}}{\omega
^{2}-v_{F}^{2}q^{2}+i0^{+}sgn\omega }  \nonumber \\
&=&\frac{N_{s}}{\pi }\frac{v_{F}q^{2}}{\left( \omega
+i0^{+}\right) ^{2}-v_{F}^{2}q^{2}}.  \label{pi_r_full}
\end{eqnarray}
\subsection{$q$ near $2k_{F}$}
We will also need the $2k_{F}$ bubble. This time, I choose to do
the calculation in Matsubara frequencies:
\begin{eqnarray*}
\Pi _{2k_{F}}\left( i\omega ,q\right) &=&\frac{N_{s}}{(2\pi
)^{2}}\int dk\int d\varepsilon G_{+}\left( i\varepsilon +i\omega
,k+q\right)
G_{-}\left( i\varepsilon ,k\right) \\
&=&-\frac{N_{s}}{(2\pi )^{2}}\int dk\int d\varepsilon
\frac{1}{\varepsilon +\omega +i\left( k+q\right)
}\frac{1}{\varepsilon -ik}.
\end{eqnarray*}
Poles in $\varepsilon _{1}=ik$ and $\varepsilon _{2}=-i\left(
k+q\right) -\omega $ have to be on different sides of the real
axis, otherwise the integral is equal to zero. Choose $q>0.$ Then
this condition is satisfied in two intervals of $k:$ $k>0$ and
$-\Lambda /2<k<-q,$ where $\Lambda $ is the ultraviolet cut-off

\begin{eqnarray}
\Pi _{2k_{F}} &=&-\frac{iN_{s}}{2\pi }\left[ \int_{0}^{\Lambda
/2}dk-\int_{-\Lambda /2}^{-q}dk\right] \frac{1}{\omega
+2iv_{F}k+iv_{F}q}
\nonumber \\
&=&-\frac{N_{s}}{4\pi }\left[ \ln \frac{i\Lambda v_{F}}{\omega
+iv_{F}q}-\ln \frac{\omega -iv_{F}q}{-i\Lambda v_{F}}\right]\notag\\
&=&-\frac{N_{s}}{4\pi }\ln \frac{\Lambda ^{2}v_{F}^{2}}{\omega
^{2}+v_{F}q^{2}}.  \label{bubble_2kf}
\end{eqnarray}
Because the result depends on $q^{2}$ there is no need for a
separate calculation for the case $q<0.$

\section{Some details of bosonization procedure}
\label{sec:bos_details}
\subsection{Anomalous commutators}

\begin{eqnarray*}
\rho \left( q\right) &=&\sum_{p}a_{p-q/2}^{\dagger }a_{p+q/2}=\rho
_{+}+\rho
_{-}; \\
\rho _{\pm } &=&\sum_{p>0(p<0)}a_{p-q/2}^{\dagger }a_{p+q/2}.
\end{eqnarray*}
The operators of full density commute. The operators of left-right
densities have non-trivial commutators. For example, let us
calculate $\left[ \rho _{+}\left( q\right) ,\rho _{+}\left(
q^{\prime }\right) \right] $

\begin{eqnarray*}
C_{++}\left( q,q^{\prime }\right)  &=&\left[ \rho _{+}\left(
q\right) ,\rho _{+}\left( q^{\prime }\right) \right]
=\sum_{p>0,k>0}\left[ a_{p-q/2}^{\dagger
}a_{p+q/2}^{{}},a_{k-q^{\prime }/2}^{\dagger
}a_{k+q^{\prime }/2}^{{}}\right]  \\
&=&\sum_{p>0,k>0}\left(
\begin{array}{c}
a_{p-q/2}^{\dagger }\underbrace{a_{p+q/2}a_{k-q^{\prime }/2}^{\dagger }}%
_{=\delta _{p+q/2,k-q^{\prime }/2}-a_{k-q^{\prime }/2}^{\dagger
}a_{p+q/2}^{{}}}a_{k+q^{\prime }/2} \\
-a_{k-q^{\prime }/2}^{\dagger }\underbrace{a_{k+q^{\prime
}/2}a_{p-q/2}^{\dagger }}_{=\delta _{k+q^{\prime
}/2,p-q/2}-a_{p-q/2}^{\dagger }a_{k+q^{\prime }/2}^{{}}}a_{p+q/2}
\end{array}
\right) .
\end{eqnarray*}
The first $\delta -$ function means that $k=p+q/2+q^{\prime }/2>0$
and the second one that $k=p-q/2-q^{\prime }/2.$
\begin{eqnarray*}
C_{++}\left( q,q^{\prime }\right)  &=&\sum_{p>0}a_{p-q/2}^{\dagger
}a_{p+q/2+q^{\prime }}\vartheta \left( p+q/2+q^{\prime }/2\right)
-a_{p-q/2-q^{\prime }}^{\dagger }a_{p+q/2}\theta \left(
p-q/2-q^{\prime
}/2\right)  \\
&&-\left[ f\left( q,q^{\prime }\right) -f\left( q^{\prime
},q\right) \right] ,
\end{eqnarray*}
where
\begin{eqnarray*}
f(q,q^{\prime }) &=&\sum_{p,k>0}a_{p-q/2}^{\dagger }a_{k-q^{\prime
}/2}^{\dagger }a_{p+q/2}^{{}}a_{k+q^{\prime }/2} \\
&=&\sum_{p,k>0}a_{p}^{\dagger }a_{k}^{\dagger
}a_{p+q}^{{}}a_{k+q^{\prime }}
\end{eqnarray*}
It is easy to show that $f\left( q,q^{\prime }\right) =f\left(
q^{\prime },q\right) .$ Indeed,
\begin{eqnarray*}
f(q^{\prime },q) &=&\sum_{p,k>0}a_{p}^{\dagger }a_{k}^{\dagger
}a_{p-q^{\prime }}^{{}}a_{k+q}^{{}} \\
&=&\underset{}{\mathrm{re-labelling}\text{ }\mathrm{k\longleftrightarrow p=}}%
\sum_{p,k>0}a_{k}^{\dagger }a_{p}^{\dagger }a_{k-q^{\prime }}^{{}}a_{p+q} \\
&=&\underset{}{\mathrm{{anticommuting}=}}\sum_{p,k>0}a_{p}^{\dagger
}a_{k}^{\dagger }a_{p+q}^{{}}a_{k+q^{\prime }}^{{}}=f\left(
q,q^{\prime }\right) .
\end{eqnarray*}
Thus
\begin{equation*}
C_{++}\left( q,q^{\prime }\right) =\sum_{p>0}a_{p-q/2}^{\dagger
}a_{p+q/2+q^{\prime }}\theta \left( p+q/2+q^{\prime }/2\right)
-a_{p-q/2-q^{\prime }}^{\dagger }a_{p+q/2}\theta \left(
p-q/2-q^{\prime }/2\right) .
\end{equation*}
Introduce a new momentum
\begin{equation*}
Q=\frac{q+q^{\prime }}{2}.
\end{equation*}
In the first sum, shift $p+q^{\prime }/2\rightarrow p$ and in the
second sum shift $p-q^{\prime }/2\rightarrow p.$ Then
\begin{eqnarray*}
C_{++}\left( q,2Q-q\right)  &=&\sum_{p>0}a_{p-Q}^{\dagger
}a_{p+Q}^{{}}\left[
\theta \left( p+q/2\right) -\theta \left( p-q/2\right) \right]  \\
&=&\sum_{p>-q/2}a_{p-Q}^{\dagger
}a_{p+Q}^{{}}-\sum_{p>q/2}a_{p-Q}^{\dagger }a_{p+Q}^{{}}
\end{eqnarray*}
If the main contribution to the sum is given by the states which
lie either deep below or far above the Fermi levels, then the
quantum fluctuations in the occupancy of these states are small,
and the operators $a_{p -Q}^{\dagger }a_{p+Q}$ can be replaced by
their expectation values $\langle a_{\pi
-Q}^{\dagger}a_{p+Q}\rangle =\delta _{Q,0}n_{p}=$ $\delta
_{Q,0}\theta \left(p_{F}-p\right)$. Doing this, we find
\begin{eqnarray*}
C_{++}\left( q,2Q-q\right) &=&\delta _{Q,0}\left(
\sum_{p>-q/2}^{p_{F}}-\sum_{p>q/2}^{p_{F}}\right) =\delta _{Q,0}\frac{L}{%
2\pi }\left( \int_{-q/2}^{p_{F}}dp-\int_{q/2}^{p_{F}}dp\right)\\ &=&\delta _{Q,0}%
\frac{qL}{2\pi }.
\end{eqnarray*}
Therefore,
\begin{equation}
\lbrack \rho _{+}\left( q),\rho _{+}(-q\right) ]=\frac{qL}{2\pi },\mathrm{%
spinless.}  \label{comm_spinless}
\end{equation}
The same procedure for fermions with spin gives
\begin{equation*}
\lbrack \rho _{+,\sigma }\left( q),\rho _{+,\sigma ^{\prime
}}(-q\right)
]=\delta _{\sigma \sigma ^{\prime }}\frac{qL}{2\pi },\mathrm{with}\text{ }%
\mathrm{spin.}
\end{equation*}
Similarly,
\begin{equation*}
\lbrack \rho _{-,\sigma }\left( q),\rho _{-,\sigma ^{\prime
}}(-q\right)
]=-\delta _{\sigma \sigma ^{\prime }}\frac{qL}{2\pi },\mathrm{with}\text{ }%
\mathrm{spin.}
\end{equation*}
and
\begin{equation*}
\lbrack \rho _{+,\sigma }\left( q),\rho _{-,\sigma }(-q\right)
]=0.
\end{equation*}
Combining these results together
\begin{equation*}
\left[ \rho _{\alpha ,\sigma }\left( q\right) ,\rho _{\alpha
^{\prime },\sigma ^{\prime }}\left( -q\right) \right] =\alpha
\delta _{\alpha ,\alpha ^{\prime }}\delta _{\sigma ,\sigma
^{\prime }}\frac{qL}{2\pi },
\end{equation*}
where $\alpha =\pm $ is the chirality index. For full charge
density and current, it means that
\begin{eqnarray*}
\left[ \rho ^{c}\left( q\right) ,\rho ^{c}\left( -q\right) \right]
&=&\left[ \rho _{+}^{c}\left( q\right) +\rho _{-}^{c}\left(
q\right) ,\rho
_{+}^{c}\left( -q\right) +\rho _{-}^{c}\left( -q\right) \right]  \\
&=&\frac{qV}{2\pi }+\frac{qV}{2\pi }-\frac{qV}{2\pi
}-\frac{qV}{2\pi }=0.
\end{eqnarray*}
Similarly,
\begin{equation*}
\left[ j^{c}\left( q\right) ,j^{c}\left( -q\right) \right] =0,
\end{equation*}
whereas\
\begin{equation*}
\left[ \rho ^{c}\left( q\right) ,j^{c}\left( -q\right) \right] =\frac{qV}{%
2\pi }{\small +}\frac{qV}{2\pi }{\small +}\frac{qV}{2\pi }{\small +}\frac{qV%
}{2\pi }{\small =}\frac{2}{\pi }qL.
\end{equation*}
In 4-notations,
\begin{equation*}
\lbrack j^{\mu }\left( q\right) ,j^{\nu }\left( -q\right)
]=\epsilon ^{\mu \nu }\frac{2}{\pi }qL,
\end{equation*}
where $\epsilon ^{00}=\epsilon ^{11}=0,$\ $\epsilon
^{01}=-\epsilon ^{10}=1.$

\subsection{Bosonic operators}
\label{sec:boson} Let's check that the representation of density
operators via standard bosonic operators does reproduce
commutation relation for density. Expand the density operators
over the normal modes
\begin{eqnarray*}
\rho _{+}\left( x\right)  &=&\frac{1}{L}\sum_{q>0}A_{q}\left(
b_{q}e^{iqx}+b_{q}^{\dagger }e^{-iqx}\right);  \\
\rho _{-}\left( x\right)  &=&\frac{1}{L}\sum_{q<0}A_{q}\left(
b_{q}e^{iqx}+b_{q}^{\dagger }e^{-iqx}\right),
\end{eqnarray*}
where, without a loss of generality, $A_{q}$ can be chosen real
and even
function of $q.$ Fourier transforming $\rho _{+}\left( x\right) $%
\begin{eqnarray}
\rho _{+}\left( q\right)  &=&\int_{-\infty }^{\infty }dxe^{-iqx}\frac{1}{L}%
\sum_{q^{\prime }>0}A_{q^{\prime }}\left( b_{q^{\prime
}}e^{iq^{\prime
}x}+b_{q^{\prime }}^{\dagger }e^{-iq^{\prime }x}\right) ,  \label{e1} \\
&=&A_{q}\left( \theta \left( q\right) b_{q}+\theta \left(
-q\right) b_{-q}^{\dagger }\right) .  \notag
\end{eqnarray}
Choose $q>0$ and substitute (\ref{e1}) into the commutation
relation
\begin{eqnarray*}
\left[ \rho _{+}\left( q\right) ,\rho _{+}\left( -q\right) \right]
&=&A_{q}^{2}\left[ b_{q},b_{q}^{\dagger }\right] =A_{q}^{2}=\frac{qL}{2\pi }%
\rightarrow  \\
A_{q} &=&\sqrt{\frac{qL}{2\pi }}.
\end{eqnarray*}

\subsubsection{Commutation relations for bosonic fields $\protect\varphi $
and $\protect\vartheta $} \label{sec:phi_theta} Using
\begin{eqnarray*}
\varphi \left( x\right) &=&-i\sum_{-\infty <q<\infty
}\frac{1}{\sqrt{2\left|
q\right| L}}{\rm sgn}q\left( b_{q}e^{iqx}-b_{q}^{\dagger }e^{-iqx}\right) ; \\
\vartheta \left( x\right) &=&i\sum_{-\infty <q<\infty }\frac{1}{\sqrt{%
2\left| q\right| L}}\left( e^{iqx}b_{q}-b_{q}^{\dagger
}e^{-iqx}\right);\\
\vartheta ^{\prime }\left( x\right) &=&-\sum_{-\infty <q<\infty }\frac{1}{%
\sqrt{2\left| q\right| L}}q\left( e^{iqx}b_{q}+b_{q}^{\dagger
}e^{-iqx}\right),
\end{eqnarray*}
we find
\begin{eqnarray*}
\left[ \varphi \left( x\right) ,\vartheta ^{\prime }\left(
x\right) \right]&=&i\sum_{q,q^{\prime }}\frac{1}{\sqrt{2\left| q\right| L}}\frac{1}{\sqrt{%
2\left| q\right| ^{\prime }L}}\left| q^{\prime }\right|\\
&&\times\underbrace{\left[ b_{q}e^{iqx}-b_{q}^{\dagger
}e^{-iqx},b_{q^{\prime }}e^{iq^{\prime
}x^{\prime }}-b_{q^{\prime }}^{\dagger }e^{iq^{\prime }x^{\prime }}\right] }%
_{=2\delta _{q,-q^{\prime }}} \\
&=&i\frac{1}{L}\sum_{q}e^{iq\left( x-x^{\prime }\right) }=i\delta
\left( x-x^{\prime }\right) .
\end{eqnarray*}

\subsection{Problem with backscattering}
\label{sec:app_back}
 As it was pointed out in the main text, straightforward bosonization
of the Hamiltonian for the spinless case encounters a problem if
one tries to account for backscattering. As backscattering
($g_{1}$) is just an exchange process to forward scattering of
fermions of opposite chiralities ($g_{2}$), the Luttinger liquid
parameters with $g_{1}\neq 0$ should be obtained from those with
$g_{1}=0$ by a simple replacement: $g_{2}\rightarrow g_{2}-g_{1}.$
However, if we do this, we cannot satisfy the Pauli principle
which says that for a contact interaction, when
$g_{2}=g_{4}=g_{1}$, all the interaction effects should disappear.
Indeed, Eqs.\ref {u_spinless}) and (\ref{K_spinless})  for $u$ and
$K$, correspondingly, change to
\begin{eqnarray*}
u^{2} &=&\left( 1+\frac{g_{4}}{2\pi }\right) ^{2}-\left( \frac{g_{2}-g_{1}}{%
2\pi }\right) ^{2}; \\
K^{2} &=&\frac{1+\frac{g_{4}-g_{2}+g_{1}}{2\pi }}{1+\frac{g_{4}+g_{2}-g_{1}}{%
2\pi }}.
\end{eqnarray*}
For contact interaction, when $g_{2}=g_{4}=g_{1},$ we get
\begin{eqnarray*}
u^{2} &=&\left( 1+\frac{g}{2\pi }\right) ^{2}\neq 1 \\
K &=&1.
\end{eqnarray*}
The charge velocity is different from 1. In addition, the product
$uK$ is renormalized from unity--this is also a problem, as it
means that the current operator is renormalized by the
interactions. How to fix this problem? Ref. \cite{starykh} shows
how to arrive at the expressions for $u$ and $K$ which satisfy all
necessary constraints just on the basis on Galilean invariance and
dimensional analysis. Ref. \cite{giamarchi} arrives at the same
result by using a careful point-splitting of the operators. Here,
I present the method of Ref. \cite{giamarchi}.

Recall that the density operator, represented in terms of bosonic
fields, contains not only the lowest harmonic ($q\rightarrow 0$),
corresponding to
long-wavelength excitations, but also harmonics oscillating at $q=2k_{F},$ $%
4k_{F},$ etc. Indeed, taking into account only the $2k_{F}-$
oscillations, we have
\begin{eqnarray*}
\rho \left( x\right) &=&\left( \psi _{+}^{\dagger }\left( x\right)
e^{-ik_{F}x}+\psi _{-}^{\dagger }\left( x\right)
e^{ik_{F}x}\right) \left( \psi _{+}\left( x\right)
e^{ik_{F}x}+\psi _{-}\left( x\right)
e^{-ik_{F}x}\right) \\
&=&\psi _{+}^{\dagger }\left( x\right) \psi _{+}\left( x\right)
+\psi _{-}^{\dagger }\left( x\right) \psi _{-}\left( x\right)
+e^{-2ik_{F}x}\psi _{+}^{\dagger }\left( x\right) \psi _{-}\left(
x\right) +H.c.
\end{eqnarray*}
The first term in this equation has to be treated using the
point-splitting procedure, because it involves two fermionic
operators at the same point. The result is an infinite constant,
$\rho _{0},$ which is just a uniform density, plus the gradient
term. The $2k_{F}$ -component can be bosonized without a problem,
as it involves products of different fermions. The result is
\begin{equation*}
\rho \left( x\right) -\rho _{0}=\frac{1}{\sqrt{\pi }}\partial _{x}\varphi +%
\frac{1}{2\pi \alpha }\exp \left[ 2\sqrt{\pi }\varphi
+2k_{F}x\right] +H.c.
\end{equation*}
Using this expression for the interaction part of $H,$ we have
\begin{eqnarray*}
H_{\text{int}} &=&\frac{1}{2}\int dx\int dx^{\prime }V\left(
x-x^{\prime }\right) \left[ \rho \left( x\right) -\rho _{0}\right]
\left[ \rho \left(
x^{\prime }\right) -\rho _{0}\right] \\
&=&H_{F}+H_{B},
\end{eqnarray*}
where the forward and backscattering parts of the Hamiltonian are
given by
\begin{eqnarray*}
H_{F} &=&\frac{1}{2\pi }\int dx\int dx^{\prime }V\left(
x-x^{\prime }\right)
\partial _{x}\varphi \partial _{x^{\prime }}\varphi ; \\
H_{B} &=&\frac{1}{2}\left( \frac{1}{2\pi a}\right) ^{2}\int dx\int
dx^{\prime }V\left( x-x^{\prime }\right)\\
 &&\times\left\{ \exp \left[ 2i\sqrt{\pi }%
\varphi \left( x\right) \right] \exp \left[ -2i\sqrt{\pi }\varphi
\left( x^{\prime }\right) \right] e^{2ik_{F}\left( x-x^{\prime
}\right) }+H.c\right\} .
\end{eqnarray*}
In $H_{B}$, we neglected the terms that oscillate with $x,x^{\prime },$and $%
x+x^{\prime }$, and kept only those terms that oscillate with
$x-x^{\prime }. $ As our potential is sufficiently short-ranged,
the oscillations of the first group of terms will average out,
whereas the second group will survive. Introducing new coordinates
\begin{eqnarray*}
R &\equiv &\frac{x+x^{\prime }}{2}; \\
r &\equiv &x-x^{\prime },
\end{eqnarray*}
and assuming that $\left| R\right| \gg \left| r\right| ,$ the
forward-scattering part of the Hamiltonian reduces to
\begin{equation*}
H_{F}=\frac{1}{2\pi }\int dR\left( \partial _{R}\varphi \right)
^{2}\int drV\left( r\right) =\frac{V(0)}{2\pi }\int dR\left(
\partial _{R}\varphi \right) ^{2}.
\end{equation*}
The product of the two exponentials needs to be evaluated with
care. Applying the Baker-Hausdorff identity
\begin{equation*}
e^{A}e^{B}=:e^{A+B}:e^{\langle
AB-\frac{1}{2}A^{2}-\frac{1}{2}B^{2}\rangle },
\end{equation*}
we get
\begin{eqnarray*}
\exp \left[ 2i\sqrt{\pi }\varphi \left( x\right) \right] \exp \left[ -2i%
\sqrt{\pi }\varphi \left( x^{\prime }\right) \right] &=&  \exp \left[ 2i\sqrt{%
\pi }\left( \varphi \left( x\right) -\varphi \left( x^{\prime
}\right) \right) \right] : \\
&&\times \exp [4\pi \langle \varphi \left( x-x^{\prime }\right)
\varphi \left( 0\right) -\varphi ^{2}\left( 0\right) ],
\end{eqnarray*}
Using the expression for the free bosonic propagator
\begin{equation*}
\langle \varphi \left( x-x^{\prime }\right) \varphi \left(
0\right) -\varphi ^{2}\left( 0\right) ]=\frac{1}{4\pi }\ln
\frac{a^{2}}{(x-x^{\prime })^{2}},
\end{equation*}
and expanding in $r=x-x^{\prime }$ under the normal-ordering sign,
we obtain
\begin{equation*}
\exp \left[ 2i\sqrt{\pi }\varphi \left( x\right) \right] \exp \left[ -2i%
\sqrt{\pi }\varphi \left( x^{\prime }\right) \right]
=-\frac{1}{2}4\pi \left( \partial _{R}\varphi \right)
^{2}r^{2}\frac{a^{2}}{r^{2}}=-2\pi \left( \partial _{R}\varphi
\right) ^{2}a^{2}.
\end{equation*}
(While expanding, we neglected the first derivative of $\varphi $
which can be always eliminated by choosing appropriate boundary
condition.). $H_{B}$ reduces to
\begin{eqnarray*}
H_{B} &=&-\frac{1}{2}\left( \frac{1}{2\pi a}\right) ^{2}2\pi
a^{2}\int dR\left( \partial _{R}\varphi \right) ^{2}\int drV\left(
r\right) 2\cos
2k_{F}r \\
&=&-\frac{1}{2\pi }\int dR\left( \partial _{R}\varphi \right)
^{2}\int
drV\left( r\right) \cos 2k_{F}r \\
&=&-\frac{V\left( 2k_{F}\right) }{2\pi }\int dR\left( \partial
_{R}\varphi \right) ^{2}.
\end{eqnarray*}
Therefore, the bosonized form of the total Hamiltonian
\begin{equation*}
H_{int}=\frac{V\left( 0\right) -V\left( 2k_{F}\right) }{2\pi }\int
dR\left(
\partial _{R}\varphi \right) ^{2}
\end{equation*}
manifestly obeys the Pauli principle. The Luttinger-liquid
parameters are now given by
\begin{equation*}
u =\sqrt{1+\frac{V\left( 0\right) -V\left( 2k_{F}\right) }{2\pi
}}; \;
K =\frac{1}{\sqrt{1+\frac{V\left( 0\right) -V\left( 2k_{F}\right) }{2\pi }}%
}.
\end{equation*}

\end{document}